\documentclass[traditabstract,a4]{aa}  

\usepackage{graphicx}
\usepackage{txfonts}

\usepackage{natbib}
\usepackage{rotating}
\usepackage{graphicx}
\usepackage{xcolor}
\usepackage{graphicx}
\usepackage{xurl}

\usepackage{booktabs} 
\usepackage{array}
\usepackage{amsmath}

\usepackage{subcaption}

\usepackage{soul}

\usepackage[colorlinks=true,linkcolor=red,citecolor=blue, breaklinks=true]{hyperref}
\usepackage{hyperref}
\newcommand{\xsky}{E_{\rm x}^{\rm sky}}
\newcommand{\ysky}{E_{\rm y}^{\rm sky}}
\newcommand{\xload}{E_{\rm x}^{\rm load}}
\newcommand{\yload}{E_{\rm y}^{\rm load}}
\newcommand{\Bonce}{B_{\rm 2}^{\rm x}}
\newcommand{\Bcato}{B_{\rm 3}^{\rm x}}
\newcommand{\Bvuno}{B_{\rm 2}^{\rm y}}
\newcommand{\Bvcuatro}{B_{\rm 3}^{\rm y}} 

\newcommand{\Bunoa}{B_{\rm a}^{\rm x}} 
\newcommand{\Bdosa}{B_{\rm a}^{\rm y}} 
\newcommand{\Oxsky}{O_{\rm x}^{\rm s}}
\newcommand{\Oysky}{O_{\rm y}^{\rm s}}
\newcommand{\Oxload}{O_{\rm x}^{\rm l}}
\newcommand{\Oyload}{O_{\rm y}^{\rm l}}
\newcommand{\Osa}{O_{\rm a}^{\rm s}}
\newcommand{\Ola}{O_{\rm a}^{\rm l}}

\newcommand{\xskyc}{E_{\rm x}^{\rm sky*}}
\newcommand{\yskyc}{E_{\rm y}^{\rm sky*}}

\newcommand{\iload}{I_{\rm load}}
\newcommand{\qload}{Q_{\rm load}}
\newcommand{\uload}{U_{\rm load}}
\newcommand{\vload}{V_{\rm load}}

\newcommand{\isky}{I_{\rm sky}}
\newcommand{\qsky}{Q_{\rm sky}}
\newcommand{\usky}{U_{\rm sky}}
\newcommand{\vsky}{V_{\rm sky}}

\newcommand{\Iw}{\rm L^w}
\newcommand{\Iirf}{\rm L^{irf}}
\newcommand{\Ifhs}{\rm L^{fh}_s}
\newcommand{\Ifhl}{\rm L^{fh}_l}
\newcommand{\Iomtluno}{\rm L^{omt}_{l1}}
\newcommand{\Iomtsuno}{\rm L^{omt}_{s1}}
\newcommand{\Iomtsdos}{\rm L^{omt}_{s2}}
\newcommand{\Iomtldos}{\rm L^{omt}_{l2}}
\newcommand{\Ihxdos}{\rm L^{hyb}_{x2}}
\newcommand{\Ihxtres}{\rm L^{hyb}_{x3}}
\newcommand{\Ihydos}{\rm L^{hyb}_{y2}}
\newcommand{\Ihytres}{\rm L^{hyb}_{y3}}

\newcommand{\Rw}{\rm R^w}
\newcommand{\Rirf}{\rm R^{\rm irf}}
\newcommand{\Rfhs}{\rm R^{\rm fh}_s}
\newcommand{\Rfhl}{\rm R^{\rm fh}_l}
\newcommand{\Romtluno}{\rm R^{\rm omt}_{l1}}
\newcommand{\Romtsuno}{\rm R^{\rm omt}_{s1}}
\newcommand{\Rload}{\rm R^{\rm load}}
\newcommand{\Romtsdos}{\rm R^{\rm omt}_{s2}}
\newcommand{\Romtldos}{\rm R^{\rm omt}_{l2}}
\newcommand{\Rhxdos}{\rm R^{\rm hyb}_{x2}}
\newcommand{\Rhxtres}{\rm R^{\rm hyb}_{x3}}
\newcommand{\Rhydos}{\rm R^{\rm hyb}_{y2}}
\newcommand{\Rhytres}{\rm R^{\rm hyb}_{y3}}
\newcommand{\Rlnauno}{\rm R^{\rm lna}_1}
\newcommand{\Rlnados}{\rm R^{\rm lna}_2}
\newcommand{\Rlnatres}{\rm R^{\rm lna}_3}
\newcommand{\Rlnacuatro}{\rm R^{\rm lna}_4}

\newcommand{\SPOw}{\rm SPO^{\rm fh}_{s,w}}
\newcommand{\SPOirf}{\rm SPO^{\rm fh}_{s,irf}}
\newcommand{\SPOload}{\rm SPO^{\rm fh}_l}

\newcommand{\hw}{\rm h^{\rm w}}
\newcommand{\hirf}{\rm h^{\rm irf}}
\newcommand{\hfhs}{\rm h^{\rm fhs}}
\newcommand{\homtsuno}{\rm h^{\rm omt}_{s1}}
\newcommand{\hfhl}{\rm h^{\rm fhl}}
\newcommand{\homtluno}{\rm h^{\rm omt}_{l1}}
\newcommand{\hload}{\rm h^{\rm load}}
\newcommand{\homtsdos}{\rm h^{\rm omt}_{s2}}
\newcommand{\homtldos}{\rm h^{\rm omt}_{l2}}

\newcommand{\betahybxdos}{\rm  h^{\rm hyb}_{x2}}
\newcommand{\betahybxtres}{\rm  h^{\rm hyb}_{x3}}
\newcommand{\betahybydos}{\rm  h^{\rm hyb}_{y2}}
\newcommand{\betahybytres}{\rm  h^{\rm hyb}_{y3}}
\newcommand{\betalnauno}{\rm  h^{\rm lna}_1}
\newcommand{\betalnados}{\rm  h^{\rm lna}_2}
\newcommand{\betalnatres}{\rm  h^{\rm lna}_3}
\newcommand{\betalnacuatro}{\rm  h^{\rm lna}_4}

\newcommand{\betasky}{ \beta^{\rm sky}}
\newcommand{\betaload}{ \beta^{\rm load}}
\newcommand{\betaskyEffective}{\beta_{\rm sky}^{\rm eff}}
\newcommand{\betaloadEffective}{\beta_{\rm load}^{\rm eff}}

\newcommand{\betaaocho}{ \beta^{\rm a8}}
\newcommand{\betaanueve}{ \beta^{\rm a9}}
\newcommand{\betaadiez}{ \beta^{\rm a10}}

\newcommand{\Tbload}{\rm T^{a}_{load}}
\newcommand{\Tbsky}{\rm T^{a}_{sky}}
\newcommand{\Tbw}{\rm T^{a}_{w}}
\newcommand{\Tbirf}{\rm T^{a}_{irf}}
\newcommand{\Tbfhs}{\rm T^{a}_{fhs}}
\newcommand{\Tbomtsuno}{\rm T^{a}_{omt_{s1}}}
\newcommand{\Tbfhl}{\rm T^{a}_{fhl}}
\newcommand{\Tbomtluno}{\rm T^{a}_{omt_{l1}}}
\newcommand{\Tbomtsdos}{\rm T^{a}_{omt_{s2}}}
\newcommand{\Tbomtldos}{\rm T^{a}_{omt_{l2}}}

\newcommand{\Tbhybxdos}{\rm T^{a}_{hyb_{x2}}}
\newcommand{\Tbhybxtres}{\rm T^{a}_{hyb_{x3}}}
\newcommand{\Tbhybydos}{\rm T^{a}_{hyb_{y2}}}
\newcommand{\Tbhybytres}{\rm T^{a}_{hyb_{y3}}}
\newcommand{\Tblnauno}{\rm T^{a}_{lna_{1}}}
\newcommand{\Tblnados}{\rm T^{a}_{lna_{2}}}
\newcommand{\Tblnatres}{\rm T^{a}_{lna_{3}}}
\newcommand{\Tblnacuatro}{\rm T^{a}_{lna_{4}}}

\newcommand{\TbAuno}{\rm T^{a}_{A_1}}
\newcommand{\TbAdos}{\rm T^{a}_{A_2}}
\newcommand{\TbBuno}{\rm T^{a}_{B_1}}
\newcommand{\TbBdos}{\rm T^{a}_{B_2}}
\newcommand{\TbCuno}{\rm T^{a}_{C_1}}
\newcommand{\TbCdos}{\rm T^{a}_{C_2}}
\newcommand{\TbDuno}{\rm T^{a}_{D_1}}
\newcommand{\TbDdos}{\rm T^{a}_{D_2}}

\newcommand{\Tnlnauno}{\rm N_1}
\newcommand{\Tnlnados}{\rm N_2}
\newcommand{\Tnlnatres}{\rm N_3}
\newcommand{\Tnlnacuatro}{\rm N_4}

\newcommand{\Tw}{\rm T_{w}}
\newcommand{\Tirf}{\rm T_{irf}}
\newcommand{\Tfhs}{\rm T_{fhs}}
\newcommand{\Tomts}{\rm T^{omt}_{s}}
\newcommand{\Tfhl}{\rm T_{fhl}}
\newcommand{\Tomtl}{\rm T^{omt}_{l}}
\newcommand{\Thybx}{\rm T^{hyb}_x}
\newcommand{\Thyby}{\rm T^{hyb}_y}

\newcommand{\Tload}{\rm T_{load}}
\newcommand{\Tsky}{\rm T_{sky}}
\newcommand{\Tenvuno}{\rm T_{env1}}
\newcommand{\Tenvdos}{\rm T_{env2}}
\newcommand{\Tcryuno}{\rm T_{cryo1}}
\newcommand{\Tcrydos}{\rm T_{cryo2}}
\newcommand{\Troom}{\rm T_{room}}

\newcommand{\TbBEM}{\rm T^{a}_{BEM}}
\newcommand{\RBEM}{\rm R^{BEM}_{amp_k}}
\newcommand{\TnBEMuno}{\rm N^{BEM}_{C_1}}

\newcommand{\TnBEMtres}{\rm N^{BEM}_{D_1}}

\newcommand{\Tbmixer}{\rm T^{a}_{mixer}}

\newcommand{\hbemfilterk}{\rm h^{BEM}_{k}}
\newcommand{\hdcfilterk}{\rm h^{DC}_{k}}

\newcommand{\hbemfilteruno}{\rm h^{BEM}_{C_1}}

\newcommand{\hbemfiltertres}{\rm h^{BEM}_{D_1}}

\newcommand{\Tbemfilter}{\rm T^{BEM}_{filter_k}}
\newcommand{\Ibemfilter}{\rm L^{BEM}_{filter_k}}
\newcommand{\Rbemfilter}{\rm R^{BEM}_{filter_k}}

\newcommand{\TbDC}{\rm T^{a}_{DC}}
\newcommand{\RDC}{\rm R^{DC}_{amp_k}}
\newcommand{\hDCfilteruno}{\rm h^{DC}_{filter_k}}
\newcommand{\TDCfilter}{\rm T^{DC}_{filter_k}}
\newcommand{\IDCfilter}{\rm L^{DC}_{filter_k}}
\newcommand{\RDCfilter}{\rm R^{DC}_{filter_k}}
\newcommand{\TnDCMuno}{\rm N^{DC}_{C_1}}

\newcommand{\TnDCMtres}{\rm N^{DC}_{D_1}}

\newcommand{\Tnsky}{\rm N^{s}}
\newcommand{\Tnload}{\rm N^{l}}
\newcommand{\Tneff}{\rm T_{\rm noise}^{\rm eff}}

\newcommand{\Toffeff}{\rm T_{offset}^{eff}}

\newcommand{\betaampBEMuno}{\rm  h^{\rm BEM}_{amp,k}}
\newcommand{\betaampDCuno}{\rm  h^{\rm DC}_{amp,k}}
\newcommand{\betafilterBEMk}{\rm  h^{\rm BEM}_{filter,k}}
\newcommand{\betafilterDCk}{\rm  h^{\rm DC}_{filter,k}}

\newcommand{\ToffFEMuno}{ \rm T^{FEM,1}_{off,k}}
\newcommand{\ToffFEMdos}{ \rm T^{FEM,2}_{off,k}}
\newcommand{\ToffBEM}{\rm T^{BEM}_{off,k}}
\newcommand{\ToffDC}{\rm T^{DC}_{off,k}}
\newcommand{\zetaFEM}{ \rm \zeta^{FEM}_k}
\newcommand{\zetaBEM}{ \rm \zeta^{BEM}_k}
\newcommand{\zetaDC}{ \rm \zeta^{DC}_k}

\newcommand{\zetaBEMuno}{ \rm \zeta^{BEM}_{C_1}}

\newcommand{\zetaBEMtres}{ \rm \zeta^{BEM}_{D_1}}

\newcommand{\zetaDCuno}{ \rm \zeta^{DC}_{C_1}}

\newcommand{\zetaDCtres}{ \rm \zeta^{DC}_{D_1}}

\newcommand{\Tofftotuno}{ \rm T^{tot,1}_{off,k}}
\newcommand{\Tofftotdos}{ \rm T^{tot,2}_{off,k}}

\newcommand{\GBEMuno}{ \rm G^{BEM}_{C_1}}

\newcommand{\GBEMtres}{ \rm G^{BEM}_{D_1}}

\newcommand{\GBEMk}{ \rm G^{BEM}_{k}}

\newcommand{\GFEMk}{ \rm G^{FEM}_{k}}
\newcommand{\GDCk}{ \rm G^{DC}_{k}}
\newcommand{\GDCuno}{ \rm G^{DC}_{C_1}}

\newcommand{\Textcry}{\rm T^{ext}_{cry}}
\newcommand{\TBEMfilter}{\rm T^{BEM}_{filter_k}}
\newcommand{\TafterfiltBEM}{\rm T^{BEM}_{filter}}
\newcommand{\TafteramplDC}{\rm T^{DC}_{amp_k}}
\newcommand{\TFPGA}{\rm T^{FPGA}}

\begin{document}

   \title{Systematic errors in spectral measurements with the Tenerife Microwave Spectrometer}

   \subtitle{}

   \author{A. M. Arriero-Lopez
          \inst{1,2}
          \and
          J. A. Rubiño-Martín\inst{1,2}
          \and
          F. Cuttaia\inst{3}
          \and
          L. Terenzi\inst{3}
          \and
          R. Hoyland\inst{1,2}
          }

   \institute{Instituto de Astrofisica de Canarias (IAC), E-38200 La Laguna, Tenerife, Spain
         \and
             Departamento de Astrofísica, Universidad de La Laguna, E-38206 La Laguna, Tenerife, Spain
        \and
            Istituto Nazionale di Astrofisica, Via Piero Gobetti 93, 40129 Bologna, Italy
             }

   \date{Received month day, year; accepted month day, year}


\abstract{We present an analytical instrument model of the TMS radiometer, a pseudo-correlation system designed for absolute sky-temperature measurements through a continuous comparison between sky and reference-load signals. The goal of this work is to quantify and understand the impact of instrumental non-idealities that are intrinsic to absolute radiometric measurements.
We used a combination of the Jones-matrix formalism, to describe non-ideal signal mixing effects in the different instrument components, and the Friis formalism, to account for noise-temperature contributions which introduce specific offset terms. The model includes all components from the cryostat entrance window onwards and allows us to propagate realistic losses, return losses, and noise figures through the system. We find that non-ideal mixing effects, such as intensity-to-polarization and sky-to-load leakage, are expected to appear at the percent level, but they could be in principle calibrated out. The total TMS intensity response exhibits a frequency-dependent offset with a band-averaged level of 6.9\,K, dominated by the cryostat window, and with smaller contributions from the infrared filter and orthomode transducers. 
Under the required TMS thermal stability of 1\,mK over one hour, the resulting variation in the output signal can reach up to 91.3\,$\mu$K, which sets the fundamental limit on the absolute sky-temperature accuracy. In contrast, relative spectral measurements across the 10--20\,GHz band are stable at the few-microkelvin level, consistent with the instrument target sensitivity of $\sim 10$\,Jy/sr. This work provides a detailed and quantitative assessment of the systematic effects affecting absolute radiometric measurements and establishes a robust framework for the calibration and performance optimization of the TMS instrument.}


   \maketitle
%

\section{Introduction}

Measurements of the Cosmic Microwave Background (CMB) constitute one of the most powerful tools in modern cosmology for testing our understanding of the Universe. Over the past decades, observational studies of CMB anisotropies carried out by various experiments have led to the robust establishment of the $\Lambda$CDM concordance model \citep[e.g.][]{WMAP9, Planck2020-VI}. 
In recent years, significant efforts have been devoted to characterizing the CMB polarization anisotropies, as they offer a new window into the inflationary phase of the Universe, which occurred approximately $10^{-35}$\,s after the Big Bang.
Experiments such as BICEP/Keck \citep{BICEPKeck2021}, ACT \citep{ACT-DR4}, SPT \citep{SPT-3G}, CLASS \citep{CLASS2024}, GroundBIRD \citep{Groundbird}, Simons Observatory \citep{SO}, or QUIJOTE \citep{QUIJOTEMFI4} are actively searching for the signature of primordial gravitational waves in CMB polarization maps. A new generation of experiments is also planned for the coming years, including ground-based facilities such as QUBIC \citep{QUBIC}, LSPE-STRIP \citep{2012SPIE.8446E..7CB}, and CMB-S4 \citep{CMB-S4}; balloon-borne missions such as LSPE-SWIPE \citep{LSPE-JCAP}; and space-based missions like LiteBIRD \citep{LiteBIRD_ptep}.

On the other hand, a wealth of largely unexplored information is also encoded in the spectral distortions of the CMB \citep{SilkChluba2014}. Since the landmark measurement of the CMB spectrum by COBE/FIRAS \citep{FixsenMather2002}, only a few spectral observations have been conducted by ground-based and balloon-borne experiments \citep[e.g.][]{1994ApJ...424..517B, 1996ApJ...458..407S,TRIS_Zannoni2008, arcade2measurementat3}. Among them, ARCADE2 stands out, having reported a still unexplained residual signal with a synchrotron-like spectrum \citep{Seiffert2011}.
New spacecraft proposals, such as PIXIE \citep{2016SPIE.9904E..0WK, 2025JCAP...04..020K} and FOSSIL\footnote{An M-class mission concept (PI: N. Aghanim) that recently advanced to Stage 2 in the M8 mission selection process as envisioned for the ESA Voyage 2050 space program.} \citep{2021ExA....51.1515C}, show great promise for detecting the largest spectral distortion signals in the coming decades, potentially improving the sensitivity limits set by FIRAS by up to three orders of magnitude.
Meanwhile, a few observational efforts are expected to provide new measurements in the near future, including L-BASS \citep{L-BASS}, COSMO \citep{COSMO}, and BISOU \citep{BISOU2024}.

The Tenerife Microwave Spectrometer (TMS) is a ground-based experiment aimed at detecting absolute spectral distortions of the sky in the 10--20\,GHz frequency range \citep{TMS2020} from the Teide Observatory (Tenerife). It is designed to measure potential deviations from a pure blackbody spectrum at the level of $\sim 10$\,Jy/sr, and to provide a precise characterization of the spectral properties of average Galactic synchrotron (including the ARCADE2 excess) and anomalous microwave emissions. As a byproduct, TMS will also deliver an absolute calibration scale for the QUIJOTE MFI \citep{QUIJOTEMFI4} and MFI2 \citep{MFI2} instruments, which operate over the same frequency range.
TMS is now in its final implementation phase, and first light is expected by the end of 2026.
\\
TMS employs a pseudo-correlation receiver architecture similar to that of the Low Frequency Instrument (LFI) onboard the Planck satellite \citep{PlanckSeiffert_2002}, which has been shown to be highly robust against $1/f$ gain variations.
TMS will observe the sky by comparing its signal to that of a stable reference load (or blackbody calibrator) cooled to cryogenic temperatures of approximately 6\,K \citep{Paz1_10.1117/12.2561353}.
The instrument imposes stringent requirements on its thermal design in order to achieve its scientific objectives. In particular, the cold structure must maintain a spatial temperature homogeneity of $\pm100$\,mK, and a temporal stability better than $\pm 1$\,mK/h.
\\   
In this paper, we present a detailed study of potential systematic errors affecting TMS measurements, covering two aspects. We first develop an analytical model of the TMS response using the Jones matrix formalism, following the approach of previous works \citep{PlanckSeiffert_2002, Hu_2003, ODea2007MNRAS.376.1767O, BischoffPhDT, CarlosLopezPhDT, Mennella2026}. 
This model is then used to characterize the impact of systematic effects arising from the non-ideal behavior of the various components in the radiometer chain on the measurements of the Stokes parameters; in particular, we do not discuss here imperfections due to non-ideal beam of the instrument. 
Second, and focusing on intensity measurements only, we extend the previous model by studying the expected contributions to the measured signal arising from return losses and insertion losses in the different components of the radiometer chain, using the Friis equations formalism. We provide specific predictions of the resulting offsets and non-ideal responses in the detected brightness temperature based on a simplified model of the TMS thermal behavior.
\\
Taken together, the results of these two analyses define a basic TMS instrument model that will be used to design and test the future TMS calibration strategy.
The paper is organized as follows. Section 2 describes the physical components of the TMS instrument. Sections 3 and 4 show the analytic model of TMS in terms of the Jones matrices formalism, indicating the expected level of non-idealities. Section 5 extends this model for the intensity signal, by considering noise temperature effects. Section 6 provides the overall TMS response, and discusses its time stability. Section 7 aims to show some key aspects to be considered in the design of the TMS calibration strategy. Finally, section 8 show the conclusions and discussion of the paper.

\section{Tenerife Microwave Spectrometer model} \label{TMSinstrument}

\subsection{Overall spectrometer design}

The TMS front-end module (FEM) uses a pseudo-correlator architecture based on that of Planck's Low Frequency Instrument (LFI), serving as the core design for its two radiometric chains \citep{2009JInst...4T2006V, 2010A&A...520A...4B}. 
One radiometric chain receives the incoming sky signal, while the second processes the signal from the reference Cold Load cooled to approximately 6\,K. Both signals are later correlated using two $180^\circ$ hybrids, ensuring a stable and balanced comparison between the sky and the reference load.
Each TMS radiometric chain amplifies the signals using four High Electron Mobility Transistor (HEMT)-based ultra-low-noise amplifiers (LNAs).
After amplification, the frequency band generated by the heterodyne correlator is divided into sub-bands, which are then used to compute the Stokes parameters of both the sky and the reference load \citep{Alonso-Arias_2024}.
The frequency division and Stokes parameters calculation is done with some filters and mixers placed in the backend module (BEM), the Down-converter (DC), and by an ultrafast Field-Programmable Gate Array or FPGA.
We note that TMS will have some polarimetric capabilities, 
although the instrument is not optimized for this purpose.

\begin{figure}
   \centering
   \includegraphics[width=7cm]{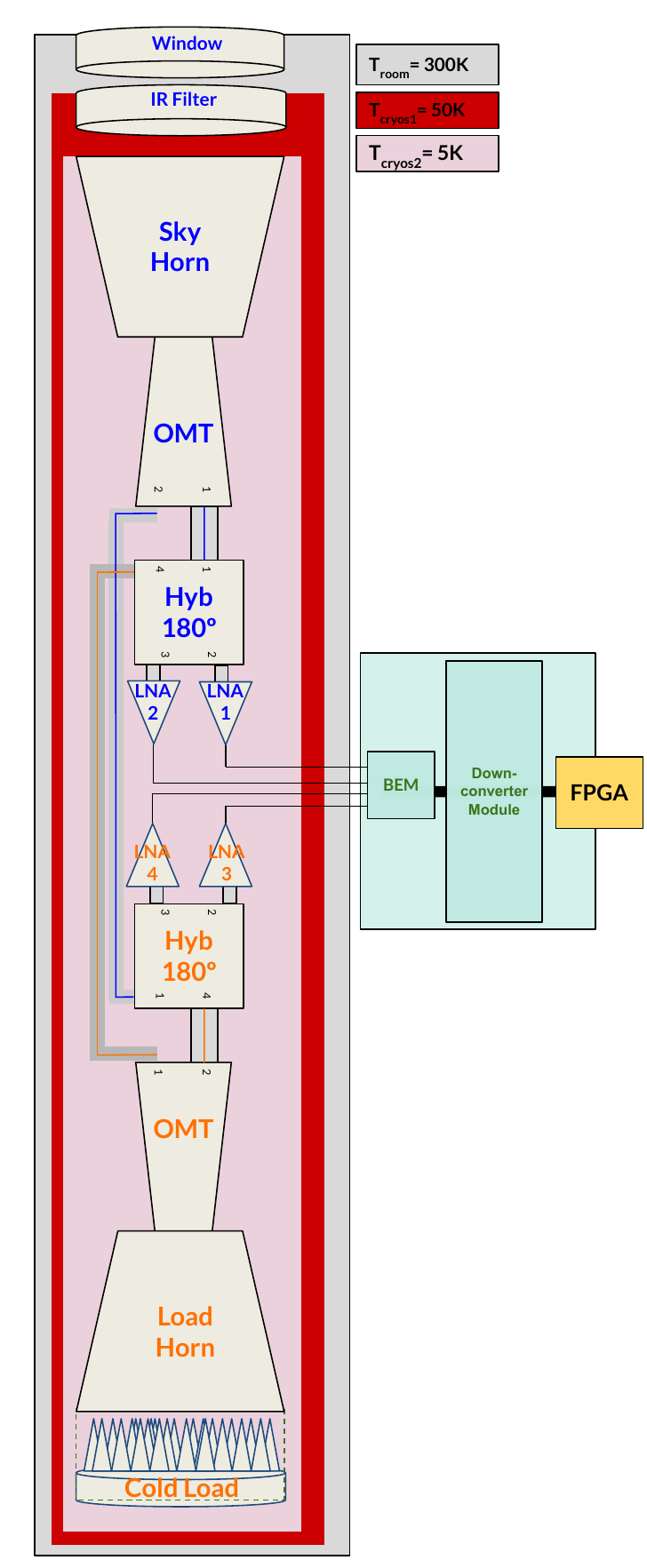}
  	\caption{Distribution of the optical, passive, and active RF components of the TMS inside the cryostat front-end module. Gray indicates room temperature (300 K), red the first cryostat stage at 50 K, and pink the second cryostat stage at 5 K. Green denotes the back-end module (BEM) and down-converter (DC), while yellow represents the digital processing inside the FPGA.}
	\label{Diagrama_cryos}
\end{figure}

Here, we provide a more detailed description of the optical and radio-frequency (RF) components of TMS, along with the internal configuration of the different elements.
The instrument consists of the cryostat window (hereafter labeled as $\rm W$), the infrared filter ($\rm IRF$), two feedhorns (FH), two orthomode transducers (OMTs), two $180^{\circ}$ hybrid combiners (H), the blackbody calibrator (hereafter referred to as Cold Load (CL), and four low-noise amplifiers (LNAs). Figure~\ref{Diagrama_cryos} shows the distribution of the optical and radio-frequency components within the cryostat.
The experimental setup operates at three different temperature stages: ambient temperature ($\Textcry \approx 300$\,K), and two cryogenic stages, a first stage at $\Tcryuno \approx 50$\,K, and the second stage at $\Tcrydos \approx 5$\,K.

\begin{figure*}
    \resizebox{\hsize}{!}
    {\includegraphics[scale=1.5]{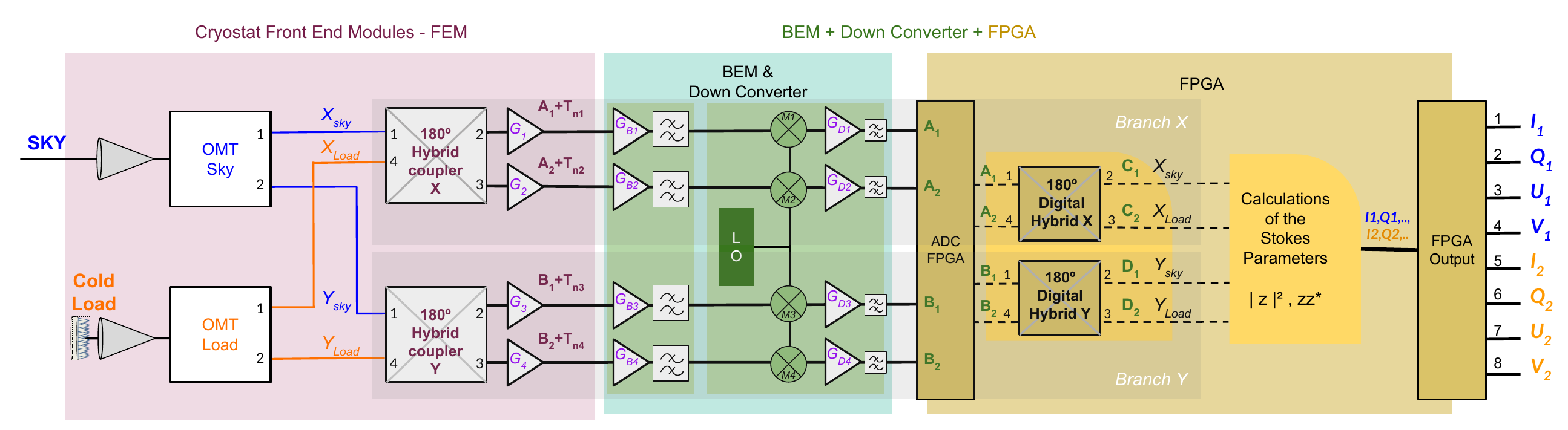}}
    \sidecaption
    \caption{ RF-schematic of the TMS instrument. The instrument is divided in four main systems: 
    Front-End (in pink), and Back-End and Down-Converter ( in green), and the FPGA (in yellow) where the stokes parameters are calculated. 
    Cable-induced phase shifts and losses between FEM and BEM are neglected. 
    }
    \label{new_model_gains}
\end{figure*}

Figure~\ref{new_model_gains} shows the propagation of the incoming RF signal through the two independent radiometric chains of TMS.
The sky chain is coupled to a feed-horn that receives radiation from the sky, which enters through the cryostat window and is subsequently guided along the RF chain.
The electric field of the incoming sky signal is decomposed into its orthogonal components, $\xsky$ and 
$\ysky$, by an OMT. These components are then independently routed to the input ports of the two $180^{\circ}$ hybrid combiners.
On the other hand, the load chain collects radiation from the internal calibrator system (CL), whose brightness temperature is expected to follow a blackbody spectrum determined solely by its physical temperature \citep{alonsoarias2023microwave}, with sub-microkelvin precision.
The signal from the CL is collected by the load feed-horn, which is connected to an OMT that separates the electric field into its orthogonal components, $\xload$ and $\yload$.
These components are then fed into a $180^{\circ}$ hybrid combiner, as in the sky chain.
At the end of the FEM, the four outputs from the two $180^{\circ}$ hybrids are connected to four LNAs.

At this stage, the signals undergo their first amplification and are subsequently conditioned for further processing, which includes additional amplification, filtering, down conversion to the base band, and sub-band division by the Back-End Module (BEM) and the Down-Converter (DC).
These steps are necessary to prepare the signals for digital processing within the FPGA, as illustrated in Figure~\ref{new_model_gains} (highlighted in dark yellow).
The FPGA module plays a key role in the overall performance of the TMS instrument, as it is responsible for executing three core processing tasks:
\begin{enumerate}
\item[i)] Ultra-fast data acquisition. 
\item[ii)] Digital implementation of a $180^{\circ}$ hybrid. A second hybrid stage is digitally programmed in the FPGA, positioned after the amplification and heterodyne detection stages. Unlike its analog counterpart, this digital hybrid introduces no losses.
\item[iii)] Computation of the Stokes parameters. After signal combination through the digital hybrid, the FPGA performs the necessary mathematical operations to compute the Stokes parameters, enabling the characterization of the polarization state of the incoming radiation.
\end{enumerate}
TMS will employ Xilinx ZCU208 UltraScale boards, which provide up to eight input channels with a bandwidth of 2.5\,GHz each. To cover the full 10--20\,GHz frequency range, the data acquisition system is designed to handle 16 output channels. For simplicity, Fig.~\ref{new_model_gains} shows only four of those outputs ($A_1$, $A_2$, $B_1$, and $B_2$), corresponding to a single 2.5\,GHz sub-band. This reduced configuration is sufficient to provide a representative description of the full TMS system.

\subsection{Analytic model of TMS}\label{Analyticmodel}

The electric field $\Vec{E}$ of a monochromatic electromagnetic wave propagating along the $\Vec{\hat{z}}$ direction can be decomposed into orthogonal components along the $\Vec{\hat{x}}$ and $\Vec{\hat{y}}$ axes.
This resulting field can be represented as a two-element Jones vector, $\Vec{E}_{\rm in}$, whose components describe the relative amplitudes and phases of the electric field in the $\Vec{\hat{x}}$ and $\Vec{\hat{y}}$ directions.
To model the effect of a medium on polarized radiation, Jones matrices \textbf{J} are employed as follows:

\begin{equation}
     \vec{E}_{\rm out}= \mathbf{J} \vec{E}_{\rm in} 
     \label{eq:b1_p_3}
\end{equation}

where $\vec{E}_{\rm out}$ is the resulting electric field after the incoming field $\vec{E}_{\rm in}$ passes through an optical element described by the Jones matrix \textbf{J}.
Hereafter, for TMS the incoming radiation from the sky, $\vec{E}_{\rm in}$, will be represented by the Jones vector $\vec{E}_{\rm sky}$, with components $(\xsky, \ysky)$; likewise, the radiation from the load will be denoted as $\vec{E}_{\rm load}$, with components $(\xload, \yload)$.

Stokes parameters are defined as linear combinations of the power measured in orthogonal polarization states \citep[e.g.][]{heiles}.
In this paper, we adopt the following conventions: 
\begin{align}
\label{eq:stokes_s}
\isky &= \left< \xsky \xskyc \right>+\left< \ysky \yskyc \right> \\
\qsky &= \left< \xsky \xskyc \right> - \left<\ysky \yskyc \right>\\
\usky &= \left< \xsky \yskyc \right>+ \left< \ysky \xskyc \right>\\
\vsky &= -i \left[\left< \xsky \yskyc \right> - \left<\ysky \xskyc \right> \right]
\end{align}
for the four Stokes parameters of the sky signal. 
Similar equations can be written for the load signal.
A Mueller vector $\mathcal{S}$, whose four elements correspond to the Stokes parameters $(I, Q, U, V)$, can be obtained as follows:
\begin{align}
\label{s_mull_sky}
\mathcal{S}_{\rm sky} &= (\isky,\qsky,\usky,\vsky ) \\
\mathcal{S}_{\rm load} &= (\iload,\qload,\uload,\vload ).
\end{align}

Using the Jones matrix formalism, the TMS instrument can be modeled by multiplying the individual matrices corresponding to its different components.
Following the schematics shown in Figs.~\ref{Diagrama_cryos} and \ref{new_model_gains}, the sky signal first pass through the cryostat window, the infrared filter and the (sky) horn, which we collectively represent by an attenuation matrix $\mathbf{J}_{{\rm att},s}$. Similarly, the load signal pass through the (load) horn, $\mathbf{J}_{{\rm att},l}$.
After these components (Fig.~\ref{new_model_gains}), the sky and load signals are separated into two orthogonal electric-field components by the OMTs. The resulting Jones vectors at the output of these elements are thus given by
\begin{align}
\Vec{E}_{\rm s}= \mathbf{J}_{\rm omt,s}\, \mathbf{J}_{{\rm att},s}\, \Vec{E}_{\rm sky},
   \hspace{0.5cm} & \Vec{E}_{\rm l}= \mathbf{J}_{\rm omt,l}\,\mathbf{J}_{{\rm att},l}\,\Vec{E}_{\rm load}.
   \label{output_omt}
\end{align}
where the subscripts ${\rm s}$ and ${\rm l}$ refer to the sky and load branches, respectively.
Next, these signals are fed into the hybrid couplers: the $\vec{x}$ components of the sky and load signals are routed to the X-hybrid, while the corresponding $\vec{y}$ components are directed to the Y-hybrid.
This signal routing and the subsequent operations can be described by the Jones vectors $\vec{E}_1$ and $\vec{E}_2$, defined as follows:
\begin{align}
     \Vec{E}_{1}= \mathbf{J}_{\rm BD,x}\, \mathbf{J}_{\rm lna,x}\, \mathbf{J}_{\rm hyb,x} \begin{pmatrix}
         \Vec{E}_{\rm s}\,\vec{\hat{x}}\\
         \Vec{E}_{\rm l}\,\vec{\hat{x}}
    \end{pmatrix}
    \equiv \begin{pmatrix}
         A_1\\
         A_2
    \end{pmatrix}
    \label{eq:A1A2}
\end{align}
\begin{align}
     \Vec{E}_{2}= \mathbf{J}_{\rm BD,y}\, \mathbf{J}_{\rm lna,y}\, \mathbf{J}_{\rm hyb,y} \begin{pmatrix}
         \Vec{E}_{\rm s}\,\vec{\hat{y}}\\
         \Vec{E}_{\rm l}\,\vec{\hat{y}}
    \end{pmatrix}
    \equiv\begin{pmatrix}
         B_1\\
         B_2
    \end{pmatrix}
    \label{eq:B1B2}
\end{align}
The resulting outputs $\rm A_1$, $\rm A_2$, and $\rm B_1$, $\rm B_2$ represent the sky and load signals after propagation through the hybrid combiners, LNAs, BEM, and DC stages, with the combined effect of the latter two denoted here as BD.

Finally, the $\vec{E}_1$ and $\vec{E}_2$ signals serve as the input to the FPGA module, highlighted in dark yellow in Figure~\ref{new_model_gains}.
As previously mentioned, this module implements an ideal digital $180^{\circ}$ hybrid, represented by the Jones matrices $\mathbf{J}_{\rm{hyb,x}}^{\rm id}$ and $\mathbf{J}_{\rm{hyb,y}}^{\rm id}$ 
for the two branches. Its mathematical form is given by
\begin{align}
        \mathbf{J}_{\rm hyb,k}^{\rm id}=\frac{1}{\sqrt{2}}
    \begin{pmatrix}
    1 & 1 \\ 
    1 & -1  
    \end{pmatrix}.
\label{ideal_jones1}
\end{align}
These digital hybrids correlate the signals from ports $\rm A_1$, $\rm A_2$, and $\rm B_1$, $\rm B_2$, yielding the output vectors $\vec{E}_3$ and $\vec{E}_4$, as follows:

\begin{align}
    \Vec{E}_{3}= \mathbf{J}_{\rm{hyb,x}}^{\rm id}\Vec{E}_{1}\equiv\frac{1}{\sqrt{2}}
    \begin{pmatrix}
        A_1+A_2\\
        A_1-A_2
    \end{pmatrix}=\begin{pmatrix}
         C_1\\
         C_2
    \end{pmatrix},
\label{output_C1}    
\end{align}
\\
\begin{align}
     \Vec{E}_{4}= \mathbf{J}_{\rm{hyb,y}}^{\rm id}\Vec{E}_{2}\equiv\frac{1}{\sqrt{2}}\begin{pmatrix}
        B_1+B_2\\
        B_1-B_2
    \end{pmatrix}=\begin{pmatrix}
         D_1\\
         D_2
    \end{pmatrix}.
    \label{output_D1}
\end{align}
At this stage, the signals have already been amplified, and under ideal conditions, the final outputs $\rm C_1$, $\rm C_2$, and $\rm D_1$, $\rm D_2$ are obtained.

\subsection{TMS ideal case: balanced and lossless spectrometer}\label{TMS ideal case}

In the case of a perfectly balanced polarimeter with ideal lossless components, and assuming perfect balanced LNAs and gains equal to 1, the TMS analytic model provides the following results for the Jones vectors: 
\begin{align}
     \Vec{E}_{1}=\begin{pmatrix}
         A_1\\
         A_2
    \end{pmatrix} = \frac{1}{\sqrt{2}}\begin{pmatrix}
         \xsky +\xload \\
         \xsky-\xload
    \end{pmatrix}
    \label{output_A1_tms}
\end{align}
\begin{align}
     \Vec{E}_{2} =\begin{pmatrix}
         B_1\\
         B_2
    \end{pmatrix}=\frac{1}{\sqrt{2}}\begin{pmatrix}
         \ysky+\yload \\
         \ysky-\yload
    \end{pmatrix} 
    \label{output_B1_tms}
\end{align}
\begin{align}
  \Vec{E}_{3}=\begin{pmatrix}
         C_1\\
         C_2
    \end{pmatrix}= \begin{pmatrix}
         \xsky \\
         \xload 
\end{pmatrix},
    \hspace{0.4cm}   & \Vec{E}_{4}=\begin{pmatrix}
         D_1 \\
         D_2 
    \end{pmatrix}= \begin{pmatrix}
         \ysky \\
         \yload 
\end{pmatrix}.
    \label{output_D1_tms}
\end{align}
Thus, in the ideal case, the four outputs directly measure the components of the electric fields from the sky and the load. 

Accordingly, the Stokes parameters of the sky signal for the ideal case can be expressed as follows:
\begin{align}
\nonumber
I_{1}&= \left< C_1 C_1^* \right>+\left< D_1 D_1^* \right>\\
\nonumber
Q_{1}&= \left< C_1 C_1^* \right> - \left< D_1 D_1^* \right>\\
\nonumber
U_{1} &= \left< C_1 D_1^* \right>+ \left< D_1 C_1^*\right>\\
V_{1} &= -i\left[\left< C_1 D_1^* \right>- \left< D_1 C_1^* \right> \right]
 \label{sky_stokes_theory}
\end{align}
and for the load signal:
\begin{align}
I_{2} & = \left< C_2 C_2^* \right>+\left< D_2 D_2^* \right> \nonumber\\
Q_{2} &= \left< C_2 C_2^* \right> - \left< D_2 D_2^* \right>\nonumber\\
U_{2} &= \left< C_2 D_2^* \right>+ \left< D_2 C_2^*\right>\nonumber\\
V_{2} & = -i\left[\left< C_2 D_2^* \right>- \left< D_2 C_2^* \right> \right].
 \label{load_stokes_theory}
\end{align}
For completeness, we may define the associated Mueller vectors for TMS as $\mathcal{S}_{1}=(I_1,Q_1,U_1,V_1 )$ and $\mathcal{S}_{2}=(I_2,Q_2,U_2,V_2 )$. 
We note that for this ideal case, we recover $\mathcal{S}_1 = \mathcal{S}_{\rm sky}$ and $\mathcal{S}_2 = \mathcal{S}_{\rm load}$, as expected.
Although these equations show that, in principle, all four Stokes parameters can be recovered, we emphasize that TMS is not optimized for polarimetric measurements. In particular, the estimation of Stokes $U$ and $V$ is expected to be affected by $1/f$ noise.

%

%

\section{TMS realistic case: including receiver errors}

The equations shown in previous section \ref{TMS ideal case} demonstrate that with a lossless and perfectly balanced spectrometer, we may obtain a clean measurement of the sky and load Stokes parameters with TMS.

Here, we present how non-idealities in the radio-frequency components affect the determination of the Stokes parameters. We employ the Jones matrix formalism \citep[e.g.][]{ODea2007MNRAS.376.1767O} to model deviations from ideal behavior in the window and infrared (IR) filters, feedhorns, OMTs, hybrids, and amplifiers.
The parametric description of all the non-ideal effects considered in this work is summarized in the following set of equations:
\begin{align}
    \mathbf{J}_{{\rm att},k}=
     \begin{pmatrix}
        A_{\rm x} & 0 \\
        0 & A_{\rm y} e^{i\phi_k}
    \end{pmatrix}    
\label{non-ideal_jones0}
\end{align}
\begin{align}
     \mathbf{J}_{{\rm omt},k}=
     \begin{pmatrix}
        (1+O^k_{\rm x}) & \vspace{0.3cm}  O^k_{\rm a}\, e^{i(\theta_2^k+\theta_3^k)}  \\
         O^k_{\rm a}\, e^{i\theta_2^k} &  (1+O^k_{\rm y})\, e^{i\theta_3^k}
    \end{pmatrix}   
\label{non-ideal_jones1}
\end{align}
\begin{align}
    \mathbf{J}_{{\rm hyb},k}=\frac{1}{\sqrt{2}}
    \begin{pmatrix}
    (1+B^k_2)e^{i\beta^k_1} & \vspace{0.3cm} (1+B^k_{\rm a})e^{i(\beta^k_{\rm a}+\beta^k_2)} \\ 
         e^{i(\beta^k_{\rm a}+\beta^k_1)}(1+B^k_{\rm a}) & - e^{i\beta^k_2}(1+B^k_3)  
    \end{pmatrix}   
\label{non-ideal_jones2}
\end{align}
\begin{align}
  \mathbf{J}_{{\rm lna},k}=
     \begin{pmatrix}
       g_i & 0 \\
        0 & g_{i+1}\, e^{i\psi_{i+1}}
    \end{pmatrix},
    \hspace{0.4cm}   & \mathbf{J}_{{\rm BD},k}=
     \begin{pmatrix}
        g^{\rm BD}_{i} & 0 \\
        0 & g^{\rm BD}_{i+1}\, e^{i\phi_{i+1}}
    \end{pmatrix}.  
\label{non-ideal_jones3}
\end{align}
Table~\ref{table:components} provides a detailed description of the parameters used in these equations, as well as the possible values of the $k$ and $i$ indices. 
We also account for the thermal noise contribution of the different LNAs in the FEM and the BEM by adding these Jones vectors to the input RF signal:
\begin{align}
  \mathbf{n}_{{\rm lna},k}=
     \begin{pmatrix}
       n_i \\
        n_{i+1} 
    \end{pmatrix}, \hspace{0.4cm}   & \mathbf{n}_{{\rm BD},k}= 
     \begin{pmatrix}
         n^{\rm BD}_{i} \\
        n^{\rm BD}_{i+1} 
    \end{pmatrix}.
\label{non-ideal_jones4}
\end{align}
When averaged during the detection process, the thermal noise terms produce an offset noise temperature contribution, which we represent here as $N_k \equiv \langle n_k^2 \rangle$ and $N^{\rm BD}_k \equiv \langle (n^{\rm BD}_k)^2 \rangle$.

%
Each component may be associated with one or more non-idealities, including attenuation, insertion loss (L) and return (R) losses, phase shifts, cross-polarization (XPD) and isolation (ISO). Regarding the cross-polarization terms in the OMTs ($O_{\rm a}^k$) and the isolation terms in the hybrids ($B^k_{\rm a}$), we adopt the simplifying assumption that their effect is identical (in both amplitude and phase) for the two orientations \citep[e.g.][]{CarlosLopezPhDT}.
%
Note that in the ideal case, equations~\ref{non-ideal_jones0}, \ref{non-ideal_jones1} and \ref{non-ideal_jones3} reduce to the identity ($\mathbf{J}_{\rm att,k}^{\rm id} = 
\mathbf{J}_{\rm omt,k}^{\rm id} =
\mathbf{J}_{\rm lna,k}^{\rm id} = \mathbf{I}$), while equation~\ref{non-ideal_jones2} reduces to the ideal hybrid (eq.~\ref{ideal_jones1}), and the noise vectors from equation~\ref{non-ideal_jones4} are zero.

In the following subsections, we address the non-ideal contributions of each subsystem individually by incorporating their realistic behavior into the analytical model. We focus on the most representative types of errors, attenuation asymmetries and polarization leakage (or isolation) terms, to illustrate their impact on the expected TMS response. The full system response, including the combined non-ideal effects of all subsystems, can subsequently be evaluated numerically within the same formalism. For conciseness, we also restrict the following discussion to the $I_1$, $I_2$, $Q_1$, and $Q_2$ parameters.

\begin{table}
\caption{Definitions of the parameters in the non-ideal Jones matrices representing the TMS instrument.}
\label{table:components}
\centering       

\begin{tabular}{c l l} 
\hline
Component & Symbol(s) &  Description\\
\hline                       
    Window,      & $\rm A_{x}$,$\rm A_{y}$ & Losses: R, L, SPO\\ 
    IR filter, and        & $\rm \phi$ &  Phase-shift\\ 
    Feedhorns & $\rm k$ & Index: sky (s) or load (l) \\
\hline

OMTs     & $\rm O^k_x$,$\rm O^k_y$ &  Losses: R, L\\ 
      & $\theta_{2}$ & Phase-shift XPD  \\
    & $\theta_{3}$ &  Phase-shift arm (2)  \\
            & $\rm O^k_{\rm a}$ & XPD \\
            & $\rm k$ & Index: sky (s) or load (l) \\
    \hline
   Hybrids & $\rm B^k_{2}$,$\rm B^k_{3}$ & Losses: R, L  \\
            & $\rm B^k_{a}$ & ISO\\
            & $\beta^k_{1}$ & Phase-shift arm (1) \\
            & $\beta^k_{2}$ & Phase-shift arm (4) \\
            & $\beta^k_{a}$ &  Phase-shift ISO \\
            & $\rm k$ &  Index: Hybrid X or Y \\
     \hline
   LNA & $g$ & Gain  \\
             Amplifiers   & $\psi$ & Phase-shift \\
                & n & Thermal noise \\
                & k &  Index: branch X or Y \\
                & i &Index: values are \\
                & & $i=1,2$ for x; $i=3,4$ for y.\\
     \hline
   BEM and DC & $g^{BD}$ & Gain \\
    amplifiers  & $\phi$ & Phase shift\\
    (BD) & $n^{BD}$ & Thermal noise \\
        & $k$ &  Index: branch X or Y \\
                & i & Index: values are \\
                & & $i=1,2$ for X; $i=3,4$ for Y.\\

\hline                                   
\end{tabular}

\end{table}

\subsection{ Slightly unbalanced radiometer and correlated noise}\label{subsec:Slightly unbalanced radiometer and correlated noise}
In the TMS design, the pseudo-correlation architecture includes four hybrids: two are implemented as physical components, and the other two are digitally synthesized in the FPGA. This configuration ensures that the final outputs exhibit identical correlated noise properties (i.e., $1/f$ noise). As a result, the comparison between the sky and load signals effectively suppresses this correlated noise component in both the total intensity and the Stokes $Q$ parameter. In practice, the effective sky intensity $I_{\rm tot}$, measured relative to the load signal, can be expressed as:
\begin{align}
\nonumber
    I_{tot}(t)=& I_1(t) - I_2(t) = \\ 
    =& \left(|C_1+n_{x}|^2+|D_1+n_{y}|^2\right)(t)- \left(|C_2+n_{x}|^2+|D_2+n_{y}|^2\right)(t) 
\label{power_tms_total}
\end{align}

In the same way, it is possible to obtain the Stokes $Q$ polarization parameter as
\begin{align}
\nonumber
    Q_{tot}(t)=& Q_1(t) - Q_2(t) = \\ 
    =& \left(|C_1+n_{x}|^2-|D_1+n_{y}|^2\right)(t)- \left(|C_2+n_{x}|^2- |D_2+n_{y}|^2\right)(t) 
    \label{power_tms_totalQ}
\end{align}
The two other Stokes parameters (U and V) can also be computed in the same way, but their timelines will not cancel out the correlated noise components, and thus will be affected by $1/f$ correlated noise.

\subsection{FEM, BEM and DC amplifiers}
\label{sec:fem_bem_dc}
\subsubsection{FEM: non-ideal LNAs}
\label{sec:lnas}

We begin by considering the effects of gain imbalance ($g_1$, $g_2$, $g_3$, $g_4$), noise voltages ($n_1$, $n_2$, $n_3$, $n_4$), and phase shifts ($\psi_{2}$, $\psi_{4}$) introduced by the LNAs in the FEM. Under these conditions, the Jones vectors $\vec{E}_3$ and $\vec{E}_4$ must include an additional contribution representing the thermal noise introduced by the LNAs:

\begin{align}
     \Vec{E}_{3} =& \mathbf{J}^{\rm id}_{\rm hyb, x}  \Vec{J}_{\rm lna, x} \left( \Vec{E}_{1}+\Vec{n_{\rm x}} \right)\\
%
    \Vec{E}_{4} =&  \mathbf{J}^{\rm id}_{\rm hyb, y} \Vec{J}_{\rm lna, y} \left( \Vec{E}_{2}+\Vec{n_{\rm y}} \right).
\end{align}

Now, using these equations and the formalism described above, we can obtain the all Stokes parameters $S_1$ and $S_2$. In particular, the $I_1$, $I_2$, $Q_1$ and $Q_2$ parameters are given by:
\begin{align}
    I_1= & \frac{\isky}{8}\left[ \sum^{4}_{k=1}g^2_k \,+2\left( g_1\,g_2 \cos{\psi_{2}} + g_3\,g_4 \cos{\psi_{4}} \right) \right] +\nonumber\\ & \frac{\iload}{8}\left[ \sum^{4}_{k=1}g^2_k\,-2\left( g_1\,g_2 \cos{\psi_{2}} + g_3\,g_4 \cos{\psi_{4}} \right) \right] +\nonumber\\ & \frac{\qload}{8}\left[  \sum^{4}_{k=1}C_k\,g^2_k \,-2 \left( g_1\,g_2 \cos{\psi_{2}} - g_3\,g_4 \cos{\psi_{4}} \right) \right] +\nonumber\\ & \frac{\qsky}{8}\left[ \sum^{4}_{k=1}C_k\,g^2_k\,+2 \left( g_1\,g_2 \cos{\psi_{2}} - g_3\,g_4 \cos{\psi_{4}} \right) \right]+ \nonumber \\& \frac{1}{2}\left[ \sum^{4}_{k=1}g^2_k\,N_k\, \right]
    \label{lna_I1_G}
\end{align}
\begin{align}
    I_2= & \frac{\iload}{8}\left[ \sum^{4}_{k=1}g^2_k \,+2 \left( g_{1 }\,g_{2 } \cos{\psi_{2}} + g_{3 }\,g_{4 } \cos{\psi_{4}} \right) \right] +\nonumber\\ & \frac{\isky}{8}\left[ \sum^{4}_{k=1}g^2_k \,-2 \left( g_{1 }\,g_{2 } \cos{\psi_{2}} + g_{3 }\,g_{4 } \cos{\psi_{4}} \right) \right] +\nonumber\\ & \frac{\qload}{8}\left[ \sum^{4}_{k=1}C_k\,g^2_k\,+2 \left( g_{1 }\,g_{2 } \cos{\psi_{2}} - g_{3 }\,g_{4 } \cos{\psi_{4}} \right) \right] +\nonumber\\ & \frac{\qsky}{8}\left[ \sum^{4}_{k=1}C_k\,g^2_k\,-2 \left( g_{1 }\,g_{2 } \cos{\psi_{2}} - g_{3 }\,g_{4 } \cos{\psi_{4}} \right) \right]+ \nonumber \\& \frac{1}{2}\left[ \sum^{4}_{k=1}g^2_k\,N_k\,\right]
\end{align}
\begin{align}
    Q_1= & \frac{\qsky}{8}\left[ \sum^{4}_{k=1}g^2_k \, +2 \left( g_{1 }\,g_{2 } \cos{\psi_{2}} + g_{3 }\,g_{4 } \cos{\psi_{4}} \right) \right]+ \nonumber \\& \frac{\qload}{8}\left[ \sum^{4}_{k=1}g^2_k \, -2 \left( g_{1 }\,g_{2 } \cos{\psi_{2}} + g_{3 }\,g_{4 } \cos{\psi_{4}} \right) \right] +\nonumber\\ & \frac{\isky}{8}\left[ \sum^{4}_{k=1}C_k\,g^2_k\,+ 2 \left( g_{1 }\,g_{2 } \cos{\psi_{2}} - g_{3 }\,g_{4 } \cos{\psi_{4}} \right) \right] +\nonumber\\ & \frac{\iload}{8}\left[ \sum^{4}_{k=1}C_k\,g^2_k\,- 2 \left( g_{1 }\,g_{2 } \cos{\psi_{2}} - g_{3 }\,g_{4 } \cos{\psi_{4}} \right) \right] +\nonumber\\ & \frac{1}{2}\left[ \sum^{4}_{k=1}g^2_k\,N_k\,\right]
\end{align}
\begin{align}
    Q_2= & \frac{\qload}{8}\left[ \sum^{4}_{k=1}g^2_k \,+2 \left( g_{1 }\,g_{2 } \cos{\psi_{2}} + g_{3 }\,g_{4 } \cos{\psi_{4}} \right) \right] +\nonumber\\ & \frac{\qsky}{8}\left[ \sum^{4}_{k=1}g^2_k \, -2 \left( g_{1 }\,g_{2 } \cos{\psi_{2}} + g_{3 }\,g_{4 } \cos{\psi_{4}} \right) \right]+\nonumber\\ & \frac{\iload}{8}\left[ \sum^{4}_{k=1}C_k\,g^2_k\,+2 \left( g_{1 }\,g_{2 } \cos{\psi_{2}} - g_{3 }\,g_{4 } \cos{\psi_{4}} \right) \right] +\nonumber\\ & \frac{\isky}{8}\left[ \sum^{4}_{k=1}C_k\,g^2_k\,-2 \left( -g_{1 }\,g_{2 } \cos{\psi_{2}} + g_{3 }\,g_{4 } \cos{\psi_{4}} \right) \right] +\nonumber \\ & \frac{1}{2}\left[ \sum^{4}_{k=1}g^2_k\,N_k\,\right].
\end{align}
In these equations, we also introduce the  coefficients $C_k$, defined as $C_k = +1$ for $k=1,2$ and $C_k = -1$ for $k=3,4$. 

These results show that the measured sky intensity signal ($I_1$) includes spurious contributions from $I_{\rm load}$ (sky-to-load leakage) and from $Q_{\rm sky}$ and $Q_{\rm load}$ (intensity-to-polarization leakage) when the amplifier gains are not perfectly balanced (and equal) between the two branches. A similar effect occurs in the case of the load intensity signal ($I_2$).

In addition, the LNAs contribute an offset signal due to thermal noise from all four amplifiers to $\mathcal{S}_1$ and $\mathcal{S}_2$, accounting for approximately 50\,\% of the total output power.

If we now compute the total intensity ($I_{\rm tot}$) and the total Q polarization ($Q_{\rm tot}$) as measured by TMS, we have
\begin{align}
    I_{\rm tot}= & \frac{\isky}{2}\left[  g_{1 }\,g_{2 } \cos{\psi_{2}} + g_{3 }\,g_{4 } \cos{\psi_{4}}  \right]- \nonumber\\ &  \frac{\iload}{2}\left[  g_{1 }\,g_{2 } \cos{\psi_{2}} +g_{3 }  g_{4 } \cos{\psi_{4}} \right] +\nonumber\\ & \frac{\qsky}{2}\left[ g_{1 }\,g_{2 } \cos{\psi_{2}} - g_{3 }\,g_{4 } \cos{\psi_{4}}  \right] -\nonumber\\ &  \frac{\qload}{2}\left[ g_{1 }\,g_{2 } \cos{\psi_{2}} - g_{3 }\,g_{4 } \cos{\psi_{4}}  \right] \\
    \label{i_tot_resta}
    Q_{\rm tot}= &  \frac{\qsky}{2}\left[  g_{1 }\,g_{2 } \cos{\psi_{2}} + g_{3 }\,g_{4 } \cos{\psi_{4}}  \right] -\nonumber\\ & \frac{\qload}{2}\left[  g_{1 }\,g_{2 } \cos{\psi_{2}} + g_{3 }\,g_{4 } \cos{\psi_{4}}  \right]  +\nonumber\\ & \frac{\isky}{2}\left[ g_{1 }\,g_{2 } \cos{\psi_{2}} - g_{3 }\,g_{4 } \cos{\psi_{4}}  \right] -\nonumber\\ & \frac{\iload}{2}\left[ g_{1 }\,g_{2 } \cos{\psi_{2}} - g_{3 }  g_{4 } \cos{\psi_{4}}  \right]. 
\end{align}
In short, gain imbalance and phase shifts introduce spurious signals, primarily appearing as intensity-to-polarization and sky-to-load cross terms. Furthermore, the measured signal is affected by the thermal noise added by each LNA. However, the process of comparing sky and load signals ensures that this thermal noise contribution cancels out assuming spectral invariance among the LNAs in the different branches.

\subsubsection{BEM and DC (BD)}
Here we describe the non-ideal effects of gain imbalance and phase shifts introduced by the BEM and DC modules. Since these two modules are placed in sequence, they can be represented by a single Jones matrix $\Vec{J}_{{\rm BD},k}$ with effective total gains $g^{\rm BD}_i$, effective phase shifts $\phi_i$, and a single noise term $\Vec{n}^{\rm BD}_k$, as
\begin{align}
     \Vec{E}_{3}=& \Vec{J}^{\rm id}_{hyb,x}\, \Vec{J}_{\rm BD,x} \left[ \Vec{J}_{\rm lna, x} \left( \Vec{E}_{1}+\Vec{n_x} \right)+\Vec{n}^{\rm BD}_x \right]\\
\label{bem_E3}
     \Vec{E}_{4}=&  \Vec{J}^{\rm id}_{hyb,y}\, \Vec{J}_{\rm BD,y} \left[ \, \Vec{J}_{\rm lna, y} \left( \Vec{E}_{2}+\Vec{n_y} \right)+\Vec{n}^{\rm BD}_y \right].
\end{align}
The final equations for the measured Stokes parameters are formally equivalent to those discussed in the previous subsection, with the substitutions $g_i \rightarrow g_i g^{\rm BD}_i$, $\psi_i \rightarrow \psi_i + \phi_i$ and $N_k \rightarrow N_k + N^{\rm BD}_k / g_i^2$. As before, for the $I_{\rm tot}$ and $Q_{\rm tot}$ parameters,the thermal noise contributions effectively cancel out under the assumption of a stable spectral response.

\subsection{Cryostat window, IR filter, and feedhorns}
For simplicity, all elements introducing attenuation of the incoming signal (as the window, IR filter and the feedhorns) are described here by a single Jones matrix, as in Eq.~\ref{non-ideal_jones0}, where each term represents an effective attenuation ($A_{\rm x}$, $A_{\rm y}$) or an effective phase shift ($\phi_k$) for the combined subsystem. 
%
Here, we discuss the case of an effective attenuation in the electric field components ($x$, $y$) but no phase shift. The outputs in this case are given by
\begin{align}
    I_1 =& \frac{\isky}{2}\left[A_{x}^2 + A_{y}^2\right] - \frac{\qsky}{2}\left[-A_{x}^2 + A_{y}^2\right] \\
%
    Q_1 =& \frac{\isky}{2}\left[A_{x}^2 - A_{y}^2\right] + \frac{\qsky}{2}\left[A_{x}^2 + A_{y}^2\right]\\
%
    I_2 =& \iload
    \label{att_I2}
\end{align}
As expected, no cross terms between sky and load signals appear at the outputs, and only overall attenuation is observed. Intensity-to-polarization terms arise if the attenuation differ for the two orthogonal polarizations. If both attenuations are equal, the output is $\rm I_1 = \isky A_{\rm x}^2$ (i.e. a global attenuation). Although not discussed here, a phase-shift term ($\phi$) governs the $U$ and $V$ sky polarization response.

\subsection{OMTs}
TMS includes two OMTs, each associated with the sky and load feedhorns. We now describe the non-ideal contributions of the individual terms in Eq.~\ref{non-ideal_jones1} separately. 

\subsubsection{Attenuation asymmetry}\label{Attenuation_asymmetry_omt}
The following set of equations illustrates the case in which the OMTs exhibits attenuation in each arm ($\Oxsky$ and $\Oysky$ for the sky arm, and $\Oxload$ and $\Oyload$ for the load arm), but the polarization leakage terms are set to zero ($O_{\rm a}^k =0$): 
\begin{align}
    I_1 =& \frac{\isky}{2} \left[\Oxsky\,(\Oxsky+2)+\Oysky\,(\Oysky+2)+2\right] +\nonumber\\& \frac{\qsky}{2}\left[\Oxsky\,(\Oxsky+2)-\Oysky\,(\Oysky+2)\right] \\
    I_2 =& \frac{\iload}{2} \left[\Oxload\,(\Oxload+2)+\Oyload\,(\Oyload+2)+2\right] +\nonumber\\& \frac{\qload}{2}\left[\Oxload\,(\Oxload+2)-\Oyload\,(\Oyload+2)\right]\\
    Q_1 =& \frac{\qsky}{2} \left[\Oxsky\,(\Oxsky+2)+\Oysky\,(\Oysky+2)+2\right] +\nonumber\\& \frac{\isky}{2}\left[\Oxsky\,(\Oxsky+2)-\Oysky\,(\Oysky+2)\right] .
\end{align}

As expected, no mixing between sky and load signals is found at the output. However, OMT asymmetry produces intensity-to-polarization leakage in the sky and load signals separately. If the losses in a given OMT arm are equal (e.g. $\Oxsky=\Oysky$), the intensity-to-polarization leakage components vanish, and we recover $I_1 = \isky(\Oxsky+1)^2$. 
Note that mismatches between the losses of the two arms may affect the amplitude of the total signal $I_{\rm tot}$. Finally, the only relevant shifting angle in this case is $\theta_3^k$, which does not affect the $I_1$ and $Q_1$ measurements, but does affect $U_1$ and $V_1$ (not shown here).

\subsubsection{Polarization leakage} 
We now show the case in which both the sky and load OMTs exhibit a certain degree of polarization leakage between their X and Y arms ($\Osa$ and $\Ola$ terms), while the attenuation asymmetry is set to zero. We have
\begin{align}
    I_1=&\isky\left((\Osa)^{2}+1\right)+2\Osa \cos{\theta^{s}_2}\left(\usky\cos{\theta^{s}_3}+\vsky\sin{\theta^{s}_3} \right)
    \label{omt_I1_lmd}
\end{align}
\begin{align}
    I_2=&\iload\left((\Ola)^2+1\right)+2\Ola \cos{\theta^{l}_2}\left(\uload\cos{\theta^{l}_3}+\vload\sin{\theta^{l}_3} \right)
    \label{omt_I2_lmd}
\end{align}
\begin{align}
    Q_1=&\qsky\left(1-(\Osa)^2\right)+2\Osa \sin{\theta^{s}_2}\left(-\usky\sin{\theta^{s}_3}+\vsky\cos{\theta^{s}_3} \right). 
    \label{omt_Q1_lmd}
\end{align}
Again, we observe that intensity-to-polarization leakage components arise, along with an overall amplitude attenuation. 

\subsection{Hybrid}
TMS has two $180^\circ$ hybrid couplers, associated with the X and Y branches of the system, as shown in Fig.~\ref{new_model_gains}. The possible non-ideal response of these subsystems is described in Eq.~\ref{non-ideal_jones2} and includes three potential sources of error: global losses, cross-polarization terms, and phase shifts. Without loss of generality, we assume here that $\beta^k_1$ is zero.

\subsubsection{Attenuation asymmetry} 
As for the OMT, we first consider the isolated effect of attenuation in the hybrid arms, assuming that the cross-polarization terms ($B^k_a$) are zero. This attenuation is described by the parameters $\Bonce$ and $\Bcato$ for the X hybrid, and $\Bvuno$ and $\Bvcuatro$ for the Y hybrid. For simplicity, we also assume zero phase-shifting angles. The measured response in this case is given by
\begin{align}
    I_1=& \frac{\isky}{8}\left[ \Bonce\left(\Bonce+4\right) +\Bvuno\left(\Bvuno+4\right)+8 \right]\nonumber\\& +\frac{\iload}{8}\left[(\Bcato)^2+(\Bvcuatro)^2\right]+\frac{\qload}{8}\left[(\Bcato)^2-(\Bvcuatro)^2\right]\nonumber\\& + \frac{\qsky}{8}\left[ \Bonce\left(\Bonce+4\right) -\Bvuno\left(\Bvuno+4\right)\right]
    \label{hyb_I1_ep}
\end{align}
\begin{align}
    I_2=& \frac{\iload}{8}\left[ \Bcato\left(\Bcato+4\right) +\Bvcuatro\left(\Bvcuatro+4\right)+8 \right]\nonumber\\& +\frac{\isky}{8}\left[(\Bonce)^2+(\Bvuno)^2\right]+\frac{\qsky}{8}\left[(\Bonce)^2-(\Bvuno)^2\right]\nonumber\\& + \frac{\qload}{8}\left[ \Bcato\left(\Bcato+4\right) -\Bvcuatro\left(\Bvcuatro+4\right)\right]
\label{hyb_I2_ep}
\end{align}
\begin{align}
    Q_1=&\frac{\qsky}{8}\left[ \Bonce\left(\Bonce+4\right) +\Bvuno\left(\Bvuno+4\right)+8 \right]\nonumber\\& +\frac{\iload}{8}\left[(\Bcato)^2-(\Bvcuatro)^2\right]+\frac{\qload}{8}\left[(\Bcato)^2+(\Bvcuatro)^2\right]\nonumber\\& + \frac{\isky}{8}\left[ \Bonce\left(\Bonce+4\right) -\Bvuno\left(\Bvuno+4\right)\right]
\label{hyb_Q1_ep}
\end{align}
Since the hybrid correlates the sky and load signals, the output Stokes parameters in this non-ideal case contain both sky-to-load and intensity-to-polarization leakage terms. The intensity-to-polarization terms vanish if both hybrid components have identical non-ideal behavior (i.e., $\Bcato = \Bvcuatro$ and $\Bonce = \Bvuno$). However, sky-to-load terms are always present and can only be factorized as an overall factor in the computation of $I_{\rm tot}$ when all attenuations are identical.

\subsubsection{Isolation} 
Second, we consider the case in which the  parameters $\Bunoa$ and $\Bdosa$ arise from non-idealities in the isolation of the two components in both X and Y hybrids, while the attenuation factors (losses) are set to zero. As before, we also assume zero phase-shifting angles.
The response is given by:
\begin{align}
    I_1=&\frac{\isky}{8}\left[ \Bunoa\left(\Bunoa+4\right) +\Bdosa\left(\Bdosa+4\right)+8 \right]\nonumber\\& +\frac{\iload}{8}\left[(\Bunoa)^2+(\Bdosa)^2\right]+\frac{\qload}{8}\left[(\Bunoa)^2-(\Bdosa)^2\right]\nonumber\\& + \frac{\qsky}{8}\left[ \Bunoa\left(\Bunoa+4\right) -\Bdosa\left(\Bdosa+4\right)\right]
\label{hyb_I1_lamb}
\end{align}
\begin{align}
    I_2=&\frac{\iload}{8}\left[ \Bunoa\left(\Bunoa+4\right) +\Bdosa\left(\Bdosa+4\right)+8 \right]\nonumber\\& +\frac{\isky}{8}\left[(\Bunoa)^2+(\Bdosa)^2\right]+\frac{\qsky}{8}\left[(\Bunoa)^2-(\Bdosa)^2\right]\nonumber\\& + \frac{\qload}{8}\left[ \Bunoa\left(\Bunoa+4\right) -\Bdosa\left(\Bdosa+4\right)\right]
\label{hyb_I2_lamb}
\end{align}
\begin{align}
    Q_1=&\frac{\qsky}{8}\left[ \Bunoa\left(\Bunoa+4\right) +\Bdosa\left(\Bdosa+4\right)+8 \right]\nonumber\\& +\frac{\iload}{8}\left[(\Bunoa)^2-(\Bdosa)^2\right]+\frac{\qload}{8}\left[(\Bunoa)^2+(\Bdosa)^2\right]\nonumber\\& + \frac{\isky}{8}\left[ \Bunoa\left(\Bunoa+4\right) -\Bdosa\left(\Bdosa+4\right)\right].
\label{hyb_Q1_lamb}
\end{align}
These expressions are analogous to those of the attenuation asymmetry. Again, both intensity-to-polarization and sky-to-load terms appear. The intensity-to-polarization terms vanish if the leakage coefficients are identical for both hybrids ($\Bunoa = \Bdosa$). In that case, the sky-to-load terms can also be factorized as an overall attenuation.


\section{Expected level of TMS non-idealities from the analytic model}
\label{sec:refparams}

Once we have established the parametric form of how the different non-idealities enter the TMS response, we next assess the relevance of the various terms. To this end, we first summarize the expected levels of non-idealities in the TMS subsystems, and in particular we quote the S-parameters describing the insertion losses (usually noted as $L$ or $S_{21}$) and return losses ($R$ or $S_{11}$) of all the components, the cross-polarization ($X_{\rm POL}$) of the OMTs, the isolation ($ISO$) of the hybrids, as well as the typical gain and noise values of the low-noise amplifier. 

\begin{figure}
\centering
\includegraphics[width=9cm]{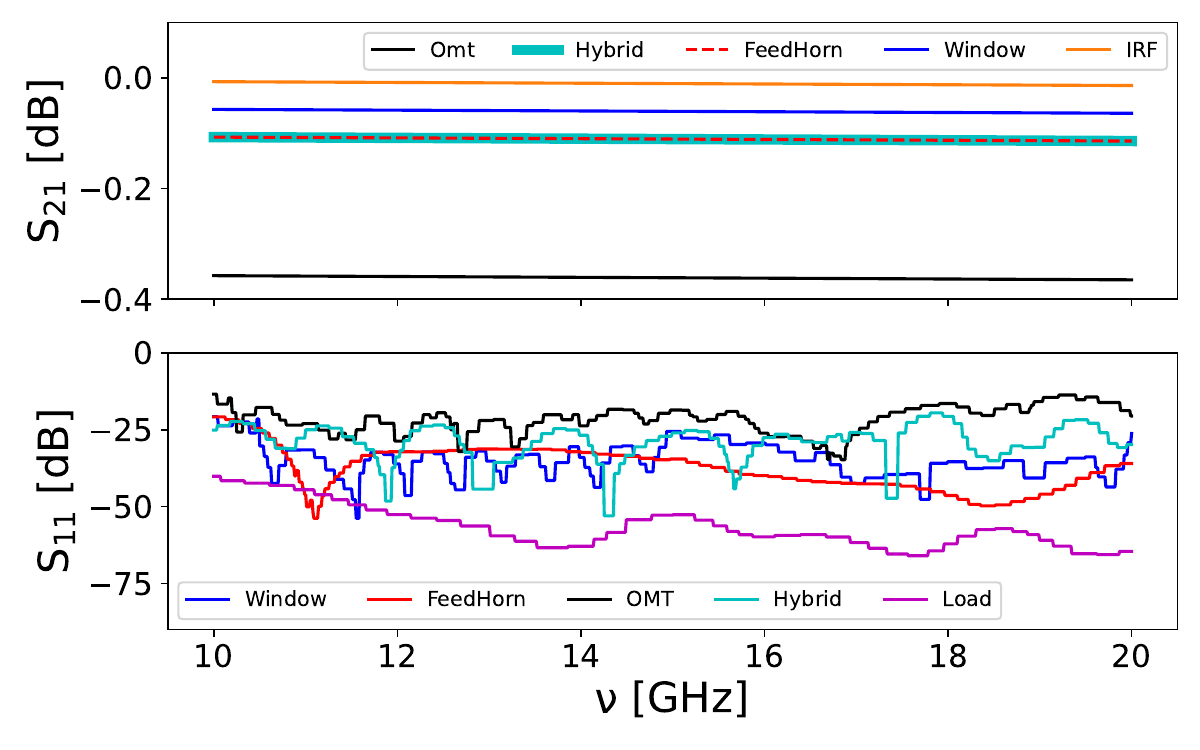}
\caption{Nominal values of insertion (L) and return (R) losses (S-parameters) of the optical and RF components of TMS. Top: insertion losses ($S_{21}$ or $L$). Bottom: return losses ($S_{11}$ or $R$). }
\label{relative_condi}
\end{figure}

\begin{figure}
\centering
\includegraphics[width=9cm]{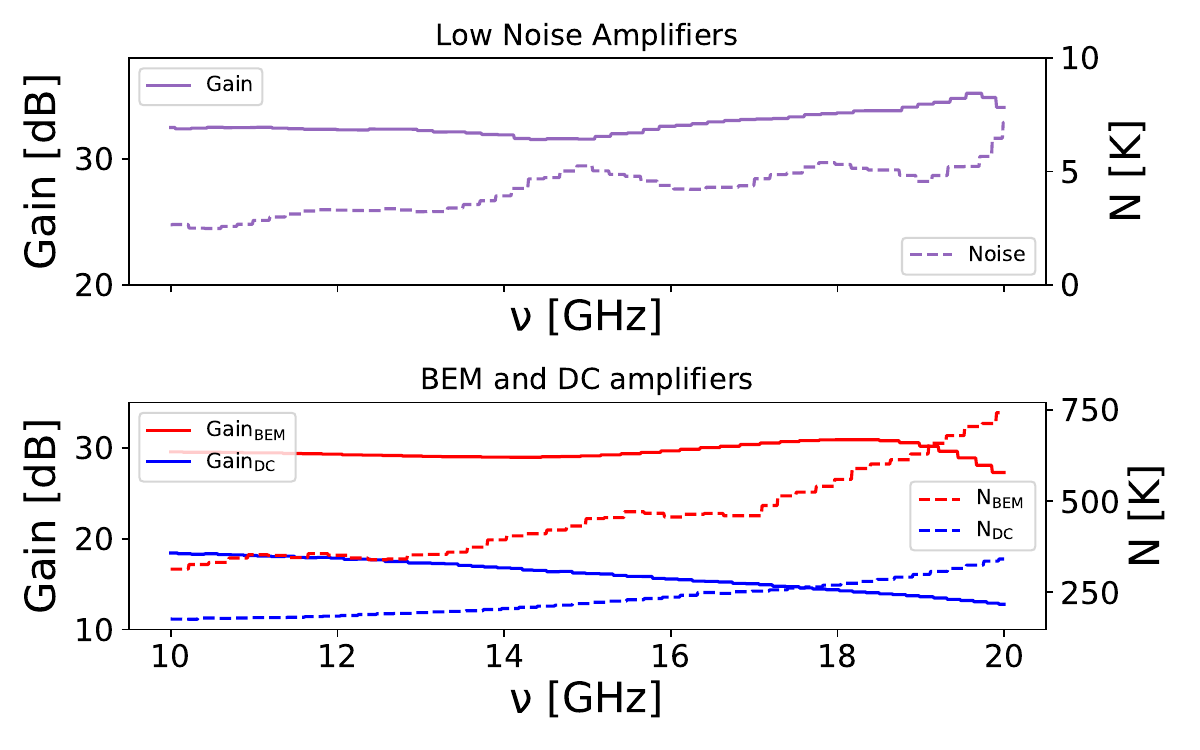}
\caption{Nominal values (gains and noise figures) of the three amplification stages of TMS. Top: parameters describing the low-noise amplifiers (FEM system), with gain shown in purple (dB) and noise temperature in yellow (K). Bottom: Gain and noise temperatures of the amplifiers located in the BEM and DC stages.
}
\label{relative_condi_amplifiers}
\end{figure}

In particular, the return losses ($R$) of components such as OMTs, hybrids, feedhorns, and the window were obtained from simulations with CST Studio and SRSR software packages, as well as from measurements performed at the IAC facilities \citep{tesisPazAlonso, 2022JInst..17P6041D}. Most of these values are also listed in \cite{Alonso-Arias_2024}. The insertion losses ($L$), cross-polarization ($X_{\rm POL}$) and isolation parameters ($ISO$) were estimated from simulations and from experience with previous microwave instruments at the Teide Observatory, using always a conservative approach.
Finally, typical gain and noise values of the LNA amplifiers were taken from the commercial data sheet of the actual amplifiers to be used in TMS\footnote{\label{nota:lna}Datasheet LNF-LNC6\_20C
6--20\,GHz Cryogenic Low Noise Amplifier. Low Noise Factory.}, 
while the values for mixers and other amplifiers were taken from the corresponding technical sheets\footnote{\label{nota:mixer}MMIC surface-mount wideband double-balanced mixer MDB-24H+; monolithic amplifier ultra-high dynamic range 5PHA-202+; data sheet QP-AMLNA-0120-01.}. The basic numbers are summarized in Figures~\ref{relative_condi}, \ref{relative_condi_amplifiers}, and listed in Table~\ref{table:B1_p_INITIAL}. 
The table also includes our estimates of the environmental temperatures at the different stages, derived using a conservative approach. More realistic and precise values will be obtained during laboratory testing of TMS.
%

\begin{table*}[t]
\caption{Summary of the nominal parameters describing the TMS components (temperatures, losses, and noise-figure values; see text for details). For completeness, we also include the input parameters used in this paper, namely the sky and load signals and the environmental temperatures in the different stages of the TMS instrument.
}
\label{table:B1_p_INITIAL}
\centering
\small

\begin{tabular*}{\textwidth}{@{\extracolsep{\fill}} l c c c c c c l @{}}
\toprule
Component  &  $\rm T\, [\rm K]$ & $\rm L$ \,[dB] &  $\rm R$\,[dB]   & SPO\,[dB]& $\rm XPD\,[dB]$ &ISO\,[dB] & Reference(s)\\

  &  & max | min & max | min   &  &  &  & \\

\midrule
$\rm Window $ & 300 & 0.064 | 0.057 & $-20.7$ | $-53.8$ & $-20$ & - & - & \cite{tesisPazAlonso}\\
$\rm IRF $ & 50 & 0.014 | 0.007 & $-40$ | $-40$ & $-20$ & - & - & Laboratory Test\\
$\rm FH_s$ & 5 & 0.114 | 0.107  & $-20.8$ | $-53.8$ & - & - & - & \cite{tesisPazAlonso}\\
$\rm OMT_s$ & 5 & 0.37 | 0.35  & $-25.5$ | $-42.5$ & - & $-35$ & $-35$ & Laboratory Test\\
$ \rm Hs_{180}$ & 5 & 0.114 | 0.107 & $-20$ | $-53$ & - & $-35$ & $-35$ & \cite{tesisPazAlonso}\\
$ \rm {Cold\,Load } $ & 8 & $0.43\times 10^{-3}$ | $1.12\times 10^{-6}$ & $-40$ | $-65.9$ & $-40$ & - & - & \cite{tesisPazAlonso}\\
$\rm FH_l$ &  5  & 0.114 | 0.107 & $-20.8$ | $-53.8$ & - & - & - & \cite{2022JInst..17P6041D}\\
$\rm OMT_l$ & 5 & 0.37 | 0.35 & $-25.5$ | $-42.5$ & - & $-35 $& $-35$ & Laboratory Test\\
$ \rm Hl_{180}$ & 5 & 0.114 | 0.107 & $-20$ | $-53$ & - & $-35$ & $-35$ & \cite{tesisPazAlonso}\\
$\rm Filter_{BEM}$ & 300 & $-2$ | $-2$ & $-25$ | $-25$ & - & - & - & Laboratory Test\\
 $\rm Filter_{DC}$ & 300 & $-2$ | $-2$ & $-25$ | $-25$ & - & - & - & Laboratory Test\\
\bottomrule
\end{tabular*}

\vspace{0.4cm}

\begin{tabular*}{\textwidth}{@{\extracolsep{\fill}} l c c c c c c l @{}}
\toprule
 Component & T [K] & $\rm T_{n}[K]$ & $\rm Gain [dB]$ & $\rm S_{11}$ $\rm [dB]$ & F $\rm [dB]$  & Reference(s) \\
&  & max | min & max | min   &  &  &  & \\

\midrule
$ \rm LNA$  & 5& 2.5 | 7.2 & 32 | 35 & $-15$ & 0.06 &  Data-sheet \ref{nota:lna} \\
$\rm BEM$   & 300& 19 | 70 & 27 | 31 & $-15$ & 4 & Data-sheet \ref{nota:mixer}\\
$\rm DC$   & 300& 176 | 341& 13 | 19 & $-15$ & 3 &  Data-sheet \ref{nota:mixer}\\
\rm Mixer  & 300& 1000 | 1000 & - & - & 6.9 &   Data-sheet \ref{nota:mixer}\\
\bottomrule
\end{tabular*}

\vspace{0.4cm}

\begin{tabular*}{\textwidth}{@{\extracolsep{\fill}} l c c c c c c l @{}}
\toprule
  $\rm Sky$ &  $\rm Load$ & $ \Textcry [\rm K]$ & $ \Tcryuno  [\rm K]$  & $ \Tcrydos [\rm K]$ & $ \Tenvuno [\rm K]$ & $ \Tenvdos [\rm K]$ & $\rm T_{PFGA} [\rm K]$  \\
\midrule
 8 & 8 & 300 & 50  & 5 & 50.1 & 5.1 & 300   \\
\bottomrule
\end{tabular*}

\end{table*}

\begin{table}
\caption{Nominal values of the non-ideal Jones matrices for the TMS components. For each component and relevant parameter, we provide the equivalence to the standard values of the return and insertion losses, cross-polarization and isolation parameters, as well as the representative value (averaged across the band). }  
\label{table:B1_p_1} 
\centering      
\tiny
\begin{tabular}{l l l l}        
\hline                
Component  &  Parameter & Equivalence & Value\\ 
\hline 
$ \rm {Window} $   & $A_{\rm x}$, $A_{\rm y}$ &  $ \sqrt{(1-R)(1-L)(1-SPO)}$ &0.975\\
        $ \rm {IRF}$   & $A_{\rm x}$, $A_{\rm y}$  & $ \sqrt{(1-R)(1-L)(1-SPO)}$ &0.987\\
       $ \rm{Feedhorn} $ &$A_{\rm x}$, $A_{\rm y}$ & $ \sqrt{(1-R)(1-L)}$ &0.974\\
       $\rm {OMT} $  & $1+O_{\rm x}^k$  &$  \sqrt{(1-R)(1-L)}$&0.910\\
        & $O_{\rm a} $ & $ \sqrt{X_{\rm POL}}$ &0.017\\
       $ \rm {Hybrid}$    & $1+B_{\rm x}^k $&$ \sqrt{(1-R)(1-L)}$&0.972\\
        & $1+B_{\rm a} $ &$ \sqrt{1-ISO}$& 0.9998\\
       \hline
    \end{tabular}
\end{table}

In order to relate these values to the coefficients in the Jones matrix formalism, we recall that Jones matrix coefficients act on complex electric field amplitudes (or equivalently voltage-like quantities), whereas insertion and return losses, as well as cross-polarization and isolation, are typically defined in terms of total power. Table~\ref{table:B1_p_1} summarizes the equivalence and lists representative (band-averaged) values for the Jones coefficients in the realistic case. Based on these values, we conclude that attenuation effects in TMS will introduce errors in the recovered Stokes parameters (particularly in the intensity) at the level of about one percent. In principle, all these factors can be calibrated by taking advantage of the parametric form of the different terms derived in the previous section.

Finally, we discuss the impact of gains and offsets introduced by the different amplification stages. Since the LNA gains are high ($\sim 30$\,dB), the relative contributions of the BEM and DC components are expected to be almost negligible, and thus the system response is essentially dominated by the FEM. The noise figures of the TMS LNAs are typically 2--6\,K, which, according to the equations in Sect.~\ref{sec:lnas}, correspond to an offset of $4$--$12$\,K in the response of the sky and load intensities. Provided that this contribution is comparable among the different LNAs, it will largely cancel out in the evaluation of $I_{\rm tot}$.

%
\section{Extending the TMS analytic model: noise temperature effects}
\label{TB_s_equation}

Although the analytic TMS model developed in the previous sections, based on the Jones matrix formalism, is quite complete in describing the non-ideal behavior of the different elements in the radiometric chain, it does not account for the additional noise temperature contributions to the antenna temperature actually measured by the instrument. In this section, we extend our model by estimating these additional noise temperature contributions, which will appear as offsets in the measured Stokes parameters. 

For this purpose, we employ the Friis formalism \citep{Friis1944, kraus1982radio}. For definiteness, we focus on the intensity Stokes parameter, which is the main target of TMS. The relevant quantities are therefore $I_1$, $I_2$, and $I_{\rm tot}$. Since we will compute noise contributions in terms of temperature, we explicitly write them as $T_{\rm sky}^{\rm a}$, $T_{\rm load}^{\rm a}$, and $\Delta T = T_{\rm sky}^{\rm a} - T_{\rm load}^{\rm a}$ throughout this section, where the notation "a" for the superscript refers to antenna temperature measurements. 

Making use of this formalism \citep[see e.g.][]{absmesua_2GHZ_1993}, the overall response of the TMS can be expressed in terms of antenna temperature by concatenating the equations corresponding to each component. For example, the antenna temperature that would be measured at the output of the first optical element in TMS, the window, is given by
%
%
\begin{equation}
\Tbw = [\Tsky (1-R)(1-L) + \Tw L + \Tenvuno R ](1-\SPOw) + \Textcry \, \SPOw.
\label{eq:tw}
\end{equation}
The first term in this equation corresponds to the attenuation effects already included in the Jones matrix formalism. The second term represents the contribution of the window physical temperature ($T_{\rm w}$) due to dielectric (Ohmic) losses ($L$). The third term accounts for internal reflection losses ($R$), this include the environmental radiation ($T_{\rm env}$) or radiation from subsequent stages of the RF chain. Finally, the last term ($\rm SPO$) accounts for the spillover arising from the interaction between the feedhorn system and a given component. 
Note that in this work, in Table~\ref{table:B1_p_INITIAL} we assigned a SPO value to the window, the IRF and the cold load, while implicitly referring to their interaction with the feedhorns. 
We also note that the value $\Textcry = 300,\mathrm{K}$ adopted in this paper is highly conservative; the actual effective temperature is likely to lie between 300\,K and the temperature of the first cold stage, which in this case is taken to be 50\,K. If we assume a configuration in which the cryostat shields are made of perfectly reflective material and are well thermally coupled to the inner shields of the cold stages, the resulting window contribution in equation~\ref{eq:tw} is expected to decrease by at least a few kelvins, depending on the thermal coupling efficiency.

Analogous expressions to equation~\ref{eq:tw} apply to all other optical elements. For RF passive components such as feedhorns, OMTs, and hybrids, the equations are similar, but spillover terms do not appear. For the RF active components (amplifiers), we should also include the noise temperature contributions. The full set of equations for all TMS components is presented in Appendix~\ref{A_appendixx}.
\\
By sequentially applying these equations to the entire TMS radiometric chain, we can rearrange the terms to obtain the following effective expression:
\begin{equation}
    \Delta T \equiv T_{\rm sky}^{\rm a} - T_{\rm load}^{\rm a} = \beta_{\rm sky}^{\rm eff} T_{\rm sky} - \beta_{\rm load}^{\rm eff} T_{\rm load} + \Toffeff +\Tneff, 
    \label{delta_temp_total}
\end{equation}
which constitutes the main equation summarizing the instrument model. 
The first two terms describe the effective attenuation of the sky and load (intensity) signals due to the non-ideal behavior of the components (provided that we have corrected for the LNA gains). These terms are fully consistent with those obtained from the Jones matrix formalism when cross-terms (e.g. polarization leakage or isolation effects) are neglected. 
The third term $\Toffeff$ is new, and accounts for an effective offset in the measured antenna temperature, arising from the cumulative contributions of dielectric losses, internal reflections and spillover. Formally, it can be expressed as a linear combination of the physical temperatures of the various components ($T_i$) and the relevant environmental contributions ($T_{{\rm env}, j}$), as
\begin{equation}
    \Toffeff = \sum_i a_i T_{i} + \sum_j b_j T_{{\rm env}, j}.
\end{equation}
Detailed equations for all these terms are given in Appendix \ref{B_appendix}. 
Finally, the fourth term $\Tneff$ also appeared in the previous formalism, and represents the aggregated contributions of the noise temperatures from the LNAs, BEM, and DC amplifiers (see  Sect.~\ref{sec:fem_bem_dc}).

\section{Predicted overall TMS intensity response}
\label{sec:overall_response}

We now estimate the expected overall TMS intensity response ($\Delta T$), accounting for all terms in equation~\ref{delta_temp_total}. 
The reference values used in this calculation are those summarized in Table~\ref{table:B1_p_INITIAL} and described above in Section~\ref{sec:refparams}. For simplicity, in this first estimate we assume that all four LNA devices have identical gains and noise figures, and we apply the same assumption to the BEM and DC amplifiers.
We also assume that both the input sky and load signals are equal to 8\,K (see Table~\ref{table:B1_p_INITIAL})\footnote{A value of 8\,K for the sky signal is reasonable for a ground based experiment as TMS. The CMB yields 2.7\,K, and the remaining $\sim 5$\,K represent a typical atmospheric contribution at elevations $EL=60^\circ$ \citep[see e.g.][]{chappard}. }. For an ideal system, this would yield $\Delta T = 0$. Throughout this section, we also make the primary assumption that the signal across the entire frequency band remains constant in time.

\begin{figure*}
\centering
\includegraphics[width=2\columnwidth]{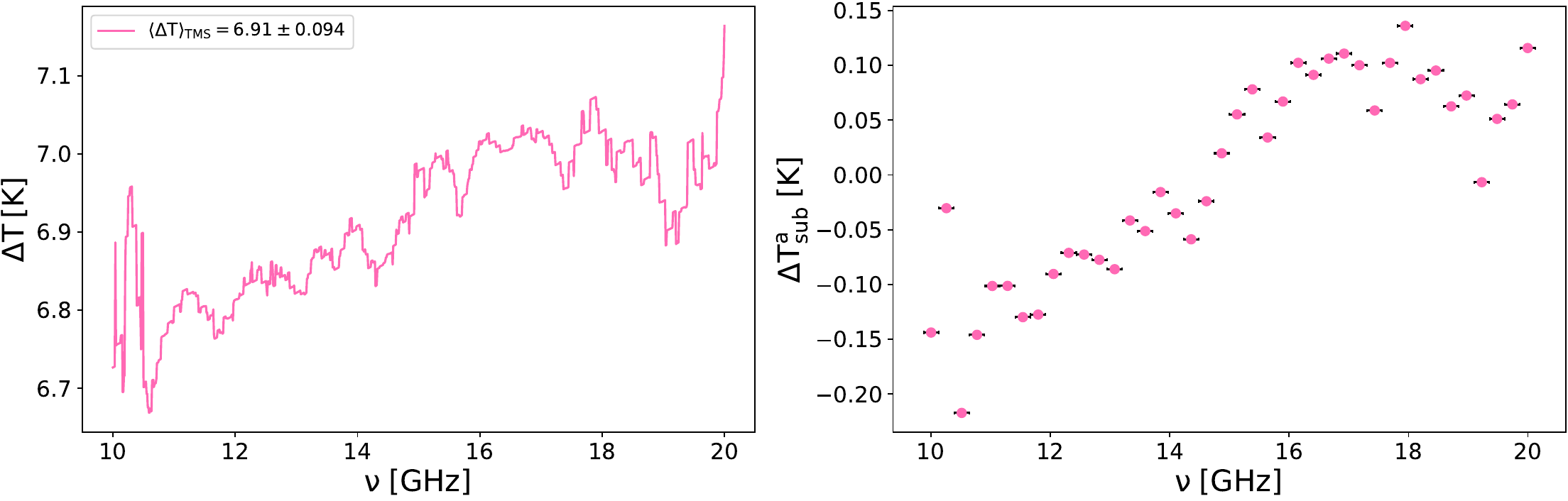}
    \caption{Left: Spectral response of the TMS output $\rm \Delta T=\Tbsky-\Tbload$ as described in equation~\ref{delta_temp_total}. Right: Averaged antenna temperature excess per sub-band ($\Delta T_{\rm sub}$) as seen by the TMS system, and relative to the mean signal $<\Delta T>$. TMS will have 40 sub-bands in the range 10--20\,GHz, with frequency widths $\Delta \nu=0.25$\,GHz.}
\label{relative_ALL_1}
\end{figure*}

\begin{figure}
\centering
   \includegraphics[width=\columnwidth]{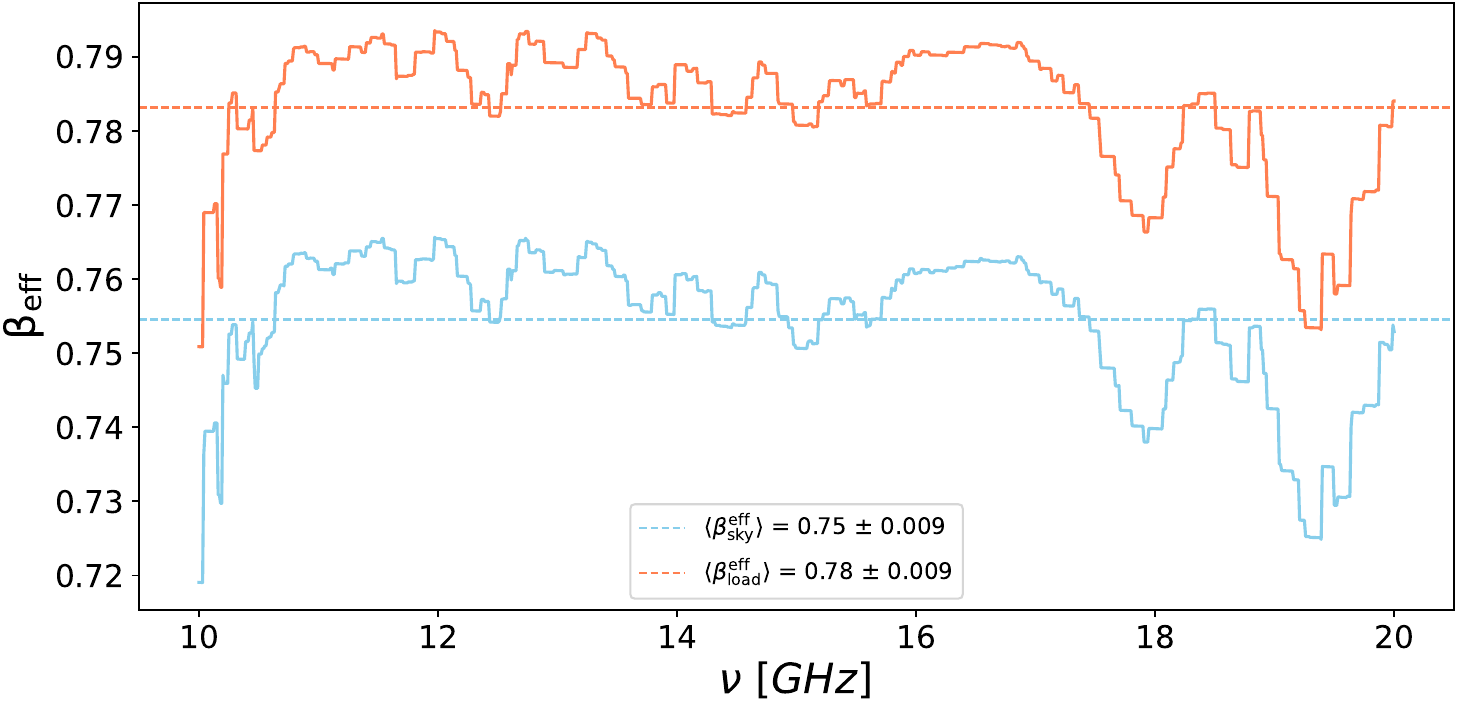}
      \caption{ Effective losses introduced by the sky and load chain components (see Eq.~\ref{eq:totalBskyBload}). The blue curve represent component losses in the sky chain ($\betaskyEffective$), while, the orange curve correspond to component losses in the load chain ($\betaloadEffective$). Losses involve R, L, and SPO terms. The high degree of similarity between the two forms suggests that they are mainly driven by common components inherent in the pseudo-correlation architecture.
              }
\label{betas_offset_t}
\end{figure}

\begin{figure}
\centering
\includegraphics[width=\columnwidth]{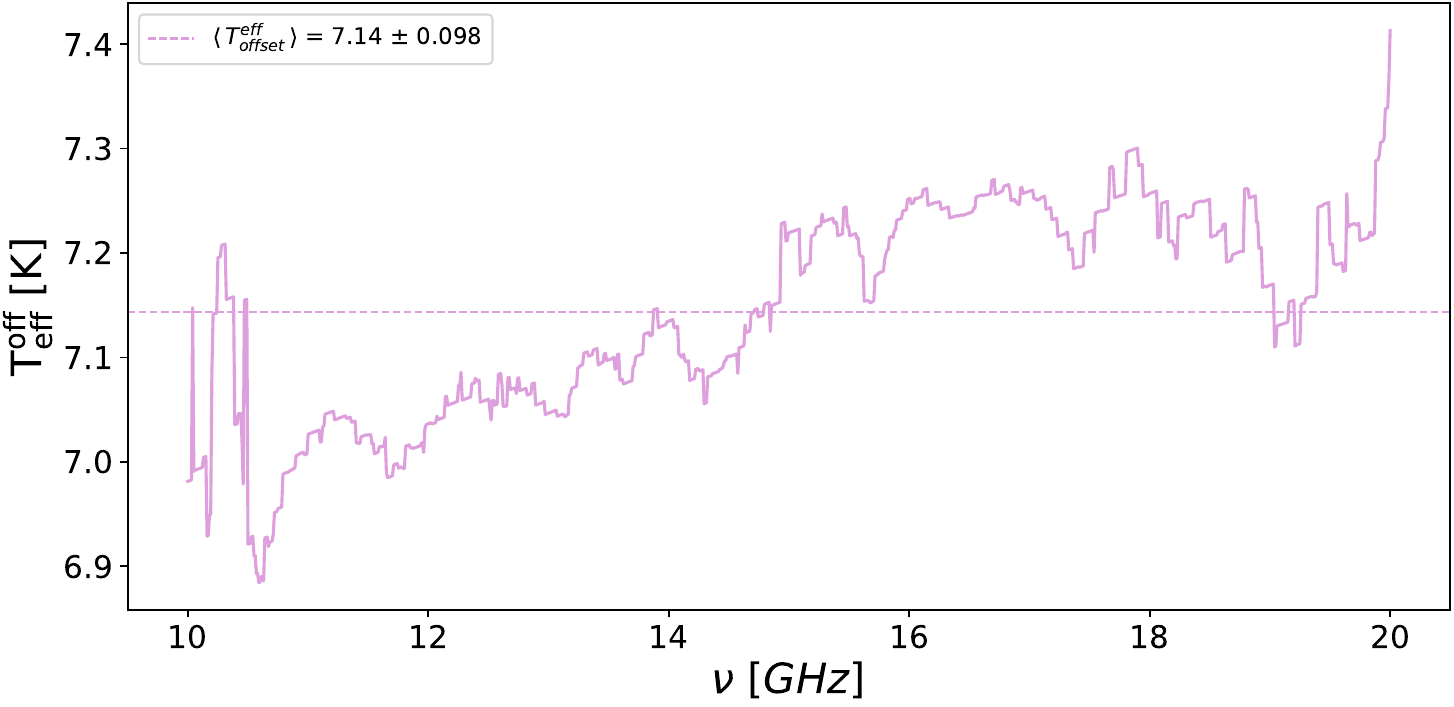}
      \caption{ Total effective offset temperature ($\Toffeff$) of the TMS system (see Eq.~\ref{tn_effect}). This contribution is generated by the losses and physical temperatures of each component. The horizontal pink line indicates its mean value.}
\label{LOSS_T_bem_DC_exc}
\end{figure}

Our result for the full system response ($\Delta T$ in eq.~\ref{delta_temp_total}) is depicted in Figure~\ref{relative_ALL_1}. For completeness, the representation of the effective attenuation coefficients ($\betaskyEffective$ and $\betaloadEffective$) is shown in Fig.~\ref{betas_offset_t}, while the offset term ($T_{\rm off}^{\rm eff}$) is shown in Fig.~\ref{LOSS_T_bem_DC_exc}. Note that due to the assumption of identical gains for the amplifiers, we have $T_{\rm noise}^{\rm eff}=0$.
The code used to compute all these terms is publicly available at the following GitHub repository\footnote{\url{https://github.com/angelaarriero/Systematic_errors_in_spectral_measurements_TMS.git}}.

The results reveal a mean offset of $6.915 \pm 0.094$\,K in the output signal, attributable to asymmetries between the sky and load signal chains caused by non-identical components (specifically the window and IR filter). In Appendix~\ref{app:interpretation}, we show that the window contributes to roughly 78 per cent of this offset, which in terms of antenna temperature corresponds to 5.386\,K from window effects alone. Within this, the most relevant contributions are the insertion loss term (3.09\,K), and the spillover term (2.22\,K). Furthermore, Fig.~\ref{relative_ALL_1} shows two pronounced peaks at the lower and upper ends of the frequency band. These features arise from the OMT response, as characterized by laboratory measurements, and reflect frequency-dependent structures intrinsic to the OMT performance.
%

\subsection{Time stability of the TMS response: absolute measurements}

The TMS requires very high thermal stability. One of the key performance requirements of the instrument is that the operating temperature of the reference load (and internal components) must remain stable to within $\pm 1$\,mK over one hour.
We therefore analyze the case in which the physical temperature of each component in the cold-structure increases by $1$\,mK. The results are summarized in Table~\ref{table:B1_p_2}. 
We find that increasing the physical temperature of each component in the cold-structure individually by $\rm 1\,mK$ results in variations in the total average response of up to $\Delta T=141$\,$\mu$K. If instead all components in the cold structure are simultaneously increased by $\rm 1\,mK$ (physical temperature), the resulting change in the output is approximately $\Delta T=-91.3$\,$\mu$K. 
This value therefore defines the limit of the TMS absolute sky–temperature accuracy, since the intrinsic precision of the instrument thermometry is itself of order 1\,mK. 

On the other hand, components such as the window, the infra-red filter and the top of the sky-horn will have temperature variations from a few tenths of a kelvin up to several hundred millikelvin. Although, these temperature variations can be monitored and thus could be corrected in the final response, it is useful to estimate the amplitude of the expected change. For example, an increase of 10\,K in the window temperature, represents an antenna temperature excess of $\langle \rm \Delta T^a \rangle \approx \rm 0.11\,K$; in the case of the IR filter, variations in the physical temperature of 100\,mK imply differences of $\langle \rm \Delta T^a \rangle \approx \rm 0.2\,mK$; and, for the sky feed-horn, physical temperature variations along the horn might be up to $\rm 200\,mK$, that implies measured temperature differences of $\langle \rm \Delta T^a \rangle \approx \rm 4\,mK$. We can obtain a representative value of the typical variation in the response by increasing simultaneously the window by 1\,K, the infrared filter and sky-feedhorn by $0.1$\,K, and all the components in the cold-structure by 1\,mK. In that case, the resulting change is $\Delta T=-12.9$\,mK.
However, we emphasize that most of this variation can be monitored and corrected.

\begin{table}
\caption{ Total increment in the average output $\rm \Delta T^a$ when the cold-structure components increase their physical temperature $\rm T_c$ by $1$\,mK.}           
\label{table:B1_p_2} 
\centering                          
\begin{tabular}{l c c}        
\hline
        Component & $\rm \langle \Delta T^a_{1\,mK}\rangle [\mu\,K]$ & \\
    
       \hline
       $\rm OMT_s$ &  141.0&\\
       $ \rm Hs_{180}$  & 45.69&\\
       
       $\rm FH_l$  & 20.22&\\
       $\rm OMT_l$ & 141.0&\\
       $ \rm Hl_{180}$ & 45.69&\\
        \hline
\end{tabular}
\end{table}

\subsection{TMS spectral response: relative measurements}
TMS will also take very precise relative measurements of the sky spectrum. The right panel in Figure~\ref{relative_ALL_1} shows the predicted average temperature in the  TMS sub-bands, relative to the mean temperature. We note that the biggest expected contribution per band reaches up an excess of antenna temperature of $\pm$ 200\,mK over the mean value. This spectral dependence will have to be calibrated in order to recover the intrinsic spectrum of the sky signal. Fortunately, this pattern is very stable. For example, if all components in the cold structure are simultaneously increased by $\rm 1\,mK$ (as we did in the previous subsection), the resulting change in the output is approximately $\Delta T=\pm 3$\,$\mu$K. Thus, TMS can in principle provide a very precise determination of the spectral dependence of a given signal across the full 10--20\,GHz band (at microkelvin precision), although the absolute calibration level will be subject to larger uncertainties.

\subsection{Contribution of the amplifiers}
TMS incorporates three amplification stages: the LNAs, the BEM, and the DC. The amplifiers are mainly characterized by three frequency-dependent parameters: the gain, the noise temperature, and the phase shift. In the analysis presented so far, we have assumed that these frequency dependencies are identical for all amplifiers within a given stage; in particular, the four LNAs were considered to exhibit the same spectral behavior. This assumption is consistent with the typical performance reported in manufacturer datasheet shown in Figure \ref{relative_condi_amplifiers}.
Nevertheless, in this subsection, we want to take a step forward and explore how variations of the LNA gains and noise temperatures may affect the final TMS output. Accordingly, exploring amplifier-to-amplifier deviations allows us to assess the robustness of the TMS spectral response under more realistic, non-ideal conditions.


For definiteness, we explore three possible non-ideal configurations for the amplifier responses in Appendix~\ref{C_appendix} (see table~\ref{tab:amp_appC} for the definitions of the parameters, and Figures~\ref{amplifiers_amplitude_1}, \ref{amplifiers_amplitude_2}, and \ref{amplifiers_shifting} for the display of the gains and noise figures). 
All three cases consider variations of 1\,dB with respect to the nominal gain in some of the branches, together with 1\,mK variations in the noise temperatures. Case 1 assumes that the gain differences in the X branch have the opposite sign to those in the Y branch, corresponding to the maximum mismatch scenario. Case 2 introduces gain variations in both branches, but they are partially compensated in the sense that $G_1$ and $G_2$ have opposite signs, as do $G_3$ and $G_4$.
Finally, case 3 only considers (partially compensated) gains variations in $G_1$ and $G_2$. 

%
%

\begin{figure*}[h]
   \centering
   \includegraphics[width=2\columnwidth]{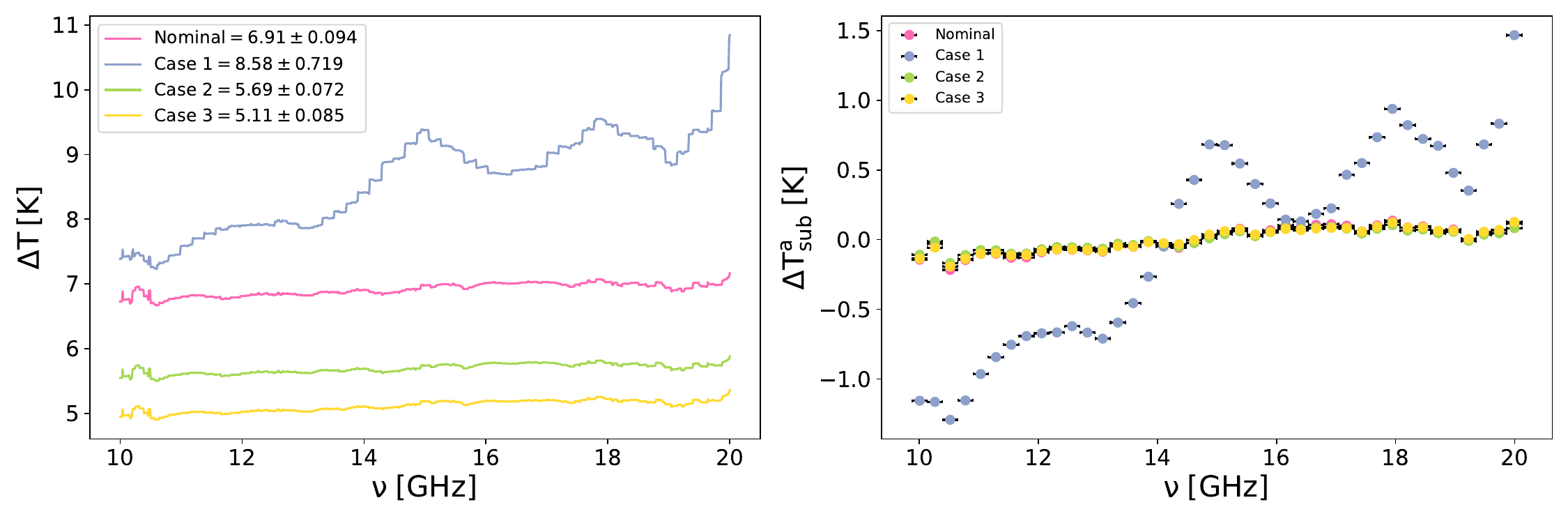}
      \caption{TMS spectral response $\rm \Delta T$ when relative differences in the response of the LNA amplifiers are included (see text for details). Left: measured spectral response. Right: averaged response in spectral bands, and relative to the mean response across the band. }
    \label{Ampli_delta_T_sub_delay_var}
\end{figure*}

Figure~\ref{Ampli_delta_T_sub_delay_var} shows the resulting spectral dependence $\Delta T$ for the three cases. As expected, the largest effect appears in Case~1, corresponding to the maximum mismatch scenario. In this case, the new emerging spectral response is largely dominated by the frequency dependence contribution of the LNA noise (see Fig.~\ref{relative_condi_amplifiers}), whereas in the other cases the effect manifests primarily as an overall offset with respect to the nominal case. For completeness, the right-hand panel of the figure shows the band-averaged response, normalized to the mean level across the band. It is evident that significant variations in the final response arise when uncompensated gain variations are present among the amplifiers within the same branch.These results will guide the selection of Front-End and Back-End LNAs to achieve a highly matched band response between the two branches or to inform the design of optimized RF rectifiers.

\section{Prospects for the TMS calibration strategy}

With the introduction of an FPGA backend in the TMS instrument it is now possible to access the data streams directly from the downconverter
outputs and not just the power output at the end of the pseudo-correlator block. This paper has detailed the various non-idealities expected in the TMS instrument and their effects on the final measurements. Access to the microwave voltages (in phase and amplitude) gives the opportunity to correct for some of these non-idealities.

Although a detailed description of the TMS calibration strategy will be part of a new paper, we anticipate here that we are considering a calibration scheme where a monochromatic stable signal is injected into the front of the TMS receiver. This can be done through an identical feedhorn to the TMS horn assuring it is centred and aligned vertically with the TMS horn. The signal could be the output of a VNA so that it can be swept across the whole band. The signal should be amplitude corrected and carefully injected through an OMT into the X or Y polar direction. In this case the output of the TMS should give a maximum value for Q and a minimum value for U (zero). There will be non-idealities along the signal path that give rise to values for U. Simple mathematical complex multiplication carried out in real time in the FPGA of each of the signal outputs with a coefficient can be done to obtain corrected outputs and thus for a Q polarized signal input the output is Q. The complex multiplier contains 2 components, on one hand there is a correction to the amplitude of the each of the branches and on the other there is a phase rotation so that the phases align as in the ideal case. This process is capable of calibrating out the non-ideal terms which multiply the final signal such as differences in insertion loss or return loss of the various microwave components in the instrument. In the language of our instrument model, those terms corresponds to $\betaskyEffective$ and $\betaloadEffective$ in eq.~\ref{delta_temp_total}.

However, this still leaves residual non-ideal offset terms, and therefore the calibration scheme and procedures need to be further completed. In principle, one possible approach to remove the relative offsets across the full band would be to place an external blackbody at the instrument sky input. Since such a source has an intrinsically flat spectrum, deviations from spectral flatness in the measured signal could be attributed to instrumental effects and potentially corrected. After removing the blackbody, the resulting data would then be calibrated to an effective flat spectral response and an idealized instrumental gain. The ultimate accuracy of this approach remains uncertain and would depend on the availability of a well-characterized and sufficiently stable external blackbody, as well as on the temporal stability of the various non-idealities. The effects associated with the window would not be corrected by this procedure; however, they could, in principle, be characterized separately by studying the relative signal contribution when additional external layers of window material are introduced. These issues will be discussed in a subsequent work.

\section{Discussion and conclusions}

This paper presents an analytical model of the TMS instrument, which employs a pseudo-correlation architecture to perform a continuous and instantaneous comparison between the sky and reference-load signals, largely suppressing correlated noise. This model accounts for potential systematic effects affecting the measurements and is developed in two complementary parts. First, we use the Jones-matrix formalism to provide a parametric description of the non-ideal behavior of all physical elements in the radiometer chain starting at the cryostat entrance window. In this approach, we do not consider imperfections introduced by the optical system upstream of the cryostat. Using realistic values for the expected non-idealities of the individual components (see Table~\ref{table:B1_p_INITIAL}), we find that the dominant errors, primarily associated with intensity-to-polarization and sky-to-load leakage, are expected to be at the percent level. The current model will be a valuable tool in the data processing and post-processing stages, to help in the modeling of those non-idealities. 

The TMS instrument model is then extended to include noise-temperature effects using the Friis formalism. For the intensity signal measured by TMS, we find that the final response, $\Delta T$, includes offset terms in addition to the non-ideal mixing effects described above by the Jones-matrix analysis. The resulting spectral response exhibits frequency-dependent variations across the band, with a mean offset level of about 6.9\,K. 
\\
We find that the insertion loss of the cryostat window introduces a spectral slope that increases the response towards higher frequencies. The return losses of the hybrids and OMTs further distort the spectral response, introducing additional departures from spectral flatness.
In addition to this tilt, the absolute brightness temperature offset is significant. 
%
Overall, the dominant contribution to this offset arises from the cryostat window (approximately 78\,\%), followed by the IR filter and the OMTs, each contributing at the level of roughly 5--7\,\%. 
\\
Using the complete model, we show that under the expected thermal stability requirements of the TMS cryostat (namely, physical temperature variations of the internal cold-structure limited to 1\,mK over one hour), the resulting change in the output signal can reach up to 91.3\,$\mu$K. This value effectively defines the limit on the absolute sky-temperature accuracy achievable by TMS. In contrast, relative spectral measurements across the band are expected to be significantly more precise, remaining stable at levels below 3\,$\mu$K (roughly equivalent to $\sim 10$\,Jy/sr at 10\,GHz) under the same conditions. These results characterize both the intrinsic accuracy and the attainable precision of the instrument under nominal operating conditions.
\\
The TMS instrument model presented here\footnote{\url{https://github.com/angelaarriero/Systematic_errors_in_spectral_measurements_TMS.git}} provides a foundation for the design and optimization of the instrument calibration strategy, which will be addressed in detail in a forthcoming paper. Some of the key aspects of that calibration strategy have been outlined here, and will make use of the flexibility and advantages provided by the FPGA-based data acquisition system. Our findings underscore that accurate control of the relative gains between the two radiometer branches will be critical for fine-tuning the system performance and minimizing residual systematic effects. 
\\
Finally, the TMS instrument model presented here may be used to explore potential mitigation strategies based on improved thermal control. For instance, one can rewrite Eq.~\ref{eq:tw} as
\begin{equation}
    \Tbw \approx \Tsky + (\Tw-\Tsky) L + (\Tenvuno - \Tsky) R + (\Textcry-\Tsky)\SPOw
\label{eq:tw}
\end{equation}

by neglecting second order correction terms. 
In this way, it is evident that the loss- and offset-related terms become negligible when the temperatures of the window ($\Tw$), the external cryostat ($\Textcry$), and the environment ($\Tenvuno$) approach the input sky temperature. However, the practical implementation of such conditions in a ground-based experiment is expected to be challenging, given the inherently less stable thermal environment compared to that of a space mission.
\begin{acknowledgements}
The TMS experiment is being developed by the Instituto de Astrofisica de Canarias (IAC), with an instrumental participation from the INAF group (Bologna, Italy), the University of Milano (Italy), and the Universidad Politecnica de Cartagena (UPCT). 
Partial financial support is provided by the Spanish Ministry of Science and Innovation (MCIN/AEI/\detokenize{10.13039/501100011033}) under the projects IACA15-BE-3707, EQC2018-004918-P, PID2020-120514GB-I00, PID2023-151567NB-I00 and the Severo Ochoa Programs SEV-2015-0548 and CEX2019-000920-S. We acknowledge financial support from the Spanish MICINN under the FEDER Agreement INSIDE-OOCC (ICTS-2019-03-IAC-12).
\end{acknowledgements}

\bibliography{my_bib}

@INPROCEEDINGS{TMS2020,
       author = {{Rubi{\~n}o Mart{\'\i}n}, Jos{\'e} Alberto and {Alonso Arias}, Paz and {Hoyland}, Roger J. and {Aguiar-Gonz{\'a}lez}, Marta and {De Miguel-Hern{\'a}ndez}, Javier and {G{\'e}nova-Santos}, Ricardo T. and {Gomez-Re{\~n}asco}, Maria F. and {Guidi}, Federica and {Fern{\'a}ndez-Izquierdo}, Patricia and {Fern{\'a}ndez-Torreiro}, Mateo and {Fuerte-Rodriguez}, Pablo A. and {Hernandez-Monteagudo}, Carlos and {L{\'o}pez-Caraballo}, Carlos H. and {Perez-de-Taoro}, Angeles and {Peel}, Michael W. and {Rebolo}, Rafael and {Zamora-Jimenez}, Antonio and {Gonz{\'a}lez-Carretero}, Eduardo D. and {Colodro-Conde}, Carlos and {P{\'e}rez-Lemus}, Cristina and {Toledo-Moreo}, Rafael and {P{\'e}rez-Liz{\'a}n}, David and {Cuttaia}, Francesco and {Terenzi}, Luca and {Franceschet}, Cristian and {Realini}, Sabrina and {Chluba}, Jens and {Murga-Llano}, Gaizka and {Sanquirce-Garcia}, Ruben},
        title = "{The Tenerife Microwave Spectrometer (TMS) experiment: studying the absolute spectrum of the sky emission in the 10-20GHz range}",
    booktitle = {Millimeter, Submillimeter, and Far-Infrared Detectors and Instrumentation for Astronomy X},
         year = 2020,
       editor = {{Zmuidzinas}, Jonas and {Gao}, Jian-Rong},
       series = {Society of Photo-Optical Instrumentation Engineers (SPIE) Conference Series},
       volume = {11453},
        month = dec,
          eid = {114530T},
        pages = {114530T},
          doi = {10.1117/12.2561309},
       adsurl = {https://ui.adsabs.harvard.edu/abs/2020SPIE11453E..0TR}
}

@article{arcade2measurementat3,
author = {Fixsen, D. and Kogut, A. and Levin, S. and Limón, Maestreq and Lubin, Philip and Mirel, P. and Seiffert, Michael and Singal, Jiniya and Wollack, Edward and Villela, Thyrso and Wuensche, Carlos},
year = {2011},
month = {05},
pages = {5},
title = {ARCADE 2 Measurement of the Absolute Sky Brightness at 3-90 GHz},
volume = {734},
journal = {The Astrophysical Journal},
doi = {10.1088/0004-637X/734/1/5}
}

@article{PlanckSeiffert_2002,
   title={$\vec{1/f}$ noise and other systematic effects in the Planck-LFI radiometers},
   volume={391},
   ISSN={1432-0746},
   url={http://dx.doi.org/10.1051/0004-6361:20020880},
   DOI={10.1051/0004-6361:20020880},
   number={3},
   journal={Astronomy \&amp; Astrophysics},
   publisher={EDP Sciences},
   author={Seiffert, M. and Mennella, A. and Burigana, C. and Mandolesi, N. and Bersanelli, M. and Meinhold, P. and Lubin, P.},
   year={2002},
   month=aug, pages={1185–1197} }

@inproceedings{Paz1_10.1117/12.2561353,
author = {Paz Alonso-Arias and Jos{\'e} Alberto Rubi{\~n}o-Mart{\'i}n and Roger J. Hoyland and Marta Aguiar-Gonz{\'a}lez and Javier de-Miguel-Hern{\'a}ndez and Ricardo T. G{\'e}nova-Santos and Maria F. Gomez-Re{\~n}asco and Federica Guidi and Mateo Fern{\'a}ndez-Torreiro and Pablo A. Fuerte-Rodr{\'i}guez and Carlos Hern{\'a}ndez-Monteagudo and Carlos H. L{\'o}pez-Caraballo and Angeles Perez-de-Taoro and Michael W. Peel and Rafael Rebolo-L{\'o}pez and Antonio Zamora-Jimenez and Eduardo D. Gonz{\'a}lez-Carretero and Carlos Colodro-Conde and Cristina Perez-Lemus and Rafael Toledo-Moreo and Francesco Cuttaia and Luca Terenzi and Cristian Franceschet and Sabrina Realini},
title = {{New technologies for the Tenerife Microwave Spectrometer and current status}},
volume = {11447},
booktitle = {Ground-based and Airborne Instrumentation for Astronomy VIII},
editor = {Christopher J. Evans and Julia J. Bryant and Kentaro Motohara},
organization = {International Society for Optics and Photonics},
publisher = {SPIE},
pages = {114476N},
year = {2020},
doi = {10.1117/12.2561353},
URL = {https://doi.org/10.1117/12.2561353}
}

@misc{alonsoarias2023microwave,
      title={A Microwave Blackbody Target for Cosmic Microwave Background Spectral Measurements in the 10-20GHz range}, author={P. Alonso-Arias and F. Cuttaia and L. Terenzi and A. Simonetto and P. A. Fuerte-Rodríguez and R. Hoyland and J. A. Rubiño-Martín},year={2023},
      eprint={2309.07320},
      archivePrefix={arXiv},
      primaryClass={astro-ph.IM}
}

@article{Hu_2003,
   title={Benchmark parameters for CMB polarization experiments},
   volume={67},
   ISSN={1089-4918},
   url={http://dx.doi.org/10.1103/PhysRevD.67.043004},
   DOI={10.1103/physrevd.67.043004},
   number={4},
   journal={Physical Review D},
   publisher={American Physical Society (APS)},
   author={Hu, Wayne and Hedman, Matthew M. and Zaldarriaga, Matias},
   year={2003},
   month=feb }

@article{heiles,
author = "Heiles, C",
title = "A Heuristic Introduction to Radioastronomical Polarization",
journal = "Single-Dish Radio Astronomy: Techniques and Applications, ASP Conference Proceedings",
year = "2002",
number = "",
pages = "131-152",
volume = "278",
note = ""
}

@ARTICLE{absmesua_2GHZ_1993,
       author = {{Bensadoun}, M. and {Bersanelli}, M. and {de Amici}, G. and {Kogut}, A. and {Levin}, S.~M. and {Limon}, M. and {Smoot}, G.~F. and {Witebsky}, C.},
        title = "{Measurements of the Cosmic Microwave Background Temperature at 1.47 GHz}",
      journal = {\apj},
         year = 1993,
        month = may,
       volume = {409},
        pages = {1},
          doi = {10.1086/172637},
       adsurl = {https://ui.adsabs.harvard.edu/abs/1993ApJ...409....1B}
}

@ARTICLE{1994ApJ...424..517B,
       author = {{Bersanelli}, M. and {Bensadoun}, M. and {de Amici}, G. and {Levin}, S. and {Limon}, M. and {Smoot}, G.~F. and {Vinje}, W.},
        title = "{Absolute Measurement of the Cosmic Microwave Background at 2 GHz}",
      journal = {\apj},
         year = 1994,
        month = apr,
       volume = {424},
        pages = {517},
          doi = {10.1086/173910},
       adsurl = {https://ui.adsabs.harvard.edu/abs/1994ApJ...424..517B}
}

@article{Alonso-Arias_2024,
doi = {10.1088/1748-0221/19/02/P02040},
url = {https://dx.doi.org/10.1088/1748-0221/19/02/P02040},
year = {2024},
month = {feb},
publisher = {IOP Publishing},
volume = {19},
number = {02},
pages = {P02040},
author = {P. Alonso-Arias and F. Cuttaia and L. Terenzi and A. Simonetto and P.A. Fuerte-Rodríguez and R. Hoyland and J.A. Rubiño-Martín},
title = {A microwave blackbody target for cosmic microwave background spectral measurements in the 10–20 GHz range},
journal = {Journal of Instrumentation}
}

@ARTICLE{ODea2007MNRAS.376.1767O,
       author = {{O'Dea}, Daniel and {Challinor}, Anthony and {Johnson}, Bradley R.},
        title = "{Systematic errors in cosmic microwave background polarization measurements}",
      journal = {\mnras},
         year = 2007,
        month = apr,
       volume = {376},
       number = {4},
        pages = {1767-1783},
          doi = {10.1111/j.1365-2966.2007.11558.x},
archivePrefix = {arXiv},
       eprint = {astro-ph/0610361},
 primaryClass = {astro-ph},
       adsurl = {https://ui.adsabs.harvard.edu/abs/2007MNRAS.376.1767O}
}

@book{kraus1982radio,
  title={Radio Astronomy},
  author={Kraus, J.D.},
  url={https://books.google.es/books?id=Hm_vAAAAMAAJ},
  year={1982},
  publisher={Cygnus-Quasar Books}
}

@INPROCEEDINGS{2016SPIE.9904E..0WK,
       author = {{Kogut}, Alan and {Chluba}, Jens and {Fixsen}, Dale J. and {Meyer}, Stephan and {Spergel}, David},
        title = "{The Primordial Inflation Explorer (PIXIE)}",
    booktitle = {Space Telescopes and Instrumentation 2016: Optical, Infrared, and Millimeter Wave},
         year = 2016,
       editor = {{MacEwen}, Howard A. and {Fazio}, Giovanni G. and {Lystrup}, Makenzie and {Batalha}, Natalie and {Siegler}, Nicholas and {Tong}, Edward C.},
       series = {Society of Photo-Optical Instrumentation Engineers (SPIE) Conference Series},
       volume = {9904},
        month = jul,
          eid = {99040W},
        pages = {99040W},
          doi = {10.1117/12.2231090},
       adsurl = {https://ui.adsabs.harvard.edu/abs/2016SPIE.9904E..0WK}
}

@ARTICLE{2025JCAP...04..020K,
       author = {{Kogut}, Alan and {Aghanim}, Nabila and {Chluba}, Jens and {Chuss}, David T. and {Delabrouille}, Jacques and {Dvorkin}, Cora and {Fixsen}, Dale and {Ghosh}, Shamik and {Hensley}, Brandon S. and {Hill}, J. Colin and {Maffei}, Bruno and {Pullen}, Anthony R. and {Rotti}, Aditya and {Sabyr}, Alina and {Switzer}, Eric R. and {Thiele}, Leander and {Wollack}, Edward J. and {Zelko}, Ioana},
        title = "{The Primordial Inflation Explorer (PIXIE): mission design and science goals}",
      journal = {\jcap},
         year = 2025,
        month = apr,
       volume = {2025},
       number = {4},
          eid = {020},
        pages = {020},
          doi = {10.1088/1475-7516/2025/04/020},
archivePrefix = {arXiv},
       eprint = {2405.20403},
 primaryClass = {astro-ph.CO},
       adsurl = {https://ui.adsabs.harvard.edu/abs/2025JCAP...04..020K}
}

@ARTICLE{L-BASS,
       author = {{Zerafa}, D.~P. and {Wilkinson}, P.~N. and {Radcliffe}, C.~J. and {Leahy}, J.~P. and {Browne}, I.~W.~A. and {Black}, P.~J.},
        title = "{L-BASS: a project to produce an absolutely calibrated 1.4 GHz sky map I {\textendash} Scientific rationale and system overview}",
      journal = {RAS Techniques and Instruments},
         year = 2025,
        month = jan,
       volume = {4},
          eid = {rzaf017},
        pages = {rzaf017},
          doi = {10.1093/rasti/rzaf017},
archivePrefix = {arXiv},
       eprint = {2505.08465},
 primaryClass = {astro-ph.IM},
       adsurl = {https://ui.adsabs.harvard.edu/abs/2025RASTI...4...17Z}
}

@ARTICLE{Planck2020-VI,
       author = {{Planck Collaboration} and {Aghanim}, N. and {Akrami}, Y. and {Ashdown}, M. and {Aumont}, J. and {Baccigalupi}, C. and {Ballardini}, M. and {Banday}, A.~J. and {Barreiro}, R.~B. and {Bartolo}, N. and {Basak}, S. and {Battye}, R. and {Benabed}, K. and {Bernard}, J. -P. and {Bersanelli}, M. and {Bielewicz}, P. and {Bock}, J.~J. and {Bond}, J.~R. and {Borrill}, J. and {Bouchet}, F.~R. and {Boulanger}, F. and {Bucher}, M. and {Burigana}, C. and {Butler}, R.~C. and {Calabrese}, E. and {Cardoso}, J. -F. and {Carron}, J. and {Challinor}, A. and {Chiang}, H.~C. and {Chluba}, J. and {Colombo}, L.~P.~L. and {Combet}, C. and {Contreras}, D. and {Crill}, B.~P. and {Cuttaia}, F. and {de Bernardis}, P. and {de Zotti}, G. and {Delabrouille}, J. and {Delouis}, J. -M. and {Di Valentino}, E. and {Diego}, J.~M. and {Dor{\'e}}, O. and {Douspis}, M. and {Ducout}, A. and {Dupac}, X. and {Dusini}, S. and {Efstathiou}, G. and {Elsner}, F. and {En{\ss}lin}, T.~A. and {Eriksen}, H.~K. and {Fantaye}, Y. and {Farhang}, M. and {Fergusson}, J. and {Fernandez-Cobos}, R. and {Finelli}, F. and {Forastieri}, F. and {Frailis}, M. and {Fraisse}, A.~A. and {Franceschi}, E. and {Frolov}, A. and {Galeotta}, S. and {Galli}, S. and {Ganga}, K. and {G{\'e}nova-Santos}, R.~T. and {Gerbino}, M. and {Ghosh}, T. and {Gonz{\'a}lez-Nuevo}, J. and {G{\'o}rski}, K.~M. and {Gratton}, S. and {Gruppuso}, A. and {Gudmundsson}, J.~E. and {Hamann}, J. and {Handley}, W. and {Hansen}, F.~K. and {Herranz}, D. and {Hildebrandt}, S.~R. and {Hivon}, E. and {Huang}, Z. and {Jaffe}, A.~H. and {Jones}, W.~C. and {Karakci}, A. and {Keih{\"a}nen}, E. and {Keskitalo}, R. and {Kiiveri}, K. and {Kim}, J. and {Kisner}, T.~S. and {Knox}, L. and {Krachmalnicoff}, N. and {Kunz}, M. and {Kurki-Suonio}, H. and {Lagache}, G. and {Lamarre}, J. -M. and {Lasenby}, A. and {Lattanzi}, M. and {Lawrence}, C.~R. and {Le Jeune}, M. and {Lemos}, P. and {Lesgourgues}, J. and {Levrier}, F. and {Lewis}, A. and {Liguori}, M. and {Lilje}, P.~B. and {Lilley}, M. and {Lindholm}, V. and {L{\'o}pez-Caniego}, M. and {Lubin}, P.~M. and {Ma}, Y. -Z. and {Mac{\'\i}as-P{\'e}rez}, J.~F. and {Maggio}, G. and {Maino}, D. and {Mandolesi}, N. and {Mangilli}, A. and {Marcos-Caballero}, A. and {Maris}, M. and {Martin}, P.~G. and {Martinelli}, M. and {Mart{\'\i}nez-Gonz{\'a}lez}, E. and {Matarrese}, S. and {Mauri}, N. and {McEwen}, J.~D. and {Meinhold}, P.~R. and {Melchiorri}, A. and {Mennella}, A. and {Migliaccio}, M. and {Millea}, M. and {Mitra}, S. and {Miville-Desch{\^e}nes}, M. -A. and {Molinari}, D. and {Montier}, L. and {Morgante}, G. and {Moss}, A. and {Natoli}, P. and {N{\o}rgaard-Nielsen}, H.~U. and {Pagano}, L. and {Paoletti}, D. and {Partridge}, B. and {Patanchon}, G. and {Peiris}, H.~V. and {Perrotta}, F. and {Pettorino}, V. and {Piacentini}, F. and {Polastri}, L. and {Polenta}, G. and {Puget}, J. -L. and {Rachen}, J.~P. and {Reinecke}, M. and {Remazeilles}, M. and {Renzi}, A. and {Rocha}, G. and {Rosset}, C. and {Roudier}, G. and {Rubi{\~n}o-Mart{\'\i}n}, J.~A. and {Ruiz-Granados}, B. and {Salvati}, L. and {Sandri}, M. and {Savelainen}, M. and {Scott}, D. and {Shellard}, E.~P.~S. and {Sirignano}, C. and {Sirri}, G. and {Spencer}, L.~D. and {Sunyaev}, R. and {Suur-Uski}, A. -S. and {Tauber}, J.~A. and {Tavagnacco}, D. and {Tenti}, M. and {Toffolatti}, L. and {Tomasi}, M. and {Trombetti}, T. and {Valenziano}, L. and {Valiviita}, J. and {Van Tent}, B. and {Vibert}, L. and {Vielva}, P. and {Villa}, F. and {Vittorio}, N. and {Wandelt}, B.~D. and {Wehus}, I.~K. and {White}, M. and {White}, S.~D.~M. and {Zacchei}, A. and {Zonca}, A.},
        title = "{Planck 2018 results. VI. Cosmological parameters}",
      journal = {\aap},
         year = 2020,
        month = sep,
       volume = {641},
          eid = {A6},
        pages = {A6},
          doi = {10.1051/0004-6361/201833910},
archivePrefix = {arXiv},
       eprint = {1807.06209},
 primaryClass = {astro-ph.CO},
       adsurl = {https://ui.adsabs.harvard.edu/abs/2020A&A...641A...6P}
}

@ARTICLE{WMAP9,
       author = {{Bennett}, C.~L. and {Larson}, D. and {Weiland}, J.~L. and {Jarosik}, N. and {Hinshaw}, G. and {Odegard}, N. and {Smith}, K.~M. and {Hill}, R.~S. and {Gold}, B. and {Halpern}, M. and {Komatsu}, E. and {Nolta}, M.~R. and {Page}, L. and {Spergel}, D.~N. and {Wollack}, E. and {Dunkley}, J. and {Kogut}, A. and {Limon}, M. and {Meyer}, S.~S. and {Tucker}, G.~S. and {Wright}, E.~L.},
        title = "{Nine-year Wilkinson Microwave Anisotropy Probe (WMAP) Observations: Final Maps and Results}",
      journal = {\apjs},
         year = 2013,
        month = oct,
       volume = {208},
       number = {2},
          eid = {20},
        pages = {20},
          doi = {10.1088/0067-0049/208/2/20},
archivePrefix = {arXiv},
       eprint = {1212.5225},
 primaryClass = {astro-ph.CO},
       adsurl = {https://ui.adsabs.harvard.edu/abs/2013ApJS..208...20B}
}

@ARTICLE{SO,
       author = {{Ade}, Peter and {Aguirre}, James and {Ahmed}, Zeeshan and {Aiola}, Simone and {Ali}, Aamir and {Alonso}, David and {Alvarez}, Marcelo A. and {Arnold}, Kam and {Ashton}, Peter and {Austermann}, Jason and {Awan}, Humna and {Baccigalupi}, Carlo and {Baildon}, Taylor and {Barron}, Darcy and {Battaglia}, Nick and {Battye}, Richard and {Baxter}, Eric and {Bazarko}, Andrew and {Beall}, James A. and {Bean}, Rachel and {Beck}, Dominic and {Beckman}, Shawn and {Beringue}, Benjamin and {Bianchini}, Federico and {Boada}, Steven and {Boettger}, David and {Bond}, J. Richard and {Borrill}, Julian and {Brown}, Michael L. and {Bruno}, Sarah Marie and {Bryan}, Sean and {Calabrese}, Erminia and {Calafut}, Victoria and {Calisse}, Paolo and {Carron}, Julien and {Challinor}, Anthony and {Chesmore}, Grace and {Chinone}, Yuji and {Chluba}, Jens and {Cho}, Hsiao-Mei Sherry and {Choi}, Steve and {Coppi}, Gabriele and {Cothard}, Nicholas F. and {Coughlin}, Kevin and {Crichton}, Devin and {Crowley}, Kevin D. and {Crowley}, Kevin T. and {Cukierman}, Ari and {D'Ewart}, John M. and {D{\"u}nner}, Rolando and {de Haan}, Tijmen and {Devlin}, Mark and {Dicker}, Simon and {Didier}, Joy and {Dobbs}, Matt and {Dober}, Bradley and {Duell}, Cody J. and {Duff}, Shannon and {Duivenvoorden}, Adri and {Dunkley}, Jo and {Dusatko}, John and {Errard}, Josquin and {Fabbian}, Giulio and {Feeney}, Stephen and {Ferraro}, Simone and {Flux{\`a}}, Pedro and {Freese}, Katherine and {Frisch}, Josef C. and {Frolov}, Andrei and {Fuller}, George and {Fuzia}, Brittany and {Galitzki}, Nicholas and {Gallardo}, Patricio A. and {Tomas Galvez Ghersi}, Jose and {Gao}, Jiansong and {Gawiser}, Eric and {Gerbino}, Martina and {Gluscevic}, Vera and {Goeckner-Wald}, Neil and {Golec}, Joseph and {Gordon}, Sam and {Gralla}, Megan and {Green}, Daniel and {Grigorian}, Arpi and {Groh}, John and {Groppi}, Chris and {Guan}, Yilun and {Gudmundsson}, Jon E. and {Han}, Dongwon and {Hargrave}, Peter and {Hasegawa}, Masaya and {Hasselfield}, Matthew and {Hattori}, Makoto and {Haynes}, Victor and {Hazumi}, Masashi and {He}, Yizhou and {Healy}, Erin and {Henderson}, Shawn W. and {Hervias-Caimapo}, Carlos and {Hill}, Charles A. and {Hill}, J. Colin and {Hilton}, Gene and {Hilton}, Matt and {Hincks}, Adam D. and {Hinshaw}, Gary and {Hlo{\v{z}}ek}, Ren{\'e}e and {Ho}, Shirley and {Ho}, Shuay-Pwu Patty and {Howe}, Logan and {Huang}, Zhiqi and {Hubmayr}, Johannes and {Huffenberger}, Kevin and {Hughes}, John P. and {Ijjas}, Anna and {Ikape}, Margaret and {Irwin}, Kent and {Jaffe}, Andrew H. and {Jain}, Bhuvnesh and {Jeong}, Oliver and {Kaneko}, Daisuke and {Karpel}, Ethan D. and {Katayama}, Nobuhiko and {Keating}, Brian and {Kernasovskiy}, Sarah S. and {Keskitalo}, Reijo and {Kisner}, Theodore and {Kiuchi}, Kenji and {Klein}, Jeff and {Knowles}, Kenda and {Koopman}, Brian and {Kosowsky}, Arthur and {Krachmalnicoff}, Nicoletta and {Kuenstner}, Stephen E. and {Kuo}, Chao-Lin and {Kusaka}, Akito and {Lashner}, Jacob and {Lee}, Adrian and {Lee}, Eunseong and {Leon}, David and {Leung}, Jason S. -Y. and {Lewis}, Antony and {Li}, Yaqiong and {Li}, Zack and {Limon}, Michele and {Linder}, Eric and {Lopez-Caraballo}, Carlos and {Louis}, Thibaut and {Lowry}, Lindsay and {Lungu}, Marius and {Madhavacheril}, Mathew and {Mak}, Daisy and {Maldonado}, Felipe and {Mani}, Hamdi and {Mates}, Ben and {Matsuda}, Frederick and {Maurin}, Lo{\"\i}c and {Mauskopf}, Phil and {May}, Andrew and {McCallum}, Nialh and {McKenney}, Chris and {McMahon}, Jeff and {Meerburg}, P. Daniel and {Meyers}, Joel and {Miller}, Amber and {Mirmelstein}, Mark and {Moodley}, Kavilan and {Munchmeyer}, Moritz and {Munson}, Charles and {Naess}, Sigurd and {Nati}, Federico and {Navaroli}, Martin and {Newburgh}, Laura and {Nguyen}, Ho Nam and {Niemack}, Michael and {Nishino}, Haruki and {Orlowski-Scherer}, John and {Page}, Lyman and {Partridge}, Bruce and {Peloton}, Julien and {Perrotta}, Francesca and {Piccirillo}, Lucio and {Pisano}, Giampaolo and {Poletti}, Davide and {Puddu}, Roberto and {Puglisi}, Giuseppe and {Raum}, Chris and {Reichardt}, Christian L. and {Remazeilles}, Mathieu and {Rephaeli}, Yoel and {Riechers}, Dominik and {Rojas}, Felipe and {Roy}, Anirban and {Sadeh}, Sharon and {Sakurai}, Yuki and {Salatino}, Maria and {Sathyanarayana Rao}, Mayuri and {Schaan}, Emmanuel and {Schmittfull}, Marcel and {Sehgal}, Neelima and {Seibert}, Joseph and {Seljak}, Uros and {Sherwin}, Blake and {Shimon}, Meir and {Sierra}, Carlos and {Sievers}, Jonathan and {Sikhosana}, Precious and {Silva-Feaver}, Maximiliano and {Simon}, Sara M. and {Sinclair}, Adrian and {Siritanasak}, Praween and {Smith}, Kendrick and {Smith}, Stephen R. and {Spergel}, David and {Staggs}, Suzanne T. and {Stein}, George and {Stevens}, Jason R. and {Stompor}, Radek and {Suzuki}, Aritoki and {Tajima}, Osamu and {Takakura}, Satoru and {Teply}, Grant and {Thomas}, Daniel B. and {Thorne}, Ben and {Thornton}, Robert and {Trac}, Hy and {Tsai}, Calvin and {Tucker}, Carole and {Ullom}, Joel and {Vagnozzi}, Sunny and {van Engelen}, Alexander and {Van Lanen}, Jeff and {Van Winkle}, Daniel D. and {Vavagiakis}, Eve M. and {Verg{\`e}s}, Clara and {Vissers}, Michael and {Wagoner}, Kasey and {Walker}, Samantha and {Ward}, Jon and {Westbrook}, Ben and {Whitehorn}, Nathan and {Williams}, Jason and {Williams}, Joel and {Wollack}, Edward J. and {Xu}, Zhilei and {Yu}, Byeonghee and {Yu}, Cyndia and {Zago}, Fernando and {Zhang}, Hezi and {Zhu}, Ningfeng and {Simons Observatory Collaboration}},
        title = "{The Simons Observatory: science goals and forecasts}",
      journal = {\jcap},
         year = 2019,
        month = feb,
       volume = {2019},
       number = {2},
          eid = {056},
        pages = {056},
          doi = {10.1088/1475-7516/2019/02/056},
archivePrefix = {arXiv},
       eprint = {1808.07445},
 primaryClass = {astro-ph.CO},
       adsurl = {https://ui.adsabs.harvard.edu/abs/2019JCAP...02..056A}
}

@ARTICLE{CMB-S4,
       author = {{Abazajian}, Kevork and {Addison}, Graeme and {Adshead}, Peter and {Ahmed}, Zeeshan and {Allen}, Steven W. and {Alonso}, David and {Alvarez}, Marcelo and {Anderson}, Adam and {Arnold}, Kam S. and {Baccigalupi}, Carlo and {Bailey}, Kathy and {Barkats}, Denis and {Barron}, Darcy and {Barry}, Peter S. and {Bartlett}, James G. and {Basu Thakur}, Ritoban and {Battaglia}, Nicholas and {Baxter}, Eric and {Bean}, Rachel and {Bebek}, Chris and {Bender}, Amy N. and {Benson}, Bradford A. and {Berger}, Edo and {Bhimani}, Sanah and {Bischoff}, Colin A. and {Bleem}, Lindsey and {Bocquet}, Sebastian and {Boddy}, Kimberly and {Bonato}, Matteo and {Bond}, J. Richard and {Borrill}, Julian and {Bouchet}, Fran{\c{c}}ois R. and {Brown}, Michael L. and {Bryan}, Sean and {Burkhart}, Blakesley and {Buza}, Victor and {Byrum}, Karen and {Calabrese}, Erminia and {Calafut}, Victoria and {Caldwell}, Robert and {Carlstrom}, John E. and {Carron}, Julien and {Cecil}, Thomas and {Challinor}, Anthony and {Chang}, Clarence L. and {Chinone}, Yuji and {Cho}, Hsiao-Mei Sherry and {Cooray}, Asantha and {Crawford}, Thomas M. and {Crites}, Abigail and {Cukierman}, Ari and {Cyr-Racine}, Francis-Yan and {de Haan}, Tijmen and {de Zotti}, Gianfranco and {Delabrouille}, Jacques and {Demarteau}, Marcel and {Devlin}, Mark and {Di Valentino}, Eleonora and {Dobbs}, Matt and {Duff}, Shannon and {Duivenvoorden}, Adriaan and {Dvorkin}, Cora and {Edwards}, William and {Eimer}, Joseph and {Errard}, Josquin and {Essinger-Hileman}, Thomas and {Fabbian}, Giulio and {Feng}, Chang and {Ferraro}, Simone and {Filippini}, Jeffrey P. and {Flauger}, Raphael and {Flaugher}, Brenna and {Fraisse}, Aurelien A. and {Frolov}, Andrei and {Galitzki}, Nicholas and {Galli}, Silvia and {Ganga}, Ken and {Gerbino}, Martina and {Gilchriese}, Murdock and {Gluscevic}, Vera and {Green}, Daniel and {Grin}, Daniel and {Grohs}, Evan and {Gualtieri}, Riccardo and {Guarino}, Victor and {Gudmundsson}, Jon E. and {Habib}, Salman and {Haller}, Gunther and {Halpern}, Mark and {Halverson}, Nils W. and {Hanany}, Shaul and {Harrington}, Kathleen and {Hasegawa}, Masaya and {Hasselfield}, Matthew and {Hazumi}, Masashi and {Heitmann}, Katrin and {Henderson}, Shawn and {Henning}, Jason W. and {Hill}, J. Colin and {Hlozek}, Ren{\'e}e and {Holder}, Gil and {Holzapfel}, William and {Hubmayr}, Johannes and {Huffenberger}, Kevin M. and {Huffer}, Michael and {Hui}, Howard and {Irwin}, Kent and {Johnson}, Bradley R. and {Johnstone}, Doug and {Jones}, William C. and {Karkare}, Kirit and {Katayama}, Nobuhiko and {Kerby}, James and {Kernovsky}, Sarah and {Keskitalo}, Reijo and {Kisner}, Theodore and {Knox}, Lloyd and {Kosowsky}, Arthur and {Kovac}, John and {Kovetz}, Ely D. and {Kuhlmann}, Steve and {Kuo}, Chao-lin and {Kurita}, Nadine and {Kusaka}, Akito and {Lahteenmaki}, Anne and {Lawrence}, Charles R. and {Lee}, Adrian T. and {Lewis}, Antony and {Li}, Dale and {Linder}, Eric and {Loverde}, Marilena and {Lowitz}, Amy and {Madhavacheril}, Mathew S. and {Mantz}, Adam and {Matsuda}, Frederick and {Mauskopf}, Philip and {McMahon}, Jeff and {McQuinn}, Matthew and {Meerburg}, P. Daniel and {Melin}, Jean-Baptiste and {Meyers}, Joel and {Millea}, Marius and {Mohr}, Joseph and {Moncelsi}, Lorenzo and {Mroczkowski}, Tony and {Mukherjee}, Suvodip and {M{\"u}nchmeyer}, Moritz and {Nagai}, Daisuke and {Nagy}, Johanna and {Namikawa}, Toshiya and {Nati}, Federico and {Natoli}, Tyler and {Negrello}, Mattia and {Newburgh}, Laura and {Niemack}, Michael D. and {Nishino}, Haruki and {Nordby}, Martin and {Novosad}, Valentine and {O'Connor}, Paul and {Obied}, Georges and {Padin}, Stephen and {Pandey}, Shivam and {Partridge}, Bruce and {Pierpaoli}, Elena and {Pogosian}, Levon and {Pryke}, Clement and {Puglisi}, Giuseppe and {Racine}, Benjamin and {Raghunathan}, Srinivasan and {Rahlin}, Alexandra and {Rajagopalan}, Srini and {Raveri}, Marco and {Reichanadter}, Mark and {Reichardt}, Christian L. and {Remazeilles}, Mathieu and {Rocha}, Graca and {Roe}, Natalie A. and {Roy}, Anirban and {Ruhl}, John and {Salatino}, Maria and {Saliwanchik}, Benjamin and {Schaan}, Emmanuel and {Schillaci}, Alessandro and {Schmittfull}, Marcel M. and {Scott}, Douglas and {Sehgal}, Neelima and {Shandera}, Sarah and {Sheehy}, Christopher and {Sherwin}, Blake D. and {Shirokoff}, Erik and {Simon}, Sara M. and {Slosar}, Anze and {Somerville}, Rachel and {Spergel}, David and {Staggs}, Suzanne T. and {Stark}, Antony and {Stompor}, Radek and {Story}, Kyle T. and {Stoughton}, Chris and {Suzuki}, Aritoki and {Tajima}, Osamu and {Teply}, Grant P. and {Thompson}, Keith and {Timbie}, Peter and {Tomasi}, Maurizio and {Treu}, Jesse I. and {Tristram}, Matthieu and {Tucker}, Gregory and {Umilt{\`a}}, Caterina and {van Engelen}, Alexander and {Vieira}, Joaquin D. and {Vieregg}, Abigail G. and {Vogelsberger}, Mark and {Wang}, Gensheng and {Watson}, Scott and {White}, Martin and {Whitehorn}, Nathan and {Wollack}, Edward J. and {Kimmy Wu}, W.~L. and {Xu}, Zhilei and {Yasini}, Siavash and {Yeck}, James and {Yoon}, Ki Won and {Young}, Edward and {Zonca}, Andrea},
        title = "{CMB-S4 Science Case, Reference Design, and Project Plan}",
      journal = {arXiv e-prints},
         year = 2019,
        month = jul,
          eid = {arXiv:1907.04473},
        pages = {arXiv:1907.04473},
          doi = {10.48550/arXiv.1907.04473},
archivePrefix = {arXiv},
       eprint = {1907.04473},
 primaryClass = {astro-ph.IM},
       adsurl = {https://ui.adsabs.harvard.edu/abs/2019arXiv190704473A}
}

@INPROCEEDINGS{MFI2,
       author = {{Hoyland}, Roger J. and {Rubi{\~n}o-Mart{\'\i}n}, Jos{\'e} Alberto and {Aguiar-Gonzalez}, Marta and {Alonso-Arias}, Paz and {Artal}, Eduardo and {Ashdown}, Mark and {Barreiro}, R.~B. and {Casas}, Francisco J. and {Colodro-Conde}, Carlos and {de la Hoz}, Elena and {Fern{\'a}ndez-Torreiro}, Mateo and {Fuerte-Rodriguez}, Pablo A. and {G{\'e}nova-Santos}, Ricardo T. and {G{\'o}mez-Re{\~n}asco}, Maria F. and {Gonz{\'a}lez-Carretero}, Eduardo D. and {Gonz{\'a}lez-Gonz{\'a}lez}, Raul and {Guidi}, Frederica and {Hern{\'a}ndez-Monteagudo}, Carlos and {Herranz}, Diego and {Lasenby}, Anthony N. and {L{\'o}pez-Caraballo}, Carlos H. and {Mart{\'\i}nez-Gonzalez}, Enr{\'\i}que and {Oria-Carreras}, Asier and {Peel}, Michael W. and {P{\'e}rez-de-Taoro}, Angeles and {P{\'e}rez-Lemus}, Cristina and {Piccirillo}, Lucio and {Rebolo}, Rafael and {Rodr{\'\i}guez-D{\'\i}az}, Jes{\'u}s Salvador and {Toledo-Moreo}, Rafael and {Vega-Moreno}, Afrodisio and {Vielva}, Patricio and {Watson}, Robert A. and {Zamora-Jimenez}, Antonio},
        title = "{The new multi-frequency instrument (MFI2) for the QUIJOTE facility in Tenerife}",
    booktitle = {Millimeter, Submillimeter, and Far-Infrared Detectors and Instrumentation for Astronomy XI},
         year = 2022,
       editor = {{Zmuidzinas}, Jonas and {Gao}, Jian-Rong},
       series = {Society of Photo-Optical Instrumentation Engineers (SPIE) Conference Series},
       volume = {12190},
        month = aug,
          eid = {1219033},
        pages = {1219033},
          doi = {10.1117/12.2640826},
       adsurl = {https://ui.adsabs.harvard.edu/abs/2022SPIE12190E..33H}
}

@ARTICLE{LiteBIRD_ptep,
       author = {{LiteBIRD Collaboration} and {Allys}, E. and {Arnold}, K. and {Aumont}, J. and {Aurlien}, R. and {Azzoni}, S. and {Baccigalupi}, C. and {Banday}, A.~J. and {Banerji}, R. and {Barreiro}, R.~B. and {Bartolo}, N. and {Bautista}, L. and {Beck}, D. and {Beckman}, S. and {Bersanelli}, M. and {Boulanger}, F. and {Brilenkov}, M. and {Bucher}, M. and {Calabrese}, E. and {Campeti}, P. and {Carones}, A. and {Casas}, F.~J. and {Catalano}, A. and {Chan}, V. and {Cheung}, K. and {Chinone}, Y. and {Clark}, S.~E. and {Columbro}, F. and {D'Alessandro}, G. and {de Bernardis}, P. and {de Haan}, T. and {de la Hoz}, E. and {De Petris}, M. and {Torre}, S. Della and {Diego-Palazuelos}, P. and {Dobbs}, M. and {Dotani}, T. and {Duval}, J.~M. and {Elleflot}, T. and {Eriksen}, H.~K. and {Errard}, J. and {Essinger-Hileman}, T. and {Finelli}, F. and {Flauger}, R. and {Franceschet}, C. and {Fuskeland}, U. and {Galloway}, M. and {Ganga}, K. and {Gerbino}, M. and {Gervasi}, M. and {G{\'e}nova-Santos}, R.~T. and {Ghigna}, T. and {Giardiello}, S. and {Gjerl{\o}w}, E. and {Grain}, J. and {Grupp}, F. and {Gruppuso}, A. and {Gudmundsson}, J.~E. and {Halverson}, N.~W. and {Hargrave}, P. and {Hasebe}, T. and {Hasegawa}, M. and {Hazumi}, M. and {Henrot-Versill{\'e}}, S. and {Hensley}, B. and {Hergt}, L.~T. and {Herman}, D. and {Hivon}, E. and {Hlozek}, R.~A. and {Hornsby}, A.~L. and {Hoshino}, Y. and {Hubmayr}, J. and {Ichiki}, K. and {Iida}, T. and {Imada}, H. and {Ishino}, H. and {Jaehnig}, G. and {Katayama}, N. and {Kato}, A. and {Keskitalo}, R. and {Kisner}, T. and {Kobayashi}, Y. and {Kogut}, A. and {Kohri}, K. and {Komatsu}, E. and {Komatsu}, K. and {Konishi}, K. and {Krachmalnicoff}, N. and {Kuo}, C.~L. and {Lamagna}, L. and {Lattanzi}, M. and {Lee}, A.~T. and {Leloup}, C. and {Levrier}, F. and {Linder}, E. and {Luzzi}, G. and {Macias-Perez}, J. and {Maciaszek}, T. and {Maffei}, B. and {Maino}, D. and {Mandelli}, S. and {Mart{\'\i}nez-Gonz{\'a}lez}, E. and {Masi}, S. and {Massa}, M. and {Matarrese}, S. and {Matsuda}, F.~T. and {Matsumura}, T. and {Mele}, L. and {Migliaccio}, M. and {Minami}, Y. and {Moggi}, A. and {Montgomery}, J. and {Montier}, L. and {Morgante}, G. and {Mot}, B. and {Nagano}, Y. and {Nagasaki}, T. and {Nagata}, R. and {Nakano}, R. and {Namikawa}, T. and {Nati}, F. and {Natoli}, P. and {Nerval}, S. and {Noviello}, F. and {Odagiri}, K. and {Oguri}, S. and {Ohsaki}, H. and {Pagano}, L. and {Paiella}, A. and {Paoletti}, D. and {Passerini}, A. and {Patanchon}, G. and {Piacentini}, F. and {Piat}, M. and {Pisano}, G. and {Polenta}, G. and {Poletti}, D. and {Prouv{\'e}}, T. and {Puglisi}, G. and {Rambaud}, D. and {Raum}, C. and {Realini}, S. and {Reinecke}, M. and {Remazeilles}, M. and {Ritacco}, A. and {Roudil}, G. and {Rubino-Martin}, J.~A. and {Russell}, M. and {Sakurai}, H. and {Sakurai}, Y. and {Sasaki}, M. and {Scott}, D. and {Sekimoto}, Y. and {Shinozaki}, K. and {Shiraishi}, M. and {Shirron}, P. and {Signorelli}, G. and {Spinella}, F. and {Stever}, S. and {Stompor}, R. and {Sugiyama}, S. and {Sullivan}, R.~M. and {Suzuki}, A. and {Svalheim}, T.~L. and {Switzer}, E. and {Takaku}, R. and {Takakura}, H. and {Takase}, Y. and {Tartari}, A. and {Terao}, Y. and {Thermeau}, J. and {Thommesen}, H. and {Thompson}, K.~L. and {Tomasi}, M. and {Tominaga}, M. and {Tristram}, M. and {Tsuji}, M. and {Tsujimoto}, M. and {Vacher}, L. and {Vielva}, P. and {Vittorio}, N. and {Wang}, W. and {Watanuki}, K. and {Wehus}, I.~K. and {Weller}, J. and {Westbrook}, B. and {Wilms}, J. and {Winter}, B. and {Wollack}, E.~J. and {Yumoto}, J. and {Zannoni}, M. and {Collaboration LiteB I R D}},
        title = "{Probing cosmic inflation with the LiteBIRD cosmic microwave background polarization survey}",
      journal = {Progress of Theoretical and Experimental Physics},
         year = 2023,
        month = apr,
       volume = {2023},
       number = {4},
          eid = {042F01},
        pages = {042F01},
          doi = {10.1093/ptep/ptac150},
archivePrefix = {arXiv},
       eprint = {2202.02773},
 primaryClass = {astro-ph.IM},
       adsurl = {https://ui.adsabs.harvard.edu/abs/2023PTEP.2023d2F01L}
}

@ARTICLE{QUIJOTEMFI4,
       author = {{Rubi{\~n}o-Mart{\'\i}n}, J.~A. and {Guidi}, F. and {G{\'e}nova-Santos}, R.~T. and {Harper}, S.~E. and {Herranz}, D. and {Hoyland}, R.~J. and {Lasenby}, A.~N. and {Poidevin}, F. and {Rebolo}, R. and {Ruiz-Granados}, B. and {Vansyngel}, F. and {Vielva}, P. and {Watson}, R.~A. and {Artal}, E. and {Ashdown}, M. and {Barreiro}, R.~B. and {Bilbao-Ahedo}, J.~D. and {Casas}, F.~J. and {Casaponsa}, B. and {Cepeda-Arroita}, R. and {de la Hoz}, E. and {Dickinson}, C. and {Fern{\'a}ndez-Cobos}, R. and {Fern{\'a}ndez-Torreiro}, M. and {Gonz{\'a}lez-Gonz{\'a}lez}, R. and {Hern{\'a}ndez-Monteagudo}, C. and {L{\'o}pez-Caniego}, M. and {L{\'o}pez-Caraballo}, C. and {Mart{\'\i}nez-Gonz{\'a}lez}, E. and {Peel}, M.~W. and {Pel{\'a}ez-Santos}, A.~E. and {Perrott}, Y. and {Piccirillo}, L. and {Razavi-Ghods}, N. and {Scott}, P. and {Titterington}, D. and {Tramonte}, D. and {Vignaga}, R.},
        title = "{QUIJOTE scientific results - IV. A northern sky survey in intensity and polarization at 10-20 GHz with the multifrequency instrument}",
      journal = {\mnras},
         year = 2023,
        month = mar,
       volume = {519},
       number = {3},
        pages = {3383-3431},
          doi = {10.1093/mnras/stac3439},
archivePrefix = {arXiv},
       eprint = {2301.05113},
 primaryClass = {astro-ph.GA},
       adsurl = {https://ui.adsabs.harvard.edu/abs/2023MNRAS.519.3383R}
}

@ARTICLE{BICEPKeck2021,
       author = {{Ade}, P.~A.~R. and {Ahmed}, Z. and {Amiri}, M. and {Barkats}, D. and {Thakur}, R. Basu and {Bischoff}, C.~A. and {Beck}, D. and {Bock}, J.~J. and {Boenish}, H. and {Bullock}, E. and {Buza}, V. and {Cheshire}, J.~R. and {Connors}, J. and {Cornelison}, J. and {Crumrine}, M. and {Cukierman}, A. and {Denison}, E.~V. and {Dierickx}, M. and {Duband}, L. and {Eiben}, M. and {Fatigoni}, S. and {Filippini}, J.~P. and {Fliescher}, S. and {Goeckner-Wald}, N. and {Goldfinger}, D.~C. and {Grayson}, J. and {Grimes}, P. and {Hall}, G. and {Halal}, G. and {Halpern}, M. and {Hand}, E. and {Harrison}, S. and {Henderson}, S. and {Hildebrandt}, S.~R. and {Hilton}, G.~C. and {Hubmayr}, J. and {Hui}, H. and {Irwin}, K.~D. and {Kang}, J. and {Karkare}, K.~S. and {Karpel}, E. and {Kefeli}, S. and {Kernasovskiy}, S.~A. and {Kovac}, J.~M. and {Kuo}, C.~L. and {Lau}, K. and {Leitch}, E.~M. and {Lennox}, A. and {Megerian}, K.~G. and {Minutolo}, L. and {Moncelsi}, L. and {Nakato}, Y. and {Namikawa}, T. and {Nguyen}, H.~T. and {O'Brient}, R. and {Ogburn}, R.~W. and {Palladino}, S. and {Prouve}, T. and {Pryke}, C. and {Racine}, B. and {Reintsema}, C.~D. and {Richter}, S. and {Schillaci}, A. and {Schwarz}, R. and {Schmitt}, B.~L. and {Sheehy}, C.~D. and {Soliman}, A. and {Germaine}, T. St. and {Steinbach}, B. and {Sudiwala}, R.~V. and {Teply}, G.~P. and {Thompson}, K.~L. and {Tolan}, J.~E. and {Tucker}, C. and {Turner}, A.~D. and {Umilt{\`a}}, C. and {Verg{\`e}s}, C. and {Vieregg}, A.~G. and {Wandui}, A. and {Weber}, A.~C. and {Wiebe}, D.~V. and {Willmert}, J. and {Wong}, C.~L. and {Wu}, W.~L.~K. and {Yang}, H. and {Yoon}, K.~W. and {Young}, E. and {Yu}, C. and {Zeng}, L. and {Zhang}, C. and {Zhang}, S. and {Bicep/Keck Collaboration}},
        title = "{Improved Constraints on Primordial Gravitational Waves using Planck, WMAP, and BICEP/Keck Observations through the 2018 Observing Season}",
      journal = {\prl},
         year = 2021,
        month = oct,
       volume = {127},
       number = {15},
          eid = {151301},
        pages = {151301},
          doi = {10.1103/PhysRevLett.127.151301},
archivePrefix = {arXiv},
       eprint = {2110.00483},
 primaryClass = {astro-ph.CO},
       adsurl = {https://ui.adsabs.harvard.edu/abs/2021PhRvL.127o1301A}
}

@ARTICLE{ACT-DR4,
       author = {{Aiola}, Simone and {Calabrese}, Erminia and {Maurin}, Lo{\"\i}c and {Naess}, Sigurd and {Schmitt}, Benjamin L. and {Abitbol}, Maximilian H. and {Addison}, Graeme E. and {Ade}, Peter A.~R. and {Alonso}, David and {Amiri}, Mandana and {Amodeo}, Stefania and {Angile}, Elio and {Austermann}, Jason E. and {Baildon}, Taylor and {Battaglia}, Nick and {Beall}, James A. and {Bean}, Rachel and {Becker}, Daniel T. and {Bond}, J. Richard and {Bruno}, Sarah Marie and {Calafut}, Victoria and {Campusano}, Luis E. and {Carrero}, Felipe and {Chesmore}, Grace E. and {Cho}, Hsiao-mei and {Choi}, Steve K. and {Clark}, Susan E. and {Cothard}, Nicholas F. and {Crichton}, Devin and {Crowley}, Kevin T. and {Darwish}, Omar and {Datta}, Rahul and {Denison}, Edward V. and {Devlin}, Mark J. and {Duell}, Cody J. and {Duff}, Shannon M. and {Duivenvoorden}, Adriaan J. and {Dunkley}, Jo and {D{\"u}nner}, Rolando and {Essinger-Hileman}, Thomas and {Fankhanel}, Max and {Ferraro}, Simone and {Fox}, Anna E. and {Fuzia}, Brittany and {Gallardo}, Patricio A. and {Gluscevic}, Vera and {Golec}, Joseph E. and {Grace}, Emily and {Gralla}, Megan and {Guan}, Yilun and {Hall}, Kirsten and {Halpern}, Mark and {Han}, Dongwon and {Hargrave}, Peter and {Hasselfield}, Matthew and {Helton}, Jakob M. and {Henderson}, Shawn and {Hensley}, Brandon and {Hill}, J. Colin and {Hilton}, Gene C. and {Hilton}, Matt and {Hincks}, Adam D. and {Hlo{\v{z}}ek}, Ren{\'e}e and {Ho}, Shuay-Pwu Patty and {Hubmayr}, Johannes and {Huffenberger}, Kevin M. and {Hughes}, John P. and {Infante}, Leopoldo and {Irwin}, Kent and {Jackson}, Rebecca and {Klein}, Jeff and {Knowles}, Kenda and {Koopman}, Brian and {Kosowsky}, Arthur and {Lakey}, Vincent and {Li}, Dale and {Li}, Yaqiong and {Li}, Zack and {Lokken}, Martine and {Louis}, Thibaut and {Lungu}, Marius and {MacInnis}, Amanda and {Madhavacheril}, Mathew and {Maldonado}, Felipe and {Mallaby-Kay}, Maya and {Marsden}, Danica and {McMahon}, Jeff and {Menanteau}, Felipe and {Moodley}, Kavilan and {Morton}, Tim and {Namikawa}, Toshiya and {Nati}, Federico and {Newburgh}, Laura and {Nibarger}, John P. and {Nicola}, Andrina and {Niemack}, Michael D. and {Nolta}, Michael R. and {Orlowski-Sherer}, John and {Page}, Lyman A. and {Pappas}, Christine G. and {Partridge}, Bruce and {Phakathi}, Phumlani and {Pisano}, Giampaolo and {Prince}, Heather and {Puddu}, Roberto and {Qu}, Frank J. and {Rivera}, Jesus and {Robertson}, Naomi and {Rojas}, Felipe and {Salatino}, Maria and {Schaan}, Emmanuel and {Schillaci}, Alessandro and {Sehgal}, Neelima and {Sherwin}, Blake D. and {Sierra}, Carlos and {Sievers}, Jon and {Sifon}, Cristobal and {Sikhosana}, Precious and {Simon}, Sara and {Spergel}, David N. and {Staggs}, Suzanne T. and {Stevens}, Jason and {Storer}, Emilie and {Sunder}, Dhaneshwar D. and {Switzer}, Eric R. and {Thorne}, Ben and {Thornton}, Robert and {Trac}, Hy and {Treu}, Jesse and {Tucker}, Carole and {Vale}, Leila R. and {Van Engelen}, Alexander and {Van Lanen}, Jeff and {Vavagiakis}, Eve M. and {Wagoner}, Kasey and {Wang}, Yuhan and {Ward}, Jonathan T. and {Wollack}, Edward J. and {Xu}, Zhilei and {Zago}, Fernando and {Zhu}, Ningfeng},
        title = "{The Atacama Cosmology Telescope: DR4 maps and cosmological parameters}",
      journal = {\jcap},
         year = 2020,
        month = dec,
       volume = {2020},
       number = {12},
          eid = {047},
        pages = {047},
          doi = {10.1088/1475-7516/2020/12/047},
archivePrefix = {arXiv},
       eprint = {2007.07288},
 primaryClass = {astro-ph.CO},
       adsurl = {https://ui.adsabs.harvard.edu/abs/2020JCAP...12..047A}
}

@ARTICLE{SPT-3G,
       author = {{Camphuis}, E. and {Quan}, W. and {Balkenhol}, L. and {Khalife}, A.~R. and {Ge}, F. and {Guidi}, F. and {Huang}, N. and {Lynch}, G.~P. and {Omori}, Y. and {Trendafilova}, C. and {Anderson}, A.~J. and {Ansarinejad}, B. and {Archipley}, M. and {Barry}, P.~S. and {Benabed}, K. and {Bender}, A.~N. and {Benson}, B.~A. and {Bianchini}, F. and {Bleem}, L.~E. and {Bouchet}, F.~R. and {Bryant}, L. and {Campitiello}, M.~G. and {Carlstrom}, J.~E. and {Chang}, C.~L. and {Chaubal}, P. and {Chichura}, P.~M. and {Chokshi}, A. and {Chou}, T. -L. and {Coerver}, A. and {Crawford}, T.~M. and {Daley}, C. and {de Haan}, T. and {Dibert}, K.~R. and {Dobbs}, M.~A. and {Doohan}, M. and {Doussot}, A. and {Dutcher}, D. and {Everett}, W. and {Feng}, C. and {Ferguson}, K.~R. and {Fichman}, K. and {Foster}, A. and {Galli}, S. and {Gambrel}, A.~E. and {Gardner}, R.~W. and {Goeckner-Wald}, N. and {Gualtieri}, R. and {Guns}, S. and {Halverson}, N.~W. and {Hivon}, E. and {Holder}, G.~P. and {Holzapfel}, W.~L. and {Hood}, J.~C. and {Hryciuk}, A. and {K{\'e}ruzor{\'e}}, F. and {Knox}, L. and {Korman}, M. and {Kornoelje}, K. and {Kuo}, C. -L. and {Levy}, K. and {Lowitz}, A.~E. and {Lu}, C. and {Maniyar}, A. and {Martsen}, E.~S. and {Menanteau}, F. and {Millea}, M. and {Montgomery}, J. and {Nakato}, Y. and {Natoli}, T. and {Noble}, G.~I. and {Ouellette}, A. and {Pan}, Z. and {Paschos}, P. and {Phadke}, K.~A. and {Pollak}, A.~W. and {Prabhu}, K. and {Raghunathan}, S. and {Rahimi}, M. and {Rahlin}, A. and {Reichardt}, C.~L. and {Rouble}, M. and {Ruhl}, J.~E. and {Schiappucci}, E. and {Simpson}, A. and {Sobrin}, J.~A. and {Stark}, A.~A. and {Stephen}, J. and {Tandoi}, C. and {Thorne}, B. and {Umilta}, C. and {Vieira}, J.~D. and {Vitrier}, A. and {Wan}, Y. and {Whitehorn}, N. and {Wu}, W.~L.~K. and {Young}, M.~R. and {Zebrowski}, J.~A.},
        title = "{SPT-3G D1: CMB temperature and polarization power spectra and cosmology from 2019 and 2020 observations of the SPT-3G Main field}",
      journal = {arXiv e-prints},
         year = 2025,
        month = jun,
          eid = {arXiv:2506.20707},
        pages = {arXiv:2506.20707},
          doi = {10.48550/arXiv.2506.20707},
archivePrefix = {arXiv},
       eprint = {2506.20707},
 primaryClass = {astro-ph.CO},
       adsurl = {https://ui.adsabs.harvard.edu/abs/2025arXiv250620707C}
}

@ARTICLE{CLASS2024,
       author = {{Eimer}, Joseph R. and {Li}, Yunyang and {Brewer}, Michael K. and {Shi}, Rui and {Ali}, Aamir and {Appel}, John W. and {Bennett}, Charles L. and {Bruno}, Sarah Marie and {Bustos}, Ricardo and {Chuss}, David T. and {Cleary}, Joseph and {Dahal}, Sumit and {Datta}, Rahul and {Denes Couto}, Jullianna and {Denis}, Kevin L. and {D{\"u}nner}, Rolando and {Essinger-Hileman}, Thomas and {Flux{\'a}}, Pedro and {Hubmayer}, Johannes and {Harrington}, Kathleen and {Iuliano}, Jeffrey and {Karakla}, John and {Marriage}, Tobias A. and {N{\'u}{\~n}ez}, Carolina and {Parker}, Lucas and {Petroff}, Matthew A. and {Reeves}, Rodrigo A. and {Rostem}, Karwan and {Valle}, Deniz A.~N. and {Watts}, Duncan J. and {Weiland}, Janet L. and {Wollack}, Edward J. and {Xu}, Zhilei and {Zeng}, Lingzhen},
        title = "{CLASS Angular Power Spectra and Map-component Analysis for 40 GHz Observations through 2022}",
      journal = {\apj},
         year = 2024,
        month = mar,
       volume = {963},
       number = {2},
          eid = {92},
        pages = {92},
          doi = {10.3847/1538-4357/ad1abf},
archivePrefix = {arXiv},
       eprint = {2309.00675},
 primaryClass = {astro-ph.CO},
       adsurl = {https://ui.adsabs.harvard.edu/abs/2024ApJ...963...92E}
}

@ARTICLE{Groundbird,
       author = {{Lee}, K. and {Choi}, J. and {G{\'e}nova-Santos}, R.~T. and {Hattori}, M. and {Hazumi}, M. and {Honda}, S. and {Ikemitsu}, T. and {Ishida}, H. and {Ishitsuka}, H. and {Jo}, Y. and {Karatsu}, K. and {Kiuchi}, K. and {Komine}, J. and {Koyano}, R. and {Kutsuma}, H. and {Mima}, S. and {Minowa}, M. and {Moon}, J. and {Nagai}, M. and {Nagasaki}, T. and {Naruse}, M. and {Oguri}, S. and {Otani}, C. and {Peel}, M. and {Rebolo}, R. and {Rubi{\~n}o-Mart{\'\i}n}, J.~A. and {Sekimoto}, Y. and {Suzuki}, J. and {Taino}, T. and {Tajima}, O. and {Tomita}, N. and {Uchida}, T. and {Won}, E. and {Yoshida}, M.},
        title = "{GroundBIRD: A CMB Polarization Experiment with MKID Arrays}",
      journal = {Journal of Low Temperature Physics},
         year = 2020,
        month = aug,
       volume = {200},
       number = {5-6},
        pages = {384-391},
          doi = {10.1007/s10909-020-02511-5},
archivePrefix = {arXiv},
       eprint = {2011.07705},
 primaryClass = {astro-ph.IM},
       adsurl = {https://ui.adsabs.harvard.edu/abs/2020JLTP..200..384L}
}

@ARTICLE{FixsenMather2002,
       author = {{Fixsen}, D.~J. and {Mather}, J.~C.},
        title = "{The Spectral Results of the Far-Infrared Absolute Spectrophotometer Instrument on COBE}",
      journal = {\apj},
         year = 2002,
        month = dec,
       volume = {581},
       number = {2},
        pages = {817-822},
          doi = {10.1086/344402},
       adsurl = {https://ui.adsabs.harvard.edu/abs/2002ApJ...581..817F}
}

@PHDTHESIS{BischoffPhDT,
       author = {{Bischoff}, Colin A.},
        title = "{Observing the Cosmic Microwave Background polarization anisotropy at 40 GHz with QUIET}",
       school = {University of Chicago},
         year = 2010,
        month = aug,
       adsurl = {https://ui.adsabs.harvard.edu/abs/2010PhDT.......115B}
}

@INPROCEEDINGS{BISOU2024,
       author = {{Maffei}, B. and {Aghanim}, N. and {Aumont}, J. and {Battistelli}, E. and {Beelen}, A. and {Besnard}, A. and {Borgo}, B. and {Calvo}, M. and {Catalano}, A. and {Chluba}, J. and {Coulon}, X. and {De Bernardis}, P. and {de Jabrun}, C. and {Douspis}, M. and {Errard}, J. and {Grain}, J. and {Guiot}, P. and {Hill}, J.~C. and {Ishino}, H. and {Kogut}, A. and {Lagache}, G. and {Macias-Perez}, J. and {Masi}, S. and {Matsumura}, T. and {Monfardini}, A. and {O'Sullivan}, C. and {Pagano}, L. and {Patanchon}, G. and {Pisano}, G. and {Pitre}, L. and {Ponthieu}, N. and {Remazeilles}, M. and {Ritacco}, A. and {Savini}, G. and {Sauvage}, V. and {Shitvov}, A. and {Stever}, S.~L. and {Tartari}, A. and {Thiele}, L. and {Trappe}, N. and {Aubrun}, J. -F. and {Bray}, N. and {Louvel}, S.},
        title = "{BISOU: a balloon pathfinder for CMB spectral distortions studies}",
    booktitle = {Millimeter, Submillimeter, and Far-Infrared Detectors and Instrumentation for Astronomy XII},
         year = 2024,
       editor = {{Zmuidzinas}, Jonas and {Gao}, Jian-Rong},
       series = {Society of Photo-Optical Instrumentation Engineers (SPIE) Conference Series},
       volume = {13102},
        month = aug,
          eid = {131020N},
        pages = {131020N},
          doi = {10.1117/12.3018371},
       adsurl = {https://ui.adsabs.harvard.edu/abs/2024SPIE13102E..0NM}
}

@INPROCEEDINGS{COSMO,
       author = {{Masi}, S. and {Battistelli}, E. and {de Bernardis}, P. and {Columbro}, F. and {Coppolecchia}, A. and {D'Alessandro}, G. and {de Petris}, M. and {Lamagna}, L. and {Marchitelli}, E. and {Mele}, L. and {Paiella}, A. and {Piacentini}, F. and {Pisano}, G. and {Bersanelli}, M. and {Franceschet}, C. and {Manzan}, E. and {Mennella}, D. and {Realini}, S. and {Cibella}, S. and {Martini}, F. and {Pettinari}, G. and {Coppi}, G. and {Gervasi}, M. and {Limonta}, A. and {Zannoni}, M. and {Piccirillo}, L. and {Tucker}, C.},
        title = "{The COSmic Monopole Observer (COSMO)}",
    booktitle = {The Sixteenth Marcel Grossmann Meeting. On Recent Developments in Theoretical and Experimental General Relativity, Astrophysics, and Relativistic Field Theories},
         year = 2023,
       editor = {{Ruffino}, Remo and {Vereshchagin}, Gregory},
        month = jul,
        pages = {1654-1671},
          doi = {10.1142/9789811269776_0131},
       adsurl = {https://ui.adsabs.harvard.edu/abs/2023mgm..conf.1654M}
}

@INPROCEEDINGS{2012SPIE.8446E..7CB,
       author = {{Bersanelli}, M. and {Mennella}, A. and {Morgante}, G. and {Zannoni}, M. and {Addamo}, G. and {Baschirotto}, A. and {Battaglia}, P. and {Ba{\'o}}, A. and {Cappellini}, B. and {Cavaliere}, F. and {Cuttaia}, F. and {Del Torto}, F. and {Donzelli}, S. and {Farooqui}, Z. and {Frailis}, M. and {Franceschet}, C. and {Franceschi}, E. and {Gaier}, T. and {Galeotta}, S. and {Gervasi}, M. and {Gregorio}, A. and {Kangaslahti}, P. and {Krachmalnicoff}, N. and {Lawrence}, C. and {Maggio}, G. and {Mainini}, R. and {Maino}, D. and {Mandolesi}, N. and {Paroli}, B. and {Passerini}, A. and {Peverini}, O.~A. and {Poli}, S. and {Ricciardi}, S. and {Rossetti}, M. and {Sandri}, M. and {Seiffert}, M. and {Stringhetti}, L. and {Tartari}, A. and {Tascone}, R. and {Tavagnacco}, D. and {Terenzi}, L. and {Tomasi}, M. and {Tommasi}, E. and {Villa}, F. and {Virone}, Gi. and {Zacchei}, A.},
        title = "{A coherent polarimeter array for the Large Scale Polarization Explorer (LSPE) balloon experiment}",
    booktitle = {Ground-based and Airborne Instrumentation for Astronomy IV},
         year = 2012,
       editor = {{McLean}, Ian S. and {Ramsay}, Suzanne K. and {Takami}, Hideki},
       series = {Society of Photo-Optical Instrumentation Engineers (SPIE) Conference Series},
       volume = {8446},
        month = sep,
          eid = {84467C},
        pages = {84467C},
          doi = {10.1117/12.925688},
archivePrefix = {arXiv},
       eprint = {1208.0164},
 primaryClass = {astro-ph.IM},
       adsurl = {https://ui.adsabs.harvard.edu/abs/2012SPIE.8446E..7CB}
}

@ARTICLE{LSPE-JCAP,
       author = {{Addamo}, G. and {Ade}, P.~A.~R. and {Baccigalupi}, C. and {Baldini}, A.~M. and {Battaglia}, P.~M. and {Battistelli}, E.~S. and {Ba{\`u}}, A. and {de Bernardis}, P. and {Bersanelli}, M. and {Biasotti}, M. and {Boscaleri}, A. and {Caccianiga}, B. and {Caprioli}, S. and {Cavaliere}, F. and {Cei}, F. and {Cleary}, K.~A. and {Columbro}, F. and {Coppi}, G. and {Coppolecchia}, A. and {Cuttaia}, F. and {D'Alessandro}, G. and {De Gasperis}, G. and {De Petris}, M. and {Fafone}, V. and {Farsian}, F. and {Ferrari Barusso}, L. and {Fontanelli}, F. and {Franceschet}, C. and {Gaier}, T.~C. and {Galli}, L. and {Gatti}, F. and {Genova-Santos}, R. and {Gerbino}, M. and {Gervasi}, M. and {Ghigna}, T. and {Grosso}, D. and {Gruppuso}, A. and {Gualtieri}, R. and {Incardona}, F. and {Jones}, M.~E. and {Kangaslahti}, P. and {Krachmalnicoff}, N. and {Lamagna}, L. and {Lattanzi}, M. and {L{\'o}pez-Caraballo}, C.~H. and {Lumia}, M. and {Mainini}, R. and {Maino}, D. and {Mandelli}, S. and {Maris}, M. and {Masi}, S. and {Matarrese}, S. and {May}, A. and {Mele}, L. and {Mena}, P. and {Mennella}, A. and {Molina}, R. and {Molinari}, D. and {Morgante}, G. and {Natale}, U. and {Nati}, F. and {Natoli}, P. and {Pagano}, L. and {Paiella}, A. and {Panico}, F. and {Paonessa}, F. and {Paradiso}, S. and {Passerini}, A. and {Perez-de-Taoro}, M. and {Peverini}, O.~A. and {Pezzotta}, F. and {Piacentini}, F. and {Piccirillo}, L. and {Pisano}, G. and {Polenta}, G. and {Poletti}, D. and {Presta}, G. and {Realini}, S. and {Reyes}, N. and {Rocchi}, A. and {Rubino-Martin}, J.~A. and {Sandri}, M. and {Sartor}, S. and {Schillaci}, A. and {Signorelli}, G. and {Siri}, B. and {Soria}, M. and {Spinella}, F. and {Tapia}, V. and {Tartari}, A. and {Taylor}, A.~C. and {Terenzi}, L. and {Tomasi}, M. and {Tommasi}, E. and {Tucker}, C. and {Vaccaro}, D. and {Vigano}, D.~M. and {Villa}, F. and {Virone}, G. and {Vittorio}, N. and {Volpe}, A. and {Watkins}, R.~E.~J. and {Zacchei}, A. and {Zannoni}, M. and {LSPE Collaboration}},
        title = "{The large scale polarization explorer (LSPE) for CMB measurements: performance forecast}",
      journal = {\jcap},
         year = 2021,
        month = aug,
       volume = {2021},
       number = {8},
          eid = {008},
        pages = {008},
          doi = {10.1088/1475-7516/2021/08/008},
archivePrefix = {arXiv},
       eprint = {2008.11049},
 primaryClass = {astro-ph.IM},
       adsurl = {https://ui.adsabs.harvard.edu/abs/2021JCAP...08..008A}
}

@ARTICLE{QUBIC,
       author = {{Hamilton}, J. -Ch. and {Mousset}, L. and {Battistelli}, E.~S. and {de Bernardis}, P. and {Bigot-Sazy}, M. -A. and {Chanial}, P. and {Charlassier}, R. and {D'Alessandro}, G. and {de Petris}, M. and {Gamboa Lerena}, M.~M. and {Grandsire}, L. and {Landau}, S. and {Mandelli}, S. and {Marnieros}, S. and {Masi}, S. and {Mennella}, A. and {O'Sullivan}, C. and {Piat}, M. and {Ricciardi}, G. and {Sc{\'o}ccola}, C.~G. and {Stolpovskiy}, M. and {Tartari}, A. and {Torchinsky}, S.~A. and {Voisin}, F. and {Zannoni}, M. and {Ade}, P. and {Alberro}, J.~G. and {Almela}, A. and {Amico}, G. and {Arnaldi}, L.~H. and {Auguste}, D. and {Aumont}, J. and {Azzoni}, S. and {Banfi}, S. and {Ba{\`u}}, A. and {B{\'e}lier}, B. and {Bennett}, D. and {Berg{\'e}}, L. and {Bernard}, J. -Ph. and {Bersanelli}, M. and {Bonaparte}, J. and {Bonis}, J. and {Bunn}, E. and {Burke}, D. and {Buzi}, D. and {Cavaliere}, F. and {Chapron}, C. and {Cobos Cerutti}, A.~C. and {Columbro}, F. and {Coppolecchia}, A. and {de Gasperis}, G. and {de Leo}, M. and {Dheilly}, S. and {Duca}, C. and {Dumoulin}, L. and {Etchegoyen}, A. and {Fasciszewski}, A. and {Ferreyro}, L.~P. and {Fracchia}, D. and {Franceschet}, C. and {Ganga}, K.~M. and {Garc{\'\i}a}, B. and {Garc{\'\i}a Redondo}, M.~E. and {Gaspard}, M. and {Gayer}, D. and {Gervasi}, M. and {Giard}, M. and {Gilles}, V. and {Giraud-Heraud}, Y. and {G{\'o}mez Berisso}, M. and {Gonz{\'a}lez}, M. and {Gradziel}, M. and {Hampel}, M.~R. and {Harari}, D. and {Henrot-Versill{\'e}}, S. and {Incardona}, F. and {Jules}, E. and {Kaplan}, J. and {Kristukat}, C. and {Lamagna}, L. and {Loucatos}, S. and {Louis}, T. and {Maffei}, B. and {Marty}, W. and {Mattei}, A. and {May}, A. and {McCulloch}, M. and {Mele}, L. and {Melo}, D. and {Montier}, L. and {Mundo}, L.~M. and {Murphy}, J.~A. and {Murphy}, J.~D. and {Nati}, F. and {Olivieri}, E. and {Oriol}, C. and {Paiella}, A. and {Pajot}, F. and {Passerini}, A. and {Pastoriza}, H. and {Pelosi}, A. and {Perbost}, C. and {Perciballi}, M. and {Pezzotta}, F. and {Piacentini}, F. and {Piccirillo}, L. and {Pisano}, G. and {Platino}, M. and {Polenta}, G. and {Pr{\^e}le}, D. and {Puddu}, R. and {Rambaud}, D. and {Rasztocky}, E. and {Ringegni}, P. and {Romero}, G.~E. and {Salum}, J.~M. and {Schillaci}, A. and {Scully}, S. and {Spinelli}, S. and {Stankowiak}, G. and {Supanitsky}, A.~D. and {Thermeau}, J. -P. and {Timbie}, P. and {Tomasi}, M. and {Tucker}, C. and {Tucker}, G. and {Vigan{\`o}}, D. and {Vittorio}, N. and {Wicek}, F. and {Wright}, M. and {Zullo}, A. and {The Qubic Collaboration}},
        title = "{QUBIC I: Overview and science program}",
      journal = {\jcap},
         year = 2022,
        month = apr,
       volume = {2022},
       number = {4},
          eid = {034},
        pages = {034},
          doi = {10.1088/1475-7516/2022/04/034},
archivePrefix = {arXiv},
       eprint = {2011.02213},
 primaryClass = {astro-ph.IM},
       adsurl = {https://ui.adsabs.harvard.edu/abs/2022JCAP...04..034H}
}

@ARTICLE{SilkChluba2014,
       author = {{Silk}, Joseph and {Chluba}, Jens},
        title = "{Next Steps for Cosmology}",
      journal = {Science},
         year = 2014,
        month = may,
       volume = {344},
       number = {6184},
        pages = {586-588},
          doi = {10.1126/science.1252724},
       adsurl = {https://ui.adsabs.harvard.edu/abs/2014Sci...344..586S}
}

@ARTICLE{1996ApJ...458..407S,
       author = {{Staggs}, S.~T. and {Jarosik}, N.~C. and {Wilkinson}, D.~T. and {Wollack}, E.~J.},
        title = "{An Absolute Measurement of the Cosmic Microwave Background Radiation Temperature at 20 Centimeters}",
      journal = {\apj},
         year = 1996,
        month = feb,
       volume = {458},
        pages = {407},
          doi = {10.1086/176825},
       adsurl = {https://ui.adsabs.harvard.edu/abs/1996ApJ...458..407S}
}

@ARTICLE{TRIS_Zannoni2008,
       author = {{Zannoni}, M. and {Tartari}, A. and {Gervasi}, M. and {Boella}, G. and {Sironi}, G. and {De Lucia}, A. and {Passerini}, A. and {Cavaliere}, F.},
        title = "{TRIS. I. Absolute Measurements of the Sky Brightness Temperature at 0.6, 0.82, and 2.5 GHz}",
      journal = {\apj},
         year = 2008,
        month = nov,
       volume = {688},
       number = {1},
        pages = {12-23},
          doi = {10.1086/592133},
archivePrefix = {arXiv},
       eprint = {0806.1415},
 primaryClass = {astro-ph},
       adsurl = {https://ui.adsabs.harvard.edu/abs/2008ApJ...688...12Z}
}

@ARTICLE{Seiffert2011,
       author = {{Seiffert}, M. and {Fixsen}, D.~J. and {Kogut}, A. and {Levin}, S.~M. and {Limon}, M. and {Lubin}, P.~M. and {Mirel}, P. and {Singal}, J. and {Villela}, T. and {Wollack}, E. and {Wuensche}, C.~A.},
        title = "{Interpretation of the ARCADE 2 Absolute Sky Brightness Measurement}",
      journal = {\apj},
         year = 2011,
        month = jun,
       volume = {734},
       number = {1},
          eid = {6},
        pages = {6},
          doi = {10.1088/0004-637X/734/1/6},
       adsurl = {https://ui.adsabs.harvard.edu/abs/2011ApJ...734....6S}
}

@PHDTHESIS{CarlosLopezPhDT,
       author = {{L{\'o}pez Caraballo}, Carlos H.},
        title = "{Estrategias de observaci{\'o}n y m{\'e}todos de an{\'a}lisis para la medida de la radiaci{\'o}n de microondas con el experimento QUIJOTE-CMBEstrategias de observaci{\'o}n y m{\'e}todos de an{\'a}lisis para la medida de la radiaci{\'o}n de microondas con el experimento QUIJOTE-CMBObservation strategies and methods of analysis for the microwave radiation measurements with the QUIJOTE-CMB experiment;}",
       school = {University of La Laguna, Spain},
         year = 2013,
        month = jan,
       adsurl = {https://ui.adsabs.harvard.edu/abs/2013PhDT.......406L}
}

@phdthesis{tesisPazAlonso,
  title = {Instrumentation for the Tenerife Microwave Spectrometer},
  author = {{Alonso-Arias}, P.},
  year = 2022,
  month = {Junio},
  address = {San Cristobal de La Laguna, Spain},
  note = {},
  school = {Universidad de La Laguna- Instituto de Astrofisica de Canarias (ULL/IAC)},
  type = {Ph.D. thesis}
}

@ARTICLE{2022JInst..17P6041D,
       author = {{De Miguel}, J. and {Franceschet}, C. and {Realini}, S. and {Fuerte-Rodr{\'\i}guez}, P.},
        title = "{A metamaterial with applications in broad band antennas used in radio astronomy and satellite communications}",
      journal = {Journal of Instrumentation},
         year = 2022,
        month = jun,
       volume = {17},
       number = {6},
          eid = {P06041},
        pages = {P06041},
          doi = {10.1088/1748-0221/17/06/P06041},
archivePrefix = {arXiv},
       eprint = {2108.05648},
 primaryClass = {physics.ins-det},
       adsurl = {https://ui.adsabs.harvard.edu/abs/2022JInst..17P6041D}
}

@ARTICLE{Friis1944,
  author={Friis, H.T.},
  journal={Proceedings of the IRE}, 
  title={Noise Figures of Radio Receivers}, 
  year={1944},
  volume={32},
  number={7},
  pages={419-422},
  doi={10.1109/JRPROC.1944.232049}
}

@ARTICLE{chappard,
       author = {{Chappard}, Apolline and {Rubi{\~n}o-Mart{\'\i}n}, Jos{\'e} Alberto and {Tanaus{\'u} G{\'e}nova Santos}, Ricardo},
        title = "{Characterising the properties of the atmospheric emission at Teide Observatory in the 10-20 GHz range with QUIJOTE data}",
      journal = {arXiv e-prints},
         year = 2025,
        month = oct,
          eid = {arXiv:2510.19878},
        pages = {arXiv:2510.19878},
          doi = {10.48550/arXiv.2510.19878},
archivePrefix = {arXiv},
       eprint = {2510.19878},
 primaryClass = {physics.ao-ph},
       adsurl = {https://ui.adsabs.harvard.edu/abs/2025arXiv251019878C}
}

@ARTICLE{Mennella2026,
author  = { {Mennella}, A. and others}, 
title   = "{ Analytical model of the LSPE-Strip polarimeters}",
journal = {JINST, in prep.},
year = 2026
}

@ARTICLE{2010A&A...520A...4B,
       author = {{Bersanelli}, M. and {Mandolesi}, N. and {Butler}, R.~C. and {Mennella}, A. and {Villa}, F. and {Aja}, B. and {Artal}, E. and {Artina}, E. and {Baccigalupi}, C. and {Balasini}, M. and {Baldan}, G. and {Banday}, A. and {Bastia}, P. and {Battaglia}, P. and {Bernardino}, T. and {Blackhurst}, E. and {Boschini}, L. and {Burigana}, C. and {Cafagna}, G. and {Cappellini}, B. and {Cavaliere}, F. and {Colombo}, F. and {Crone}, G. and {Cuttaia}, F. and {D'Arcangelo}, O. and {Danese}, L. and {Davies}, R.~D. and {Davis}, R.~J. and {de Angelis}, L. and {de Gasperis}, G.~C. and {de La Fuente}, L. and {de Rosa}, A. and {de Zotti}, G. and {Falvella}, M.~C. and {Ferrari}, F. and {Ferretti}, R. and {Figini}, L. and {Fogliani}, S. and {Franceschet}, C. and {Franceschi}, E. and {Gaier}, T. and {Garavaglia}, S. and {Gomez}, F. and {Gorski}, K. and {Gregorio}, A. and {Guzzi}, P. and {Herreros}, J.~M. and {Hildebrandt}, S.~R. and {Hoyland}, R. and {Hughes}, N. and {Janssen}, M. and {Jukkala}, P. and {Kettle}, D. and {Kilpi{\"a}}, V.~H. and {Laaninen}, M. and {Lapolla}, P.~M. and {Lawrence}, C.~R. and {Lawson}, D. and {Leahy}, J.~P. and {Leonardi}, R. and {Leutenegger}, P. and {Levin}, S. and {Lilje}, P.~B. and {Lowe}, S.~R. and {Lubin}, P.~M. and {Maino}, D. and {Malaspina}, M. and {Maris}, M. and {Marti-Canales}, J. and {Martinez-Gonzalez}, E. and {Mediavilla}, A. and {Meinhold}, P. and {Miccolis}, M. and {Morgante}, G. and {Natoli}, P. and {Nesti}, R. and {Pagan}, L. and {Paine}, C. and {Partridge}, B. and {Pascual}, J.~P. and {Pasian}, F. and {Pearson}, D. and {Pecora}, M. and {Perrotta}, F. and {Platania}, P. and {Pospieszalski}, M. and {Poutanen}, T. and {Prina}, M. and {Rebolo}, R. and {Roddis}, N. and {Rubi{\~n}o-Martin}, J.~A. and {Salmon}, M.~J. and {Sandri}, M. and {Seiffert}, M. and {Silvestri}, R. and {Simonetto}, A. and {Sjoman}, P. and {Smoot}, G.~F. and {Sozzi}, C. and {Stringhetti}, L. and {Taddei}, E. and {Tauber}, J. and {Terenzi}, L. and {Tomasi}, M. and {Tuovinen}, J. and {Valenziano}, L. and {Varis}, J. and {Vittorio}, N. and {Wade}, L.~A. and {Wilkinson}, A. and {Winder}, F. and {Zacchei}, A. and {Zonca}, A.},
        title = "{Planck pre-launch status: Design and description of the Low Frequency Instrument}",
      journal = {\aap},
         year = 2010,
        month = sep,
       volume = {520},
          eid = {A4},
        pages = {A4},
          doi = {10.1051/0004-6361/200912853},
archivePrefix = {arXiv},
       eprint = {1001.3321},
 primaryClass = {astro-ph.IM},
       adsurl = {https://ui.adsabs.harvard.edu/abs/2010A&A...520A...4B}
}

@ARTICLE{2009JInst...4T2006V,
       author = {{Valenziano}, L. and {Cuttaia}, F. and {De Rosa}, A. and {Terenzi}, L. and {Brighenti}, A. and {Cazzola}, G.~P. and {Garbesi}, A. and {Mariotti}, S. and {Orsi}, G. and {Pagan}, L. and {Cavaliere}, F. and {Biggi}, M. and {Lapini}, R. and {Panagin}, E. and {Battaglia}, P. and {Butler}, R.~C. and {Bersanelli}, M. and {D'Arcangelo}, O. and {Levin}, S. and {Mandolesi}, N. and {Mennella}, A. and {Morgante}, G. and {Morigi}, G. and {Sandri}, M. and {Simonetto}, A. and {Tomasi}, M. and {Villa}, F. and {Frailis}, M. and {Galeotta}, S. and {Gregorio}, A. and {Leonardi}, R. and {Lowe}, S.~R. and {Maris}, M. and {Meinhold}, P. and {Mendes}, L. and {Stringhetti}, L. and {Zonca}, A. and {Zacchei}, A.},
        title = "{Planck-LFI: design and performance of the 4 Kelvin Reference Load Unit}",
      journal = {Journal of Instrumentation},
         year = 2009,
        month = dec,
       volume = {12},
       number = {12},
        pages = {T12006},
          doi = {10.1088/1748-0221/4/12/T12006},
archivePrefix = {arXiv},
       eprint = {1001.4778},
 primaryClass = {astro-ph.IM},
       adsurl = {https://ui.adsabs.harvard.edu/abs/2009JInst...4T2006V}
}

@ARTICLE{2021ExA....51.1515C,
       author = {{Chluba}, J. and {Abitbol}, M.~H. and {Aghanim}, N. and {Ali-Ha{\"\i}moud}, Y. and {Alvarez}, M. and {Basu}, K. and {Bolliet}, B. and {Burigana}, C. and {de Bernardis}, P. and {Delabrouille}, J. and {Dimastrogiovanni}, E. and {Finelli}, F. and {Fixsen}, D. and {Hart}, L. and {Hern{\'a}ndez-Monteagudo}, C. and {Hill}, J.~C. and {Kogut}, A. and {Kohri}, K. and {Lesgourgues}, J. and {Maffei}, B. and {Mather}, J. and {Mukherjee}, S. and {Patil}, S.~P. and {Ravenni}, A. and {Remazeilles}, M. and {Rotti}, A. and {Rubi{\~n}o-Martin}, J.~A. and {Silk}, J. and {Sunyaev}, R.~A. and {Switzer}, E.~R.},
        title = "{New horizons in cosmology with spectral distortions of the cosmic microwave background}",
      journal = {Experimental Astronomy},
         year = 2021,
        month = jun,
       volume = {51},
       number = {3},
        pages = {1515-1554},
          doi = {10.1007/s10686-021-09729-5},
archivePrefix = {arXiv},
       eprint = {1909.01593},
 primaryClass = {astro-ph.CO},
       adsurl = {https://ui.adsabs.harvard.edu/abs/2021ExA....51.1515C}
}

%
%
\begin{appendix}
\section{Noise temperature effects: Friis equations for individual TMS components}\label{A_appendixx}

Following the Friis formalism outlined in Section~\ref{TB_s_equation}, the overall TMS intensity response can be expressed in terms of antenna temperature by concatenating the corresponding equations for each individual component. Taking as a reference the TMS radiometric scheme shown in Figs.~\ref{Diagrama_cryos} and \ref{new_model_gains} for the sequence of components as well as for the notation, here we provide the full set of equations. 

We first define a set of  effective attenuation coefficients ($h$) that account for return losses (R), insertion losses (L), and spillover (SPO). Sky and load chains are identified by the indices s and l. Additionally, the system employs two hybrids, X and Y, alongside four LNA amplifiers, denoted by numbers 1 to 4. The $h$ coefficients are given by:
\begin{align}
    &\textit{Sky}\nonumber\\&\hw=(1-\Rw)\,(1-\Iw)\,(1-\SPOw)\nonumber\\&
    \hirf=(1-\Rirf)\,(1-\Iirf)\,(1-\SPOirf)\nonumber\\&
    \hfhs=(1-\Rfhs)\,(1-\Ifhs)\nonumber\\&
    \homtsuno=(1-\Romtsuno)\,(1-\Iomtsuno)\nonumber\\&
    \textit{Load}\nonumber\\&\hfhl=(1-\Rfhl)\,(1-\Ifhl)\nonumber\\&
     \homtluno=(1-\Romtluno)\,(1-\Iomtluno)\nonumber\\&
    \hload= (1-\Rload)\,(1-\SPOload)\nonumber\\ &
    \textit{Hybrids}\nonumber\\&\betahybxdos= (1-\Rhxdos)\,(1-\Ihxdos);\betahybxtres= (1-\Rhxtres)\,(1-\Ihxtres)  \nonumber\\&
    \betahybydos= (1-\Rhydos)\,(1-\Ihydos);
    \betahybytres= (1-\Rhytres)\,(1-\Ihytres) \nonumber\\&
    \textit{LNAs}\nonumber\\&\betalnauno= (1-\Rlnauno) \nonumber;\betalnados=(1-\Rlnados);\nonumber\\&\betalnatres=(1-\Rlnatres);\betalnacuatro=(1-\Rlnacuatro)\nonumber\\&
    \textit{BEM and DC}\nonumber\\&\betaampBEMuno= (1-\RBEM)\nonumber\\&
    \betafilterBEMk=(1-\Rbemfilter)\,(1-\Ibemfilter) \nonumber\\&
    \betaampDCuno= (1-\RDC)\nonumber\\&
    \betafilterDCk= (1-\RDCfilter)\,(1-\IDCfilter).
    \label{eq:h_losses}
\end{align}
The sky intensity signal $\Tsky$ propagates through the following sequence of components.
\\
\noindent
\textbf{Window:} 
\begin{equation}
\Tbw=  \Tsky \, \hw + \left [ \Tw \, \Iw+ \Tenvuno \, \Rw \right] \,  \left (1-\SPOw \right )+ \Textcry \, \SPOw
\label{app_window}
\end{equation}
\noindent
Note that this first equation corresponds to eq.~\ref{eq:tw}, which was used as example in the main text.

\noindent
\textbf{IR filter:}
\begin{align}
   \Tbirf= \Tbw \, \hirf +\left [  \Tirf \, \Iirf+ \nonumber  \right.\\ \left. \Tenvdos \, \Rirf \right] \,  \left (1-\SPOirf \right )+ \Tcryuno \, \SPOirf
   \label{app_irfilter}
\end{align}

\noindent
\textbf{(Sky) feedhorn:}
\begin{align}
  \Tbfhs= \left [\Tbirf \, \hfhs +  \Tfhs \, \Ifhs+ \Tenvdos\, \Rfhs \right]
   \label{app_feedhorns}
\end{align}

\noindent
\textbf{(Sky) OMT}
\begin{align}
   \Tbomtsuno= \left [0.5\, \Tbfhs \, \homtsuno +  \Tomts\, \Iomtsuno+ \nonumber \Tenvdos\, \Romtsuno \right]
   \label{app_omts1}
\end{align}
\begin{align}
   \Tbomtsdos= \left [0.5\, \Tbfhs \, \homtsdos +  \Tomts\, \Iomtsdos + \nonumber \Tenvdos\, \Romtsdos \right]
\end{align}

On the other side of the cryostat, the  reference load signal $T_{\rm load}$ propagates through the following sequence of components.

\noindent
\textbf{Cold Load:}
\begin{align}
   \Tbload=  \Tload \, \hload + \left [ \Tenvdos\, \Rload \right ] \left (1 - \SPOload \right ) + \nonumber \\  \Tcrydos\, \SPOload
\end{align}

\noindent
\textbf{(Load) feedhorn: }
\begin{align}
   \Tbfhl= \left [ \Tload \, \hfhl +  \Tfhl \, \Ifhl+ \Tenvdos\, \Rfhl \right]
   \label{app_feedhload}
\end{align}

\noindent
\textbf{(Load) OMT:}
\begin{align}
   \Tbomtluno= \left [0.5\, \Tbfhl \,  \homtluno +   \Tomtl\, \Iomtluno +  \Tenvdos\, \Romtluno \right]
   \label{app_omtl1}
\end{align}
\begin{align}
  \Tbomtldos= \left [0.5\, \Tbfhl \, \homtldos +   \Tomtl \, \Iomtldos +  \Tenvdos\, \Romtldos \right]
   \label{app_omtl2}
\end{align}

The output signals of both OMTs (sky and load) are then combined using two hybrids, as follows:

\noindent
\textbf{X-Hybrid}
\begin{align}
   \Tbhybxdos= \left [ \frac{1}{2} \left ( \Tbomtsuno+ \Tbomtluno \right ) \, \betahybxdos  + \Thybx \, \Ihxdos +  \Tenvdos\, \Rhxdos\right]
   \label{app_hybX1}
\end{align}
\begin{align}
   \Tbhybxtres= \left [ \frac{1}{2} \left ( \Tbomtsuno-\Tbomtluno \right ) \, \betahybxtres  + \Thybx \, \Ihxtres+  \Tenvdos\, \Rhxtres \right]
   \label{app_hybX2}
\end{align}

\noindent
\textbf{Y-Hybrid}
\begin{align}
  \Tbhybydos= \left [ \frac{1}{2} \left ( \Tbomtsdos+ \Tbomtldos \right ) \betahybydos  + \Thyby \, \Ihydos+  \Tenvdos\, \Rhxdos \right]
   \label{app_hybY1}
\end{align}
\begin{align}
   \Tbhybytres= \left [ \frac{1}{2} \left ( \Tbomtsdos- \Tbomtldos \right ) \, \betahybytres  + \Thyby\, \Ihytres+  \Tenvdos\, \Rhytres \right]
   \label{app_HYBy2}
\end{align}

After the hybrids, we have the (cold) amplification stage with the four LNAs, where (N) represents the noise temperature of each amplifier:
\begin{align}
    \Tblnauno=\Tbhybxdos\, \betalnauno + \Tnlnauno + \Tenvdos\, \Rlnauno
    \label{app_LAN1}
\end{align}
\begin{align}
    \Tblnados= \Tbhybxtres\,\betalnados + \Tnlnados + \Tenvdos\, \Rlnados
     \label{app_LAN2}
\end{align}
\begin{align}
    \Tblnatres=\Tbhybydos\,\betalnatres + \Tnlnatres + \Tenvdos\, \Rlnatres
     \label{app_LAN4}
\end{align}
\begin{align}
    \Tblnacuatro=\Tbhybytres\, \betalnacuatro + \Tnlnacuatro  + \Tenvdos\, \Rlnacuatro
\label{app_lna4}
\end{align}

This completes the FEM part. Now, the signals are routed to the BEM and the DC modules, as:
\begin{align}
    \TbBEM= &\bigg [\Tblnauno \,(1-\RBEM)   +\TBEMfilter \, \RBEM + \TnBEMuno \bigg ]
    \label{equ:BEM}
\end{align}
%

%
\begin{align}
    \TbDC= & \bigg [\bigg [\bigg ( \TbBEM\,\hbemfilteruno + \Tbemfilter\, \Ibemfilter+ \TafterfiltBEM\,\Rbemfilter \bigg ) \nonumber \\ & + \Tbmixer \bigg ] \,(1- \RDC)+ \TafteramplDC \RDC +\TnDCMuno \bigg]\,\hDCfilteruno+  \nonumber \\ &  \TDCfilter\, \IDCfilter + \TFPGA \,\RDCfilter
    \label{equ:DC2}
\end{align}

The combination of all previous equations, in series, provides the overall TMS response (equation~\ref{delta_temp_total}).

\section{Noise temperature effects: derivation of the effective global equation}\label{B_appendix}

Once we have obtained the individual Friis equations for each one of the TMS components, here we describe the procedure to derive the effective global equation~\ref{delta_temp_total}, which constitutes the TMS analytic model. Throughout this appendix, we also follow the notation and the sequence of subsystems shown in Figs.~\ref{Diagrama_cryos} and \ref{new_model_gains}. To facilitate the interpretation of all equations in this appendix, the parameters are described and summarized in Table~\ref{table:P_k_expl}. 

\begin{table}
\caption{Description of the parameters employed in the derivation of the effective TMS global equation. }       %
\label{table:P_k_expl}
\centering                        
\begin{tabular}{l l} 
\hline\hline   
Symbol & Explanation \\
\hline                       
    $G_k$    &  LNA gain, k:1,2,3,4\\ 
    $\rm h_{hyb}$    & R and L of hybrids \\ 
    $\rm h_{lna}$    &LNA losses \\ 
    $\Tsky $  & Sky temperature\\
    $\betasky$    &  Effective R and L of the sky components \\ 
    
   $ \Tload $  & Load temperature \\
    $\betaload$    &  Effective R and L of the load components \\ 
    
    $T^c_{s} \, \bar{L}^c_{s}$ & Physical temperatures times the  \\
      &  effective sky L   \\ 
    $ T^c_{l}\, \bar{L}^c_{l}$  & Physical temperatures times the  \\
      &  effective load L  \\ 
    $\Tenvdos$    & Environment temperature of the  \\
      & cryostat 2nd stage. \\
    $\bar{R}^c_{s} $   & Effective sky return losses  \\
    $\bar{R}^c_{l} $   & Effective load return losses \\ 
    $\overline{SPO}^c_{s}$    & Effective spill-over due to the sky  \\ & components \\ 
    $\overline{SPO}^c_{l}$    & Effective spill-over due to the load  \\ & components \\ 
    $\rm N_k$    & Noise temperature \\
\hline                                   
\end{tabular}
\end{table}

As a first step, we derive the equations describing the signals before the FPGA process ($\TbAuno$, $\TbAdos$, $\TbBuno$, and $\TbBdos$). 
\begin{align}
    \rm T^{a}_{k}=&\zetaFEM\,\zetaBEM \,\zetaDC\, \left( \Tsky\, \betasky+\Tload\, \betaload \right )+\Tofftotuno
    \label{general_temp_TMS1}
\end{align}
where $k$ takes the values A1 and B1, and 
\begin{align}
    \rm T^{a}_{k}=&\zetaFEM\,\zetaBEM \,\zetaDC\, \left( \Tsky\, \betasky-\Tload\, \betaload \right )+ \Tofftotdos
    \label{general_temp_TMS2}
\end{align}
where $k$ takes the values A2 and B2. 
\subsection{Effective attenuation coefficients}
The effective attenuation coefficients for the hybrids/LNAs, BEM, and DC are represented in both equations by $\zetaFEM$, $\zetaBEM$, and $\zetaDC$. Additionally,  $\betasky$ and $\betaload$ denote the attenuation for the sky and load chain components, respectively. Their values are defined as follows:
\begin{align}
    &\zetaFEM= (0.5\, \rm h^{lna}_{k}\,\rm h^{hyb}_{k})\,\GFEMk\, \nonumber \\& 
    \zetaBEM=(\betaampBEMuno\,\hbemfilteruno)\,\GBEMk\,\nonumber \\& 
     \zetaDC=(\betaampDCuno\,\hDCfilteruno)\,\GDCk
\end{align}

\begin{align}
        &\betasky=0.5\,\hw \,  \hirf \, \hfhs \, \homtsuno   \nonumber \\ &
    \betaload= 0.5\,\hfhl\,  \homtluno\, \hload . \nonumber\\
\end{align}

\subsection{Offset temperatures}
Equations~\ref{general_temp_TMS1} and \ref{general_temp_TMS2} have  effective offset temperature terms ($\Tofftotuno$ and $\Tofftotdos$ ) generated by the FEM, BEM and DC modules, and given by:
\begin{align}
    \Tofftotuno =\zetaBEM\,\zetaDC\,\ToffFEMuno+ \zetaDC\,\ToffBEM+\ToffDC
\label{eq:Tofftotuno}
\end{align}
\begin{align}
    \Tofftotdos=\zetaBEM\,\zetaDC\,\ToffFEMdos+ \zetaDC\,\ToffBEM+\ToffDC.
\label{eq:Tofftotdos}
\end{align}
where the effective (antenna temperature) offset generated in each branch of the FEM module $\ToffFEMuno$ and $\ToffFEMdos$ are described with the equations:

\small
\begin{align}
    \rm \ToffFEMuno=&\zetaFEM \, \left [ \left (\rm \sum T^c_{s} \, \bar{L}_{s} + \sum T^c_{l} \, \bar{L}_{l} \right ) + \Tenvdos \left (\rm \sum \bar{R}_{s}+ \sum \bar{R}_{l} \right )+\nonumber \right.\\& \left. \left (\rm  \overline{SPO}_{s}  +  \overline{SPO}_{l}  \right )\right ] + \GFEMk \left [\left (\rm T^{hyb}_{k} \, L^{hyb}_{k}+\Tenvdos\, R^{hyb}_{k}\right ) \,\rm h^{lna}_{k} + \nonumber \right.\\& \left.  \Tenvdos\, R^{lna}_{k} \right ]
    \label{eq:ToffBem1}
\end{align}
\begin{align}
    \rm \ToffFEMdos=&\zetaFEM \, \left [\left (\rm \sum T^c_{s} \, \bar{L}_{s} - \sum T^c_{l} \, \bar{L}_{l} \right ) + \Tenvdos \left (\rm \sum \bar{R}_{s}- \sum \bar{R}_{l} \right ) +  \nonumber \right.\\& \left. \left ( \overline{SPO}_{s}  - \overline{SPO}_{l}  \right ) \right ]+ \GFEMk \left [ \left [\rm T^{hyb}_{k} \, L^{hyb}_{k}+\Tenvdos\, R^{hyb}_{k}\right ] \, \rm h^{lna}_{k} + \nonumber \right.\\& \left.\Tenvdos\, R^{lna}_{k}\right ]
    \label{eq:ToffBem2}
\end{align}

\normalsize
Equations \ref{eq:ToffBem1}-\ref{eq:ToffBem2} use the  notation $\rm \bar{L}$ and $\rm \bar{R}$ to represent the effective insertion and return losses of the sky and load components, this terms are defined as:
\begin{align}
    &\rm \bar{L}_{s}=\sum_j {L}^s_{j}\, H^{L_s}_j\nonumber \\ &
    \rm \bar{L}_{l}=\sum_j {L}^l_{j}\, H^{L_l}_j\nonumber \\ &
    \rm \bar{R}_{s}=\sum_j {R}^s_{j}\, H^{R_s}_j\nonumber \\ &
    \rm \bar{R}_{l}=\sum_j {R}^l_{j}\, H^{R_l}_j\nonumber \\ 
    \label{effective_losses_Eq}
\end{align}
where the expressions $\rm H^{L_k}_{j}$ and $\rm H^{R_k}_{j}$ represents the effective losses (L or R) of each component of the sky or load chain. These terms are related to the $h$ coefficients defined in eq.~\ref{eq:h_losses}, and are given by

\begin{align}
    &\textit{Effective Insertion loss}\nonumber \\ &\rm H^{L_s}_{w}= 0.5\, \hirf \, \hfhs \, \homtsuno\, (1-\SPOw)\nonumber \\ &
    \rm H^{L_s}_{irf}= 0.5\, \hfhs \, \homtsuno\, (1-\SPOirf)\nonumber \\ &
    \rm H^{L_s}_{fhs}= 0.5\, \homtsuno\,\nonumber \\ &
    \rm H^{L_s}_{omts}= 1 \nonumber\\ &
    \rm H^{L_l}_{fhl}= 0.5\, \homtluno\,\nonumber \\ &
    \rm H^{L_l}_{omtl}= 1 \\ & \textit{Effective Return loss}\nonumber \\ &
    \rm H^{R_s}_{w}= 0.5\, \hirf \, \hfhs \, \homtsuno\, (1-\SPOw)\,\frac{\Tenvuno}{\Tenvdos}\nonumber \\ &
    \rm H^{R_s}_{irf}= 0.5\, \hfhs \, \homtsuno\, (1-\SPOirf)\nonumber \\ &
    \rm H^{R_s}_{fhs}= 0.5\, \homtsuno\,\nonumber \\ &
    \rm H^{R_s}_{omts}= 1 \nonumber\\ &
    \rm H^{R_l}_{fhl}= 0.5\, \homtluno\,\nonumber \\ &
    \rm H^{R_l}_{omtl}= 1 
\label{H_mayus_L}
\end{align}

Finally, the term $\overline{SPO}^c_{k}$ represent the effective spillover of the optical components in both branches (s,l), and is given by
\begin{align}
    & \rm \overline{SPO}^c_{s}= \Textcry \,\betaaocho+\Tcryuno\,\betaanueve \\&
    \rm \overline{SPO}^c_{l}= \Tcrydos \,\betaadiez
\end{align}
where the expressions $\betaaocho$, $\betaanueve$, and $\betaadiez$ are related with the effective attenuation coefficients shown in equation~\ref{eq:h_losses} as:
\begin{align}
    &\betaaocho=0.5\,\hirf\, \hfhs \, \homtsuno\,\SPOw \nonumber\\&
    \betaanueve=0.5 \,\hfhs\, \homtsuno\,  \SPOirf\nonumber\\&
    \betaadiez=0.5\,\hfhl\,  \homtluno \, \SPOload \nonumber \\
    \label{equ:h_TMS}
\end{align}

As a remark, we note that in Eqs.~\ref{general_temp_TMS1}–\ref{general_temp_TMS2} there is a change of sign between the sky and load terms inside the parentheses. This arises because the signals propagate through a hybrid coupler, which at the voltage level performs a summation and subtraction of the input signals. Although the Friis formalism is strictly defined in terms of noise temperatures, explicitly introducing opposite signs in these equations allows us to reproduce the expected system output in terms of antenna temperature. 

\subsection{BEM and DC offset temperatures}

Now, we may add equations \ref{equ:BEM}-\ref{equ:DC2} to introduce the non-idealities given by additional amplifiers, the remaining filters and mixers from the BEM and DC modules. This non-idealities are supplied by the terms $\ToffBEM$ and $\ToffDC$ in the equations \ref{eq:Tofftotuno}-\ref{eq:Tofftotdos}, and are given by

\begin{align}
    \ToffBEM= &\Troom\,\RBEM\,\hbemfilteruno \GBEMk+ \Troom\,\Ibemfilter+\Troom\,\RBEM
\end{align}

\begin{align}
    \ToffDC=&\Tbmixer\,\betaampDCuno\,\hDCfilteruno \GDCk+\Troom\,\RDC\,\hDCfilteruno \GDCk+ \nonumber \\ & \TDCfilter\,\IDCfilter+\Troom\,\RDCfilter
\end{align}

\subsection{Digital hybrids}
To obtain the representative $\Tbsky$ and $\Tbload$ of the system, the signals $\TbAuno$, $\TbAdos$, $\TbBuno$, and $\TbBdos$ should be processed into the FPGA to obtain $\TbCuno$, $\TbCdos$, $\TbDuno$, and $\TbDdos$ as result of the digital hybrid. This is described as follows:
\begin{align}
    \TbCuno=0.5\,
    [\TbAuno+\TbAdos] \nonumber\\ 
    \TbCdos=0.5\,[\TbAuno-\TbAdos] \nonumber\\ 
    \TbDuno=0.5\,[\TbBuno+\TbBdos]\nonumber \\ 
    \TbDdos=0.5\,[\TbBuno-\TbBdos] .
\label{eq:tb_D2}
\end{align}

Now, to obtain $\Tbsky$ and $\Tbload$, we use equations~\ref{eq:tb_D2} and we add the noise temperature ($\rm N_s$, $\rm N_l$) generated by the FEM, BEM and DC amplifiers. The result is:
\begin{align}
    \Tbsky=& \TbCuno+\TbDuno+N_s
    \label{eq:fpga_total_21}
\end{align}
\begin{align}
    \Tbload=& \TbCdos+ \TbDdos +N_l
    \label{eq:fpga_total_22}
\end{align}
\begin{align}
    &\rm N^s=N^{FEM}_{s}+N^{BEM}_{s}+N^{DC}_{s}\nonumber \\&
    \rm N^l=N^{FEM}_{l}+N^{BEM}_{l}+N^{DC}_l
\end{align}
where $\rm N_s$ represents the sky components placed in the branch $\rm C_1$ and $\rm D_1$, and $\rm N_l$ the load components placed in the branch $\rm C_2$ and $\rm D_2$. Taking expression $\rm N_s$ as an example, the terms on the right-hand side of that equation are given by:
\begin{align}
    &\rm N^{FEM}_{s}= \GBEMuno\,\zetaBEMuno\,\zetaDCuno\,\Tnlnauno+\GBEMtres\,\zetaBEMtres\,\zetaDCtres\,\Tnlnatres \nonumber \\&
    \rm N^{BEM}_{s}= \hbemfilteruno \GBEMuno\,\zetaDCuno\,\TnBEMuno+\hbemfiltertres \GBEMtres\zetaDCtres\,\TnBEMtres \nonumber \\&
    \rm N^{DC}_{s}= \hDCfilteruno \GDCuno\TnDCMuno+\TnDCMtres\hDCfilteruno \GDCuno .
\end{align}

\subsection{TMS global equation}
Equation~\ref{delta_temp_total} is derived by subtracting equation~\ref{eq:fpga_total_22} from equation~\ref{eq:fpga_total_21}, and dividing the result by the global system gain $\rm G_{tot}$. These operations give us the final equation:
\begin{equation}
    \Delta T = \beta_{\rm sky}^{\rm eff} T_{\rm sky} - \beta_{\rm load}^{\rm eff} T_{\rm load} + \Toffeff +\Tneff .
    \label{total_skyload_temp}
\end{equation}
The expressions for $\betaskyEffective$ and $\betaloadEffective$ can be written as: 
\begin{align}
    &\beta_{\rm sky}^{\rm eff}= \frac{1}{\rm G_{tot}}\left[2\,\beta^{\rm Ts}_{\rm A2}+2\,\beta^{\rm Ts}_{\rm B2}\right] \nonumber\\ &
    \beta_{\rm load}^{\rm eff}=\frac{1}{\rm G_{tot}}\left[2\,\beta^{\rm Tl}_{\rm A2}+2\,\beta^{\rm Tl}_{\rm B2}\right]
    \label{eq:totalBskyBload}
\end{align}
where $\beta^{T (s,l)}_k$ denotes the aggregate loss at TMS outputs $\rm A_2$ and $\rm B_2$, after combining the sky/load chain attenuation, the hybrid (X,Y) coupler losses, and FEM LNA imperfections. The detailed expression for those terms are given by 
\begin{align}
    &\beta^{Ts}_{A2}=0.5\,\betasky\,(\betalnados\,\betahybxtres)\,\zetaBEM\zetaDC \nonumber  \\ &
    \beta^{Tl}_{A2}=0.5\,\betaload\,(\betalnados\,\betahybxtres)\,\zetaBEM\zetaDC \nonumber \\&  
    \beta^{Ts}_{B2}=0.5\,\betasky\,(\betalnacuatro\,\betahybytres)\,\zetaBEM\zetaDC \nonumber  \\ &
    \beta^{Tl}_{B2}=0.5\,\betaload\,(\betalnacuatro\,\betahybytres)\,\zetaBEM\zetaDC. 
    \label{betas_total}
\end{align}

In equation~\ref{total_skyload_temp} it is also shown the total effective offset temperature $\Toffeff$ introduced by the output $\rm A_2$ and $\rm B_2$ represented by the Equation~\ref{eq:ToffBem2}, this total effective offset is defined as
\begin{align}
    \Toffeff=\frac{1}{G_{tot}}\left[2\,T^{tot,2}_{off,A2} + 2\,T^{tot,2}_{off,B2}\right] .
\end{align}

Furthermore, the total effective noise temperature $\Tneff$ added due to the amplification stages in the FEM, BEM, and DC is computed as
\begin{align}
    \Tneff=& \frac{1}{\rm G_{tot}}\left[\Tnsky-\Tnload\right] .
    \label{tn_effect}
\end{align}

Finally, the total gain term is computed as
\begin{align}
    \rm G_{tot}=\hbemfilterk\,\hdcfilterk\,\GFEMk\,\GBEMk\,\GDCk . 
\end{align}

\subsection{Stokes parameters and Friss equations}


The following set of equations provides the correspondence between the Jones matrix coefficients and the effective 
$h$ coefficients introduced in Eq.~\ref{eq:h_losses}. These relations allow one to trace the connection between the Jones matrix formalism and the Friis equations. Nomenclature is shown in Table \ref{table:components}.

\begin{align}
    &A \rightarrow \sqrt{\hw}\nonumber \\&
    (1+O^s_{x}) \rightarrow \sqrt{ \homtsuno}\nonumber \\&
    (1+O^s_{y}) \rightarrow \sqrt{ \homtsdos}\nonumber \\&
    (1+B^X_2) \rightarrow \sqrt{  \betahybxdos }\nonumber \\&
    (1+B^X_3) \rightarrow \sqrt{  \betahybxtres }\nonumber . \\&
\end{align}


\section{Another interpretation of the offset terms in the instrument model}
\label{app:interpretation}

Equation~\ref{delta_temp_total} constitutes the TMS analytic model. This model contains two type of terms: multiplicative (due to attenuation) and additive (offset terms). In this section, we provide another interpretation of the effective additive terms with a small subset of equations. We begin by isolating the individual contribution of each component in the FEM, following the equations and formalism outlined in Appendix~\ref{A_appendixx}, ignoring those terms with return losses ($R$):
\begin{align}
    & \Delta\Tbw =\Tw \, \Iw \, (1-\SPOw) + \Textcry\, \SPOw\nonumber\\&
    \Delta\Tbirf =\Tirf \, \Iirf \, (1-\SPOirf)+ (\Tcryuno\, \SPOirf)/\hw \nonumber\\&
     \Delta\Tbfhs=(\Tfhs\, \Ifhs)/ (\hw \, \hirf)\nonumber\\&
     \Delta\Tbomtsuno =(\Tomts \, \Iomtsuno)/(\hw \, \hirf \, \hfhs) \nonumber\\&
    \Delta\Tbhybxdos=(\Thybx \, \Ihxdos)/(\hw \, \hirf \, \hfhs \, \homtsuno) \nonumber\\&
    \Delta\Tbfhl=(\Tfhs\, \Ifhs)/ (\hload) \nonumber\\&
     \Delta\Tbomtldos =(\Tomts \, \Iomtsuno)/(\hfhl \, \hload) \nonumber\\&
    \Delta\Tbhybydos=(\Thybx \, \Ihxdos)/(\hfhl \, \homtldos \, \hload)
    \label{eq:tem_gen}
\end{align}
The numerical evaluation of these equations using the values from Table~\ref{table:B1_p_INITIAL} is presented in Table~\ref{table:B1_p_1b}. We find that the window contributes to the 78 per cent ($\approx 7$\,K) to the total output, while the infrared filter and the OMTs contribute approximately 5--7\,\% (around half a Kelvin each).

\begin{table}
\caption{Average excess temperature across the band, $\langle \Delta T^a_c \rangle$, and corresponding standard deviation for the individual opto-mechanical components of the TMS FEM system. The last column provides the fractional contribution (FC) of each component to the total excess temperature.}
\label{table:B1_p_1b} 
\centering                          
\begin{tabular}{l l l}        
\hline                
Component  & $\rm \langle \Delta T^a_c \rangle \ [\rm K]$ & FC [\%]\\ 
\hline 
$\rm \Delta Sky_{chain}$ & 8.445 & \\
\hline 
        $ \rm \Delta_{Window} $   & 7.112 $\pm 0.14$ & 77.95 \\
        $ \rm \Delta_{IRF}$   & $\rm 0.631\pm 0.02$ & 6.92\\
       $ \rm \Delta_{FHs} $   & $\rm 0.131\pm (2.5\times 10^{-3})$ & 1.44\\
       $\rm \Delta_{OMTs} $  & $\rm 0.425\pm (2.7\times 10^{-3})$ & 4.66\\
       $ \rm \Delta_{Hyb\,(X)}$    & $\rm 0.147\pm (3.8\times 10^{-3}$) & 1.61\\
       
       \hline
       \hline
       
$\rm \Delta Load_{chain}$  & 0.677  & \\
\hline 
       $ \rm \Delta_{FHl}$  &  $\rm 0.126\pm (2.3\times 10^{-3})$&1.38 \\
       $\rm \Delta_{OMTl}$  & $\rm 0.409\pm (2.3\times 10^{-3}$) & 4.48\\
       $ \rm \Delta_{Hyb\,(Y)}$  & $\rm 0.142\pm (3.6\times 10^{-3}$)& 1.56\\
       
       \hline
       \hline
       $\rm \Delta Sky_{chain}+ \Delta Load_{chain} $   & 9.123& 100 \\
       \hline
\end{tabular}
\end{table}

It is important to note that the total offset obtained here (9.123\,K; Table~\ref{table:B1_p_1b}) may initially appear inconsistent with the average offset shown in Figure~\ref{relative_ALL_1}. However, the two numbers are fully compatible once the return losses and gains of the individual components are taken into account. From Figure~\ref{betas_offset_t}, we have that the mean losses in the sky chain are $0.75 \pm 0.01$, while in the load chain they are $0.78 \pm 0.01$. When these factors are applied, we recover the expected effective average offset, as reported in Table~\ref{table:beta_results}.

Thus, with this limited set of equations,  and using the effective attenuations, we can reconstruct the same average output signal which can be obtained with the complete  equation~\ref{delta_temp_total} as:
\begin{equation}
\Delta T = (0.75\times 8)-(0.78\times 8)+7.14 + 0= 6.9\,K,
\end{equation}
where the $7.14$\,K is taken from Figure~\ref{LOSS_T_bem_DC_exc} which is the total effective offset temperature of the systems $\Toffeff$.

\begin{table}
\caption{Band-averaged excess temperature after correction for average losses. Sky and load values taken from Table~\ref{table:B1_p_1b}.} 
\label{table:beta_results} 
\centering                    
\begin{tabular}{lc}       
\hline           
       Signal&  $\Delta T$ [K]\\
       \hline  
        $\rm \Delta Sky_{chain}\times\betaskyEffective $   & 6.373 \\
        $\rm \Delta Load_{chain}\times \betaloadEffective$    & 0.530 \\
        \hline
        \hline
        Total    & 6.903\\
        \hline
        \end{tabular}
\end{table}

%

\section{Response due to unbalanced low-noise amplifiers}
\label{C_appendix}

\begin{figure}[h]
   \centering
   \includegraphics[width=9cm]{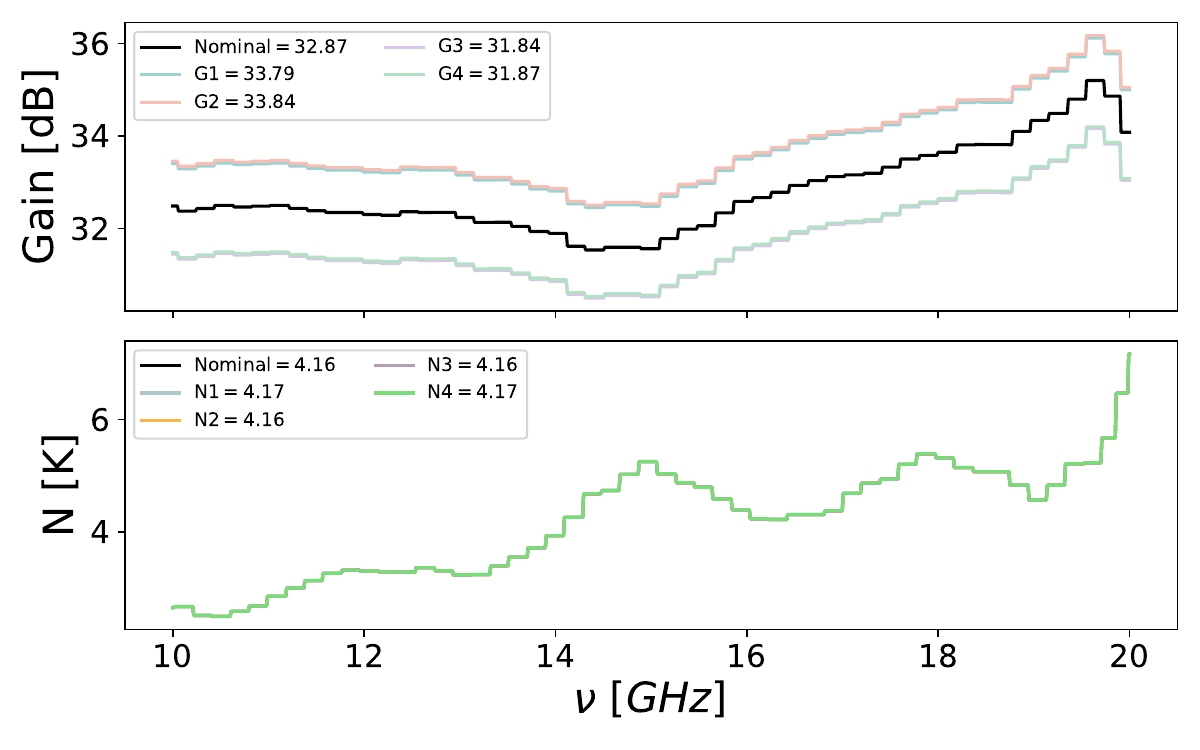}
      \caption{ Definition of case~1, illustrating an imbalance among the four LNAs. Top: Gain variations adopted for the four amplifiers. The nominal (reference) LNA gains from the manufacturer data sheet is shown in black. In this case, the X branch gains ($\rm G_1$ and $\rm G_2$) exhibit an average gain offset of $+1$\,dB, with a relative difference of 0.05\,dB between them. The Y-branch amplifiers ($\rm G_3$ and $\rm G_4$) show an average offset of $-1$\,dB, with a relative difference of 0.03\,dB. This configuration corresponds to a maximum mismatch scenario among the four amplifiers.
      Bottom: Amplitudes of the noise temperature fluctuations, as specified in Table~\ref{tab:amp_appC}.
              }
    \label{amplifiers_amplitude_1}
\end{figure}

\begin{figure}[h]
   \centering
   \includegraphics[width=9cm]{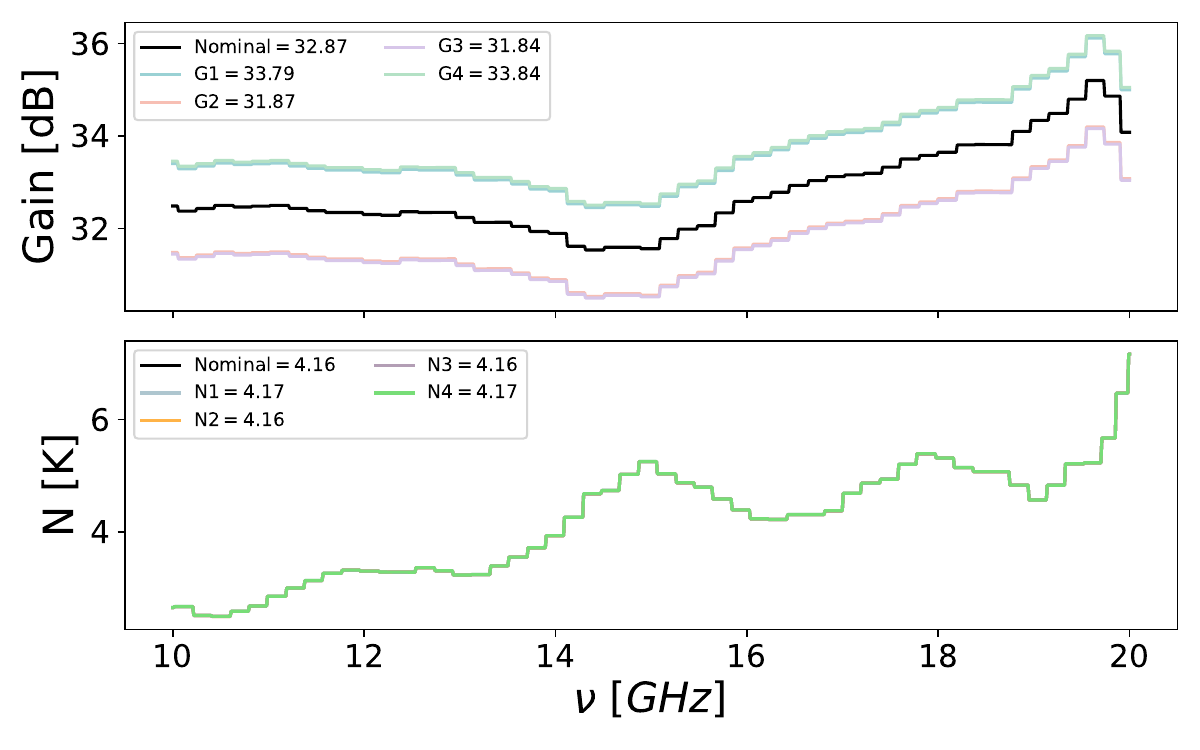}
      \caption{Similar to Fig.~\ref{amplifiers_amplitude_1}, but for the case~2. Here, the proposed variations in $\rm G_1$ and $\rm G_2$ have opposite signs, as well as for $\rm G_3$ and $\rm G_4$.
              }
    \label{amplifiers_amplitude_2}
\end{figure}

\begin{figure}[h]
   \centering
   \includegraphics[width=9cm]{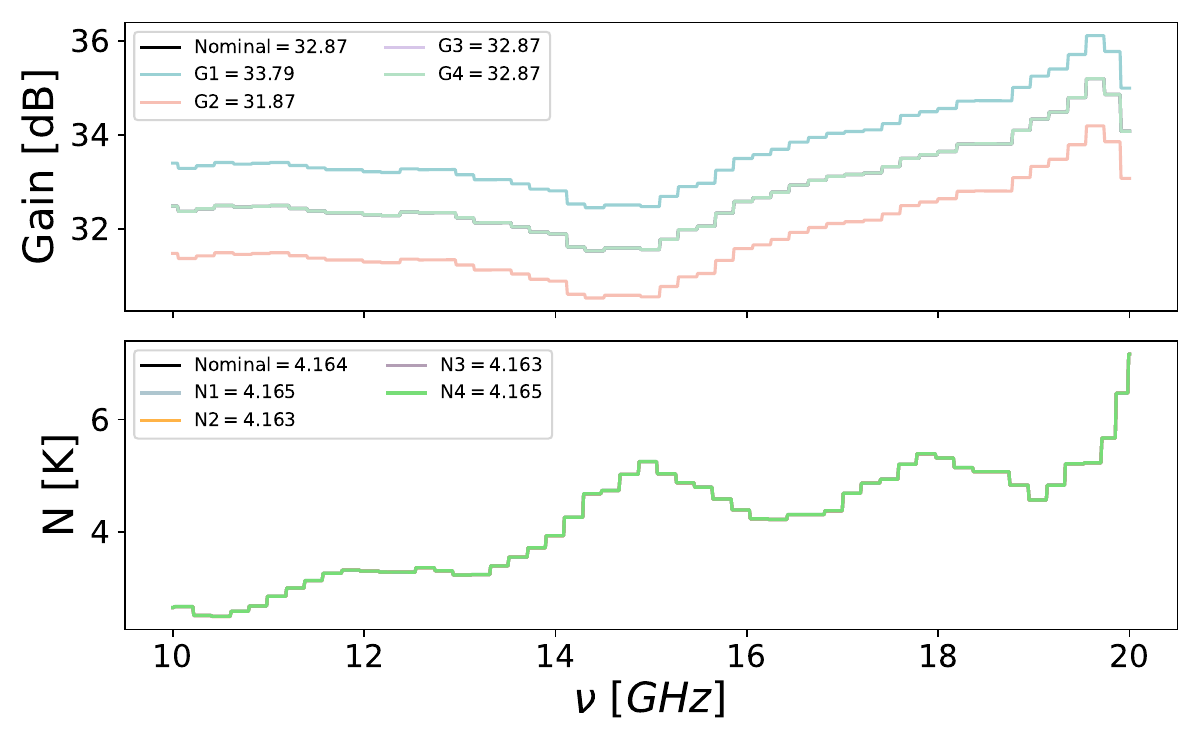}
      \caption{ Similar to Fig.~\ref{amplifiers_amplitude_1}, but for case~3. Here, only changes (with opposite signs) are introduced in $\rm G_1$ and $\rm G_2$, while $\rm G_3$ and $\rm G_4$ are kept at their nominal gains. 
              }
    \label{amplifiers_shifting}
   \end{figure}

Here we define the three cases to illustrate the TMS intensity response when variations of the LNA properties are present with respect to their common (i.e. balanced) and nominal gain and noise temperatures. Table~\ref{tab:amp_appC} summarizes the parameters adopted in each of the three cases, and Figs.~\ref{amplifiers_amplitude_1}, \ref{amplifiers_amplitude_2} and \ref{amplifiers_shifting} depict those values.

\begin{table}
\centering
\tiny
\begin{tabular}{c|cccc|cc}
\hline
Case & $\Delta G_1$ &  $\Delta G_2$ & $\Delta G_3$ & $\Delta G_4$ & $\Delta N_{1,4}$ & $\Delta N_{2,3}$\\
& (dB) &  (dB) & (dB) & (dB) & (mK) & (mK) \\
\hline
1  & $+0.915$ & $+0.969$ & $-1.038$ & $-1.0037$ & +1 & -1\\
2  & $+0.915$ & $-1.004$ & $-1.038$ & $+0.969$ & +1 & -1\\
3  & $+0.915$ & $-1.004$ & $0$ & $0$ & +1 & -1\\
\hline
\end{tabular}
\caption{Definition of the cases used to simulate non-ideal and unbalanced low-noise amplifiers. For each case, we show the gain variations adopted in each of the four LNAs with respect to their nominal response, as well as the change of the noise temperature. }
\label{tab:amp_appC}
\end{table}

\begin{figure}
    \centering
    \includegraphics[width=\columnwidth]{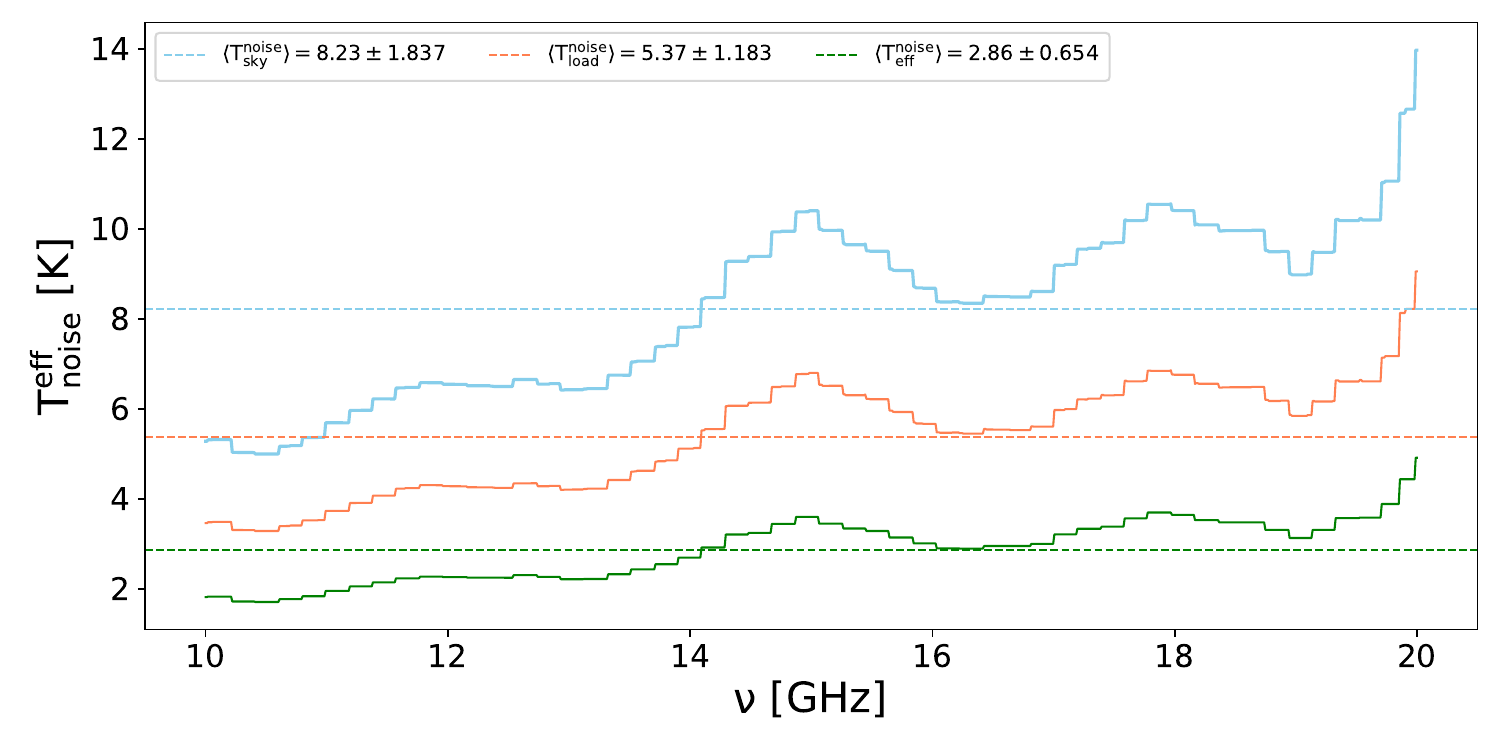}
    \includegraphics[width=\columnwidth]{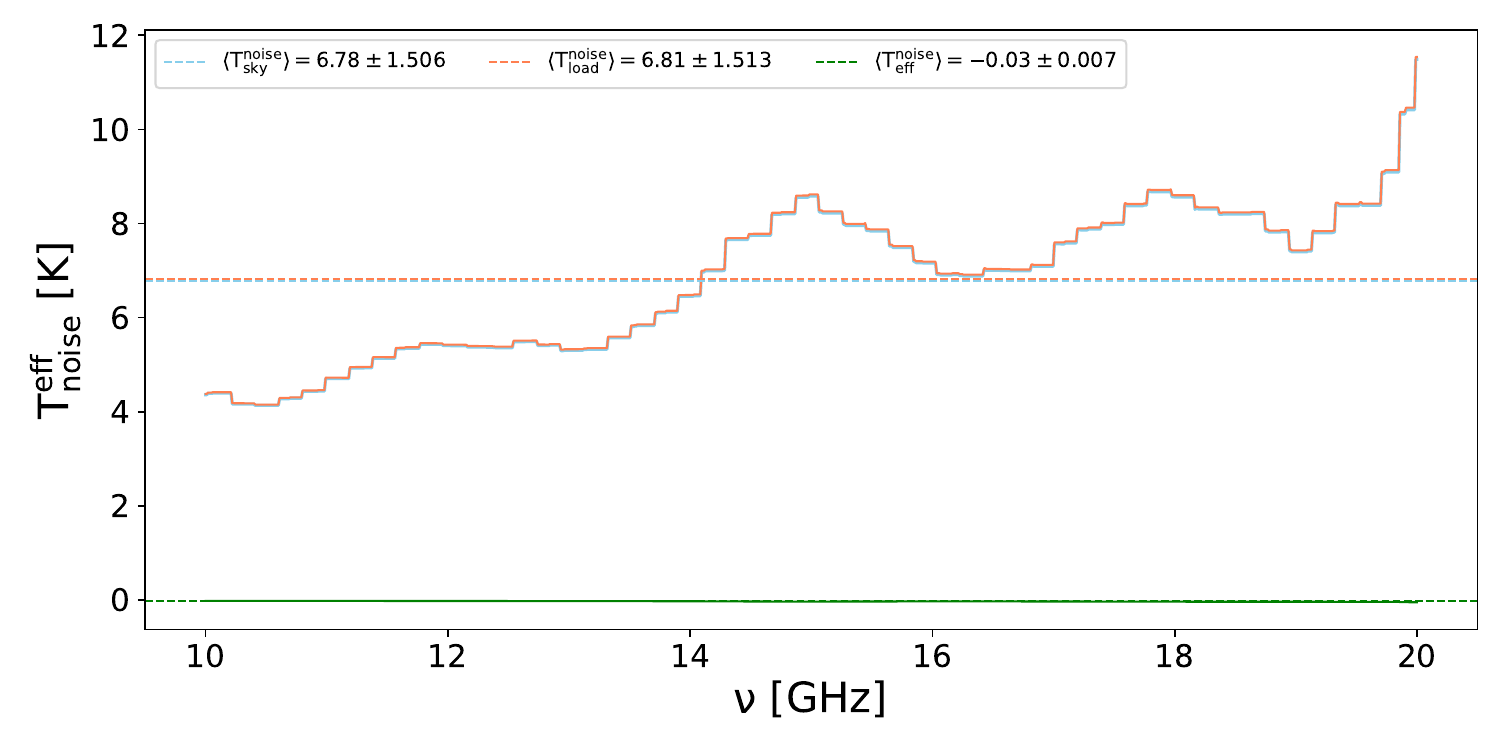}
    \includegraphics[width=\columnwidth]{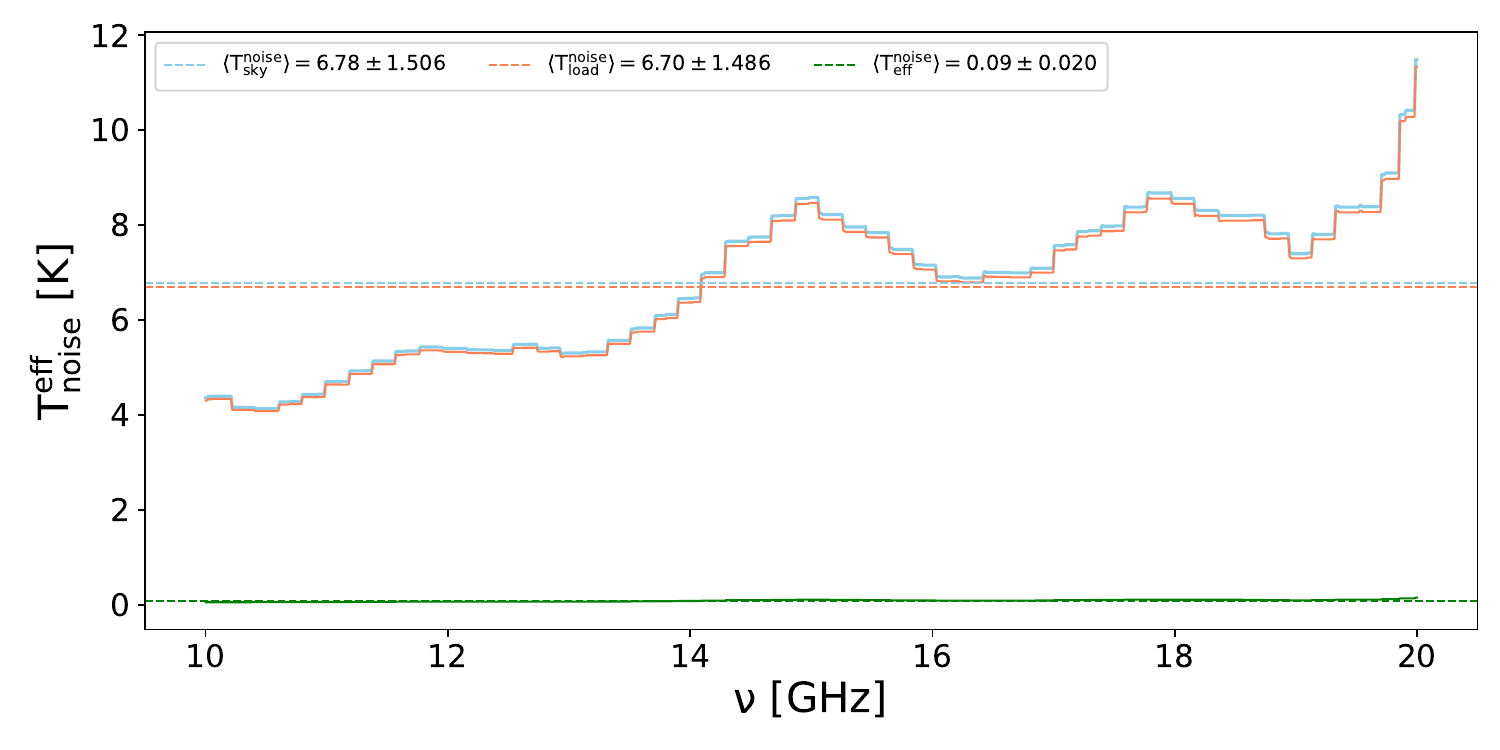}
    \caption{Noise temperature term, $T_{\rm noise}^{\rm eff}$, for the three cases with LNA imbalance considered in the text. Top: case 1. Middle: case 2. Bottom: case 3. }
    \label{fig:Tnoise_var_res}
\end{figure}

Figure~\ref{Ampli_delta_T_sub_delay_var} in the main text presented the TMS intensity response for the three cases described in this appendix. For completeness, Fig.~\ref{fig:Tnoise_var_res}
displays only the noise temperature terms ($T_{\rm noise}^{\rm eff}$) for the same three cases. As discussed in the main text, a significant contribution to the noise term only appears in the case 1 (maximum imbalance among the amplifiers), while in the other two cases, the effective noise terms are almost zero.

\end{appendix}

%
%
\end{document}